\documentclass[final,3p,authoryear]{elsarticle}
\usepackage[T1]{fontenc}
\usepackage{lmodern}
\usepackage{amsmath,amssymb,amsthm,mathtools}
\usepackage{graphicx,booktabs,array,threeparttable,multirow}
\usepackage{microtype}
\usepackage{tikz}
\usetikzlibrary{arrows.meta,calc,decorations.pathmorphing,positioning}
\usepackage{float}
\usepackage{caption}
\usepackage[hidelinks]{hyperref}

\journal{Computer Methods in Applied Mechanics and Engineering}
\newtheorem{theorem}{Theorem}
\newtheorem{proposition}{Proposition}
\theoremstyle{remark}
\newtheorem{remark}{Remark}

\newcommand{\A}{\mathbf{A}}
\newcommand{\K}{\mathbf{K}}
\newcommand{\uvec}{\mathbf{u}}
\newcommand{\fvec}{\mathbf{f}}
\newcommand{\gvec}{\mathbf{g}}
\newcommand{\F}{\mathbf{F}}
\newcommand{\Id}{\mathbf{I}}
\newcommand{\V}{\mathbf{V}}
\newcommand{\J}{\mathbf{J}}
\newcommand{\W}{\mathbf{W}}
\newcommand{\G}{\mathbf{G}}
\newcommand{\D}{\mathbf{D}}
\newcommand{\Pm}{\mathbf{P}}
\newcommand{\Cj}{\boldsymbol{\Delta}}

\newcommand{\R}{\mathbf{R}}
\newcommand{\M}{\mathbf{M}}
\newcommand{\Hf}{\mathcal{H}}
\newcommand{\Th}{\mathcal{T}}
\newcommand{\Un}{\mathcal{U}}
\newcommand{\Pa}[1]{\mathsf{#1}}
\newcommand{\norm}[1]{\lVert #1\rVert}

\begin{document}
\begin{frontmatter}

\title{A quantum-compatible assembly-free finite element formulation for solids}

\author[inst1]{Annwesha Dutta}
\author[inst2]{Saikat Sarkar\corref{cor}}
\cortext[cor]{Corresponding author. E-mail: saikat@iitd.ac.in}
\address[inst1]{School of Computational and Integrative Studies, Jawaharlal Nehru University, New Delhi, India}
\address[inst2]{Computational Solid Mechanics Lab., Structural Engineering Group, Department of Civil \& Environmental Engineering, Indian Institute of Technology Delhi, New Delhi, India}

\begin{abstract}

Conventional finite element (FE) analysis assembles a global stiffness matrix, which is a representation suitable for classical computation due to its sparsity. However, the cost of quantum linear-system methods depends not only on matrix sparsity and conditioning but also on the structure and complexity of the operator representation. In this work, we ask how FE formulations can be designed to suit quantum algorithms and in this pursuit we develop an assembly-free formulation of solids that retains element-local displacement degrees of freedom and enforces compatibility and essential boundary conditions through penalization. On uniform Cartesian grids, the Pauli expansion contains $\Theta(dN^{1/d})$ strings for $N$ encoded degrees of freedom in $d$ dimensions, compared with $\Theta(N)$ for assembled operators in standard binary global-node encodings. Directional exchange permutations yield at most 13 and 81 structured unitaries for clamped bilinear quadrilateral (Q4) and trilinear hexahedral (H8) discretizations, respectively, independent of refinement, and at most 23 for dyadically masked Q4 domains given the mask predicate. Commuting exchanges identify the conforming subspace and diagonalize the penalty spring stiffness through a node-local Walsh–Hadamard transformation. Exact spring-mode scaling combined with Bramble–Pasciak–Xu preconditioning yields mesh-independent positive-spectrum conditioning on complete Cartesian hierarchies, uniformly for sufficiently stiff penalties. For one-face-clamped hierarchies, the rectangular preconditioned factor admits a block encoding with normalization $O(\sqrt{\log m})$ for $m$ elements per direction. A complementary-energy argument establishes an asymptotically sharp compliance penalty-error bound. Tests on two- and three-dimensional cantilevers, L-shaped and perforated brackets, curved quarter-annular brackets, and graded and distorted meshes demonstrate the superior performance of the proposed formulation. This lays a foundation for quantum-compatible analysis of large-scale structural systems, complex geometries and heterogeneous materials.

\end{abstract}

\begin{keyword}
Finite element method \sep Solid mechanics \sep Assembly-free formulation \sep Penalty
method \sep Quantum computing \sep BPX preconditioning
\end{keyword}

\end{frontmatter}

\section{Introduction}\label{sec:intro}

The finite element method (FEM) is routinely used to compute displacements and stresses in engineering structures.
In a conventional analysis, the stiffness matrices of individual elements are assembled to obtain the global equilibrium equations,
\begin{equation}\label{eq:standard}
\K\uvec=\fvec,\qquad \K=\sum_{e}\Pm_e^{\mathsf T}\K_e\Pm_e ,
\end{equation}
in which the element stiffness matrices $\K_e$ are added into a global matrix through the
connectivity matrices $\Pm_e$ of the direct stiffness method \citep{Hughes2000}. Here $\K$ is the assembled stiffness matrix, $\uvec$ is the nodal displacement vector and $\fvec$ is the applied load. In classical computing, this formulation is well established. Quantum linear-system
algorithms can, in principle, solve a sparse system of dimension $N$ with a number of qubits
that grows only with $\log_2 N$
\citep{HHL2009,Childs2017,Gilyen2019,Costa2022}. Whether this becomes useful for engineering
analysis depends on four conditions that must be checked problem by problem: the matrix must
be made available to the quantum computer efficiently, its condition number must be
controlled, the load vector must be prepared, and the quantity of interest must be read out
\citep{Aaronson2015,Montanaro2016}. Computational mechanics has begun to examine these
conditions for homogenization \citep{Liu2024,Balazi2026,Wang2026}, boundary-value problems
\citep{Febrianto2026}, FE analysis on hybrid
hardware \citep{Arora2025}, complexity and preconditioning
\citep{Montanaro2016,Deiml2026,Alkadri2025,Fressart2026}, explicit oracles for elastic
structures \citep{Danz2026}, topology optimization \citep{Holscher2026}, immersed boundaries
\citep{Perera2026}, structural vibration \citep{LiuY2024} and hyperelasticity
\citep{Kouskiya2026}.

Many quantum block-encoding constructions
access a matrix as a weighted sum of simple reversible operations, i.e.\ length-preserving
(unitary) matrices such as permutations and reflections (a \emph{linear combination of
unitaries}, Section~\ref{sec:quantum}), and the cost of the encoding grows with the number of terms and with
the sum of their weights \citep{ChildsWiebe2012,LowChuang2019}. For an assembled FE matrix
written in the standard Pauli basis on uniform Cartesian grids with standard binary node
indexing, the expansion has $\Theta(N)$ terms (proved in the Supplementary Material; this
is not a lower bound for other encodings), and the number was found to grow in proportion to $N$ on
the curved meshes; fast algorithms compute this expansion
\citep{Hantzko2024,Koska2024,Georges2025,Philo2026} but cannot shorten it, because the
expansion of a given matrix is unique. Decompositions with a mesh-independent number of terms
are known for finite-difference Laplacians
\citep{Sturm2025,Kharazi2025,Hogancamp2026,Mahmud2026}, scalar
finite elements \citep{Alkadri2025}, periodic plane-stress elasticity \citep{Suresh2026} and
voxel meshes \citep{Wu2026}; ladder-operator constructions with $O(\log n)$ terms for $n$ grid points per direction are
available for Dirichlet Laplacians \citep{LiuHL2021}, related constructions exist for structured
sparse matrices \citep{Gnanasekaran2026}, and sparse-access and arithmetic constructions provide further
options \citep{Camps2024,Sunderhauf2024,Pechan2026,Yano2026,Danz2026}.

Most existing approaches retain the conforming nodal unknowns of Eq.~\eqref{eq:standard} and
concentrate on representing the resulting operator, sometimes using element-local auxiliary
structures \citep{Alkadri2025,Deiml2026,SatoKondo2026}. For
supported vector elasticity we are not aware of a construction that provides both a
mesh-independent description and a mesh-independent condition number. The compact descriptions are
periodic \citep{Suresh2026}, scalar \citep{Alkadri2025,Perera2026} or intended for variational
solvers without an analysis of the conditioning \citep{Wu2026}; the end-to-end algorithm of
\citet{Holscher2026} for plane-stress compliance uses one term per element and no preconditioner;
\citet{SatoKondo2026} block-encode the assembled plane-stress matrix through element square-root
factors and an element restriction operator; and the condition number, which grows as $h^{-2}$
with the element size $h$, has so far been controlled within the quantum algorithm for scalar
elliptic problems, by Bramble--Pasciak--Xu (BPX) multilevel preconditioning
\citep{BramblePasciakXu1990,Deiml2026,Fressart2026}.

The present work takes a different starting point. The assembled form of
Eq.~\eqref{eq:standard} is one representation of the discrete equilibrium equations, particularly
suited to classical sparse solvers: the direct stiffness method reduces the number of degrees of
freedom by accumulating the contributions of neighbouring elements in shared nodal entries. In
doing so it combines three distinct ingredients in one matrix: the element physics ($\K_e$), the
mesh topology (which elements meet at which node) and the global numbering of the nodes. The
resulting matrix is sparse, but sparsity alone does not make a matrix inexpensive to represent on
a quantum computer. Instead of asking how $\K$ can be encoded, we therefore ask which finite
element formulation of the same mechanics a quantum computer can represent most economically,
and we keep these ingredients separate. In the present work, each element retains its own
copy of every nodal displacement, and compatibility between the copies, together with the clamped
boundary condition, is enforced by stiff penalty springs.

\begin{figure}[!htbp]
\centering
\resizebox{0.95\textwidth}{!}{%
\begin{tikzpicture}[font=\small,
  blk/.style={draw,rounded corners=2pt,align=center,minimum height=2.7cm,text width=3.9cm,inner sep=4pt},
  arr/.style={-{Stealth[length=2.2mm]},thick}]
\node[blk] (a) at (0,0) {Assembly-free formulation:\\each element retains\\its own nodal displacements,\\coupled by penalty springs\\[2pt]{\footnotesize\itshape Section~\ref{sec:formulation}}};
\node[blk] (b) at (5,0) {Fixed element stiffness\\+ one exchange per direction\\+ one support reflection:\\13 (Q4) / 81 (H8) terms\\[2pt]{\footnotesize\itshape Section~\ref{sec:structure}}};
\node[blk] (c) at (10,0) {Copies at a node split into\\one physical mode\\+ artificial spring modes\\(sum/difference transform)\\[2pt]{\footnotesize\itshape Section~\ref{sec:cw}}};
\node[blk] (d) at (0,-3.7) {Spring modes scaled\\exactly; BPX on the\\physical displacements:\\$\kappa_+$ bounded\\[2pt]{\footnotesize\itshape Section~\ref{sec:uniform-preconditioning}}};
\node[blk] (e) at (5,-3.7) {Numerical verification:\\counts, penalty error,\\conditioning, geometry\\[2pt]{\footnotesize\itshape Section~\ref{sec:numerics}}};
\node[blk] (f) at (10,-3.7) {Quantum compatibility:\\factor $\Hf^{\mathsf T}\Hf=\F^{\mathsf T}\A\F$,\\normalization $O(\sqrt{\log m})$\\[2pt]{\footnotesize\itshape Section~\ref{sec:circuit-resources}}};
\draw[arr] (a) -- (b); \draw[arr] (b) -- (c);
\draw[arr] (c.south) -- ++(0,-0.5) -| (d.north);
\draw[arr] (d) -- (e); \draw[arr] (e) -- (f);
\end{tikzpicture}}
\caption{Structure of the approach. Retaining element copies separates the fixed element
stiffness from the mesh topology, which becomes one exchange operation per coordinate
direction. Because the copies remain explicit, a local sum/difference transformation splits
them into the physical displacement and artificial spring modes; the spring modes are scaled
exactly, and only the physical part requires multilevel (BPX) preconditioning. The exchanges and the
support reflection commute, and the sum/difference transformation is their common eigenbasis
(Proposition~\ref{prop:algebra}); their weighted linear combination defines the spring stiffness. Section~\ref{sec:quantum} summarizes the quantum ingredients and
Section~\ref{sec:discussion} discusses scope and limitations.}
\label{fig:overview}
\end{figure}
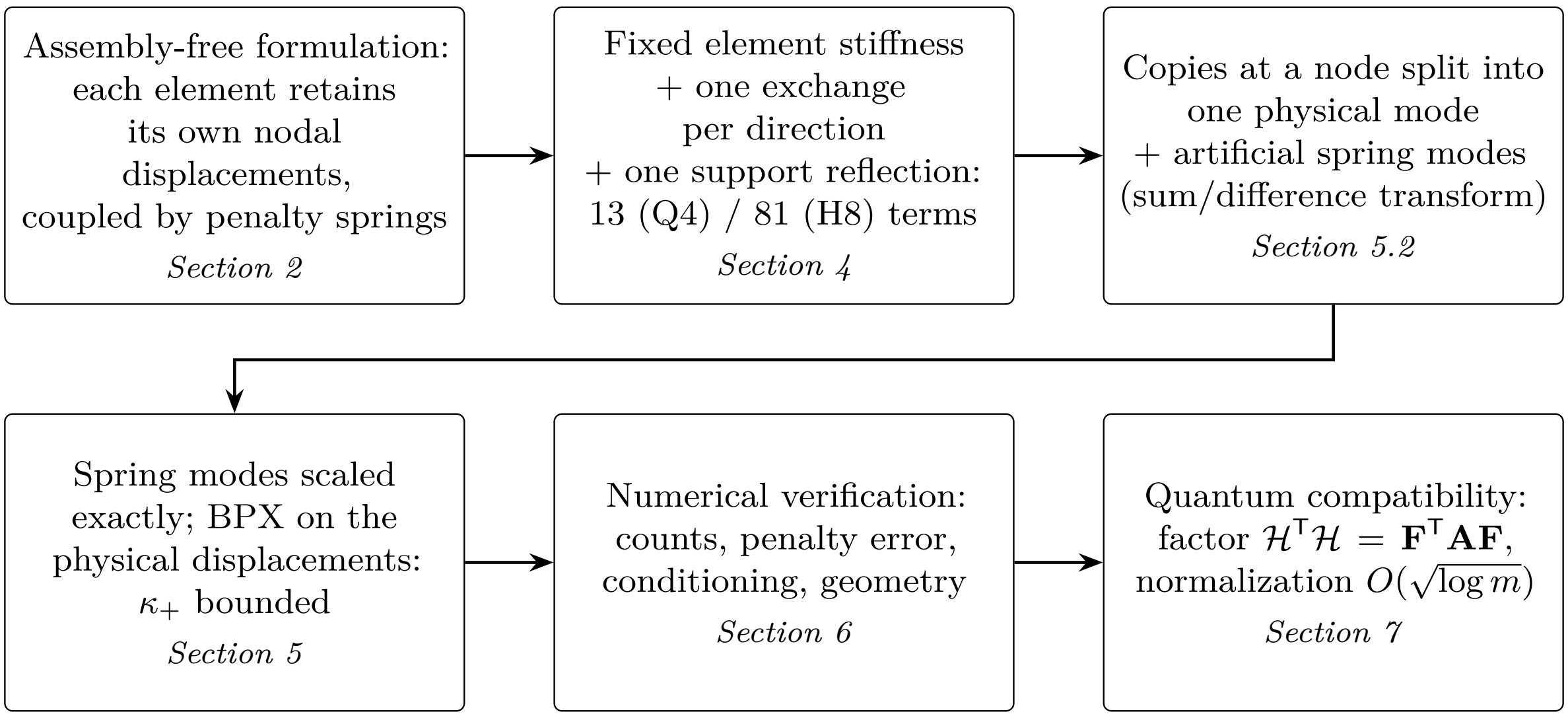

The formulation thus \emph{deliberately increases the number of degrees of freedom in order to
reduce the complexity of the operator}. For conventional classical sparse solvers, this generally
 increases storage requirements. On a quantum computer, this leads to 
only two additional qubits for Q4 and three for H8. In return, the formulation achieves massive advantages as would be demonstrated in due course in the manuscript. The resulting sequence of simplifications is shown in Fig.~\ref{fig:overview}, which also serves as the outline
of the paper.

The main contributions
are:
\begin{enumerate}
\item an assembly-free elasticity formulation in which element-wise copies of the nodal
displacements are the unknowns; on uniform meshes each directional coupling is a single exchange, so
that clamped Q4 and H8 penalty operators are sums of at most 13 and 81 structured unitaries,
independent of mesh refinement (at most 23 for dyadically masked Q4 domains, given the mask predicate), with
$\Theta(dN^{1/d})$ Pauli strings against a proved $\Theta(N)$ for the assembled operators in the
standard binary global-node encoding on comparable grids;
\item a unified treatment of essential boundary conditions: clamped supports are imposed by the
same springs as compatibility and enter the operator as a single reflection, without elimination
of degrees of freedom, Lagrange multipliers or a separate encoding, and they are covered by the
same conditioning bound;
\item the observation that, on complete Cartesian copy graphs, the exchanges and the support
reflection commute: their weighted linear combination is the spring stiffness, their product with
the element stiffness the conforming stiffness on the compatible subspace (diagonally congruent to the assembled matrix), and
their common eigenbasis, a node-local sum/difference transformation, diagonalizes the spring
stiffness exactly; scaling the spring modes and applying BPX preconditioning and Korn's inequality
to the physical displacements gives a mesh-independent condition-number bound, uniform for
sufficiently stiff springs;
\item a computable two-sided bound on the penalty error in the compliance, obtained from the
conforming solution by a complementary-energy argument and asymptotically sharp;
\item a factorization of the preconditioned operator, $\Hf^{\mathsf T}\Hf=\F^{\mathsf T}\A\F$, whose
block encoding has normalization $O(\sqrt{\log m})$, which establishes quantum compatibility at the
operator level without committing the formulation to a particular solver;
\item numerical investigations of two- and three-dimensional benchmarks, including masked, graded
and curved meshes, together with comparisons with the assembled operator and with the projected
conforming form that identify where the penalty form is advantageous and where it is not.
\end{enumerate}
Retaining element-local degrees of freedom is not itself new: element-by-element solution
algorithms \citep{HughesLevitWinget1983}, matrix-free cell-based operator evaluation
\citep{KronbichlerKormann2012} and tearing-type domain decomposition \citep{FarhatRoux1991} all work
with element- or subdomain-local data, and interior-penalty methods couple discontinuous element
fields by penalties \citep{Arnold2002}. The novelty here lies in the particular exchange-based
representation of the couplings and in its consequences for a quantum-compatible operator
structure and for preconditioning. To the best of our knowledge, existing quantum finite element constructions have not combined an
element-copy penalty formulation for supported vector elasticity with an exact spring-mode
decomposition and a mesh-independent preconditioning bound. The closest alternatives
(Table~\ref{tab:related}) differ as follows: element-local factorizations
\citep{Alkadri2025,Deiml2026,SatoKondo2026} keep the conforming nodal unknowns; compact descriptions
of elasticity are periodic \citep{Suresh2026} or intended for variational solvers \citep{Wu2026},
and openings by masks have been treated for scalar problems \citep{Perera2026}; quantum multilevel
preconditioners have been applied to scalar problems \citep{Deiml2026,Fressart2026}, and our bound
reuses BPX theory and Korn's inequality, the new ingredient being the exact, penalty-robust treatment
of the spring modes; and the elasticity algorithms of \citet{Holscher2026} and
\citet{SatoKondo2026}, which carry the solution through to the compliance or strain energy, work
with the assembled matrix and, to the best of our knowledge, without a mesh-independent conditioning
bound.

The unitary counts describe the operator, not its gate cost: each exchange $\V_i$ requires $O(\log
m)$ index arithmetic, a mask requires a predicate circuit, and the short penalty linear combination
is not the multilevel factor of Section~\ref{sec:circuit-resources} (Section~\ref{sec:broken-role}).
The formulation exchanges dimension for structure, which can be advantageous for quantum operator
representations but not for classical storage; for periodic problems assembled operators can also be very compact
\citep{Suresh2026}. The compactness of the penalty representation does not necessarily translate
into an advantage for every quantum solution algorithm: quantum linear-system methods based on block
encodings may benefit from projecting onto the conforming subspace, which avoids the penalty error
and reduces the normalization (Section~\ref{sec:projection}). The principal contribution of the
present work is a finite element formulation that exposes operator structure and permits an exact
treatment of the artificial compatibility modes, not the optimization of a particular quantum
solution algorithm; no quantum solver is implemented or costed here, and the other limitations are
stated in Section~\ref{sec:limits}.

\section{Assembly-free discretization}\label{sec:formulation}

\subsection{Linear-elasticity model}
Let $\Omega\subset\mathbb R^d$ ($d=2$ or $3$) be an elastic body with zero displacement on a
clamped boundary $\Gamma_D$ of positive measure. For small strains and isotropic material, the
equilibrium displacement minimizes the potential energy
\begin{equation}\label{eq:continuum}
\Pi(u)=\frac12\int_\Omega\boldsymbol\varepsilon(u):\mathbb C:\boldsymbol\varepsilon(u)
\,\mathrm d\Omega-W(u),\qquad
\boldsymbol\varepsilon(u)=\tfrac12\bigl(\nabla u+\nabla u^{\mathsf T}\bigr),
\end{equation}
where $\mathbb C$ is defined by Young's modulus $E$ and Poisson's ratio $\nu$ (plane stress in
two dimensions, unit thickness unless stated otherwise) and $W$ is the work of the applied
loads. We use four-node bilinear quadrilaterals (Q4) and eight-node trilinear hexahedra (H8)
with the usual strain--displacement and constitutive matrices and full Gauss integration. The
element stiffness matrices $\K_e$ are exactly those of a standard FE code. A summary of the mathematical symbols and their definitions is provided in Appendix A.

\subsection{Element-local degrees of freedom coupled by penalty springs}
\begin{figure}[!htbp]
\centering
\begin{tikzpicture}[scale=0.83, every node/.style={font=\small}]
 \fill[blue!9] (0,0) rectangle (3.0,2.3); \fill[orange!11] (4.1,0) rectangle (7.1,2.3);
 \draw[thick] (0,0) rectangle (3.0,2.3); \draw[thick] (4.1,0) rectangle (7.1,2.3);
 \node at (1.5,1.35) {element $e$}; \node at (5.6,1.35) {element $e+1$};
 \filldraw[blue!65!black] (3,1.15) circle (3pt); \filldraw[orange!80!black] (4.1,1.15) circle (3pt);
 \draw[thick,decoration={coil,segment length=4pt,amplitude=3pt},decorate] (3.08,1.15)--(4.02,1.15);
 \draw[->] (2.93,1.05)--(2.3,.48) node[below] {$u_p$}; \draw[->] (4.2,1.05)--(4.85,.48) node[below] {$u_q$};
 \node[above] at (3.55,2.44) {penalty spring, stiffness $\alpha$};
 \node[align=center] at (3.55,-.42) {two copies of one physical displacement};
 \fill[blue!9] (9.9,0) rectangle (12.3,2.3); \draw[thick] (9.9,0) rectangle (12.3,2.3);
 \node at (11.1,1.35) {element $e'$};
 \filldraw[blue!65!black] (9.9,1.15) circle (3pt);
 \draw[thick,decoration={coil,segment length=4pt,amplitude=3pt},decorate] (9.82,1.15)--(8.95,1.15);
 \draw[thick] (8.9,0.6)--(8.9,1.7);
 \foreach \y in {0.65,0.85,...,1.65} \draw (8.9,\y)--(8.7,\y-0.15);
 \node[above] at (9.4,2.44) {support spring, stiffness $\beta$};
 \node at (10.4,-.42) {clamped copy};
\end{tikzpicture}
\caption{Mechanical interpretation of the assembly-free formulation for one displacement component.
Left: the copies $u_p$ and $u_q$ of a shared node are joined by a spring with energy
$\tfrac12\alpha(u_p-u_q)^2$. Right: a copy on the clamped boundary is tied to the ground with
energy $\tfrac12\beta u_p^2$.}
\label{fig:concept}
\end{figure}
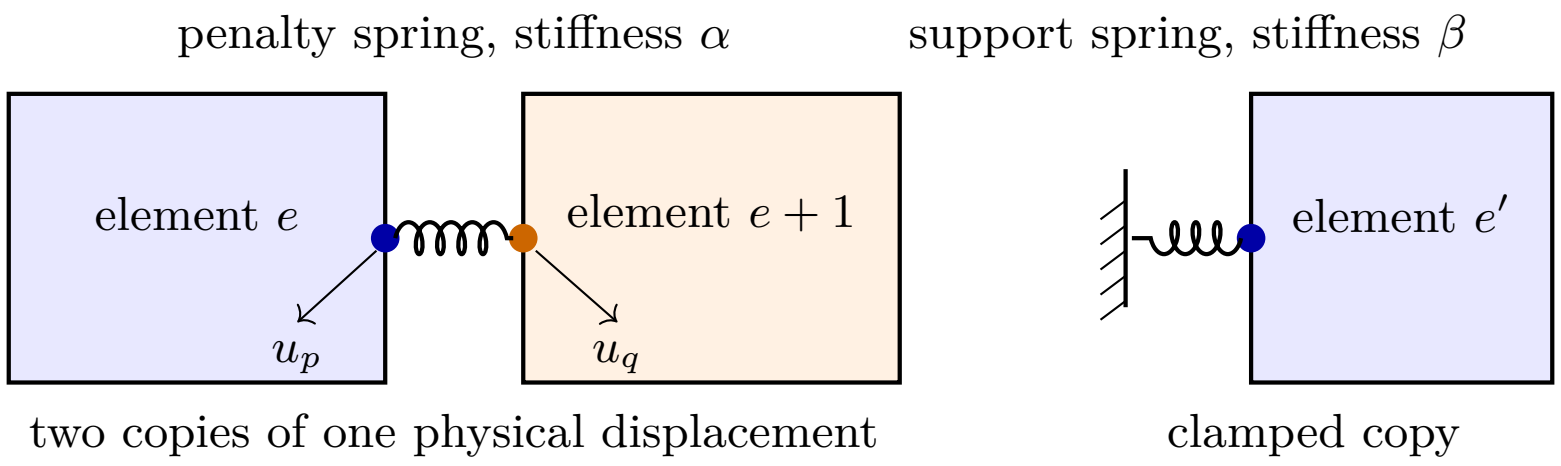

In the standard method two elements that meet at a node share its displacement. Here every
element retains its own \emph{copy} of each nodal displacement component; a node shared by $n$
elements therefore has $n$ copies of each component. Two copies form a \emph{bond}
$(p,q)\in\mathcal B$ if they represent the same nodal component and belong to elements that
share a face (an edge in 2D). Copies at nodes on $\Gamma_D$ form the set $\mathcal D$.
Fig.~\ref{fig:concept} illustrates the mechanical interpretation: each bond is a stiff spring between two
copies, and each clamped copy is tied to the ground by another stiff spring.

The discrete potential energy is
\begin{equation}\label{eq:penalty_energy}
\Pi_h(\uvec)=\frac12\sum_{e}\uvec_e^{\mathsf T}\K_e\uvec_e
+\frac\alpha2\sum_{(p,q)\in\mathcal B}(u_p-u_q)^2
+\frac\beta2\sum_{p\in\mathcal D}u_p^2-\fvec^{\mathsf T}\uvec ,
\end{equation}
and its minimization gives $\A\uvec=\fvec$ with
\begin{equation}\label{eq:split}
\A=\underbrace{\operatorname{blockdiag}(\K_1,\ldots,\K_{n_e})}_{\A_{\rm el}}
+\underbrace{\alpha\,\Cj^{\mathsf T}\Cj}_{\A_{\rm c}}
+\underbrace{\beta\,\Cj_D^{\mathsf T}\Cj_D}_{\A_{\rm bc}} .
\end{equation}
Here each row of the jump matrix $\Cj$ contains the entries $+1$ and $-1$ at the two degrees of
freedom of one bond, so that $\Cj\uvec$ collects the jumps $u_p-u_q$, and $\Cj_D$ selects the
clamped copies. No conforming
global nodal stiffness matrix is assembled; the global operator is applied in block-local and
exchange form: the element part $\A_{\rm el}$ consists of independent element blocks,
connectivity enters only through the spring matrix $\A_{\rm c}$, and supports only through
$\A_{\rm bc}$. Provided that the spring graph connects all copies of each physical node and the
conforming stiffness is positive definite, the minimizer converges to the conforming FE solution
with exactly imposed supports as $\alpha,\beta\to\infty$ \citep{Babuska1973}; at finite penalties the essential boundary
conditions, like compatibility, are enforced only approximately, and exact clamping holds in the
limiting conforming problem. 

\subsection{Rationale for retaining the element-local degrees of freedom}\label{sec:why}
Retaining the copies has one cost and two benefits. The cost is the dimension: a Q4 mesh with
$m\times m$ elements has $8m^2$ copies instead of about $2(m+1)^2$ nodal degrees of freedom, and an H8 mesh $32m^3$
slots (including one padding slot per copy, Section~\ref{sec:encoding}) instead of $3(m+1)^3$.
Under the register conventions of Section~\ref{sec:encoding} and approximately matched mesh
resolutions, the duplication costs two additional qubits for Q4 and three for H8; with exactly
$m=2^a$ elements per direction in both formulations, an assembled binary node register needs
$a+1$ bits per direction for its $m+1$ nodes, and the difference then depends on the padding and
numbering convention.

The first benefit is the separation of physics and topology. All element data are contained in
the block-diagonal $\A_{\rm el}$, which on a uniform mesh is a single $8\times8$ (Q4) matrix, or for H8 the physical $24\times24$
matrix embedded in a $32\times32$ padded block (Section~\ref{sec:encoding}), repeated over the
elements. Which copies must agree is determined by the springs alone; whether an element is
present is determined by the springs together with a small regularization of the void element
blocks (Proposition~\ref{thm:shift}b), without modification of the other element matrices.

The second benefit is that the copies make the artificial modes created by the springs
explicit. Consider two copies $u_1,u_2$ joined by one spring. In terms of the
symmetric (sum) and antisymmetric (difference) coordinates
$u_{\rm s}=(u_1+u_2)/\sqrt2$ and $u_{\rm d}=(u_1-u_2)/\sqrt2$, obtained with the orthogonal sum/difference
(Hadamard) matrix $\mathbf S_2=\frac1{\sqrt2}\bigl[\begin{smallmatrix}1&1\\1&-1\end{smallmatrix}\bigr]$,
\begin{equation}\label{eq:two-copy}
\mathbf S_2\,\alpha\begin{bmatrix}1&-1\\-1&1\end{bmatrix}\mathbf S_2^{\mathsf T}
=\begin{bmatrix}0&0\\0&2\alpha\end{bmatrix}:
\end{equation}
the spring acts only on the difference mode, while the sum mode is the normalized compatible
displacement mode ($\sqrt2$ times the nodal displacement when the two copies agree). Nodes shared by four or eight elements lead to the same structure with a
$4\times4$ or $8\times8$ sum/difference (Walsh--Hadamard) transformation
(Section~\ref{sec:cw}). After assembly these difference modes no longer exist as degrees of freedom. The duplication,
which appears to be the principal drawback of the formulation, is therefore precisely what
enables the preconditioner of Section~\ref{sec:uniform-preconditioning}.

\subsection{Choice of the penalty and penalty error}\label{sec:extrap}
We scale both springs with the element stiffness,
\begin{equation}\label{eq:gamma}
\alpha=\beta=\gamma k_0,\qquad k_0=Eh^{d-2}\ \ (k_0=Et\ \text{in plane stress}),
\end{equation}
and call $\gamma$ the dimensionless penalty. With $E=1$, unit thickness and unit element size,
$\alpha=\beta=\gamma$. Under the assumptions of the penalty expansion in the Supplementary
Material (Proposition~S5: connected copies and a positive definite conforming stiffness), the
solution is analytic in $1/\gamma$ near $1/\gamma=0$,
$\uvec(\gamma)=\uvec_0+\gamma^{-1}\uvec_1+O(\gamma^{-2})$, where $\uvec_0$ is the conforming FE
solution written in copy form (every copy carries its nodal value). The penalty error is
therefore $O(1/\gamma)$ on every mesh; for the compliance it is bounded from above, for every
$\gamma$, by a quantity computable from the conforming solution (Proposition~\ref{prop:penalty-bound}
in Section~\ref{sec:loads}). With the springs scaled as in Eq.~\eqref{eq:gamma} it was
found to depend only weakly on the mesh size (Section~\ref{sec:num-penalty}),
and Richardson
extrapolation from two penalties \citep{Richardson1927} removes its leading term.
Because the preconditioner of Section~\ref{sec:uniform-preconditioning} removes the adverse
effect of stiff springs on the condition number, the penalty can be chosen on the basis of the
required accuracy for the ideal preconditioned model, within the range $\eta<1$ of
Theorem~\ref{thm:full-pre-final} and under its assumptions; the unpreconditioned linear combination remains sensitive to
$\gamma$ through its 1-norm. The expansion, its proof and the cost of extrapolation are given in the
Supplementary Material.

\subsection{Loads and outputs}\label{sec:loads}
Let $\J_0$ map a vector of free nodal displacements to its copies (each copy receives its nodal
value), and let $\mathbf M_{\rm copy}=\J_0^{\mathsf T}\J_0$ be the diagonal matrix that counts the copies of
each node. A nodal load
$\fvec_{\rm phys}$ is shared equally among the copies,
\begin{equation}\label{eq:compatible-load}
\fvec=\J_0\mathbf M_{\rm copy}^{-1}\fvec_{\rm phys},\qquad \J_0^{\mathsf T}\fvec=\fvec_{\rm phys},
\end{equation}
so that it does the same work as the physical load on every compatible displacement field; the
copies must not be loaded as if they were independent nodes. The compliance
$\mathcal C=\fvec^{\mathsf T}\uvec$, i.e.\ twice the stored energy at equilibrium (springs included), is the
output used in
Sections~\ref{sec:num-penalty} and~\ref{sec:circuit-resources}; for a single point force $P$ it
equals $P$ times the average displacement of the copies of the loaded node (the displacement itself
when the node has one copy, as at the loaded corners used below). The penalty error in the
compliance admits a two-sided bound.

\begin{proposition}[Penalty error in the compliance]\label{prop:penalty-bound}
Consider the active physical displacement sector, i.e.\ exclude the decoupled and unloaded H8
padding and void slots. Assume that on this sector the copies of every physical node are connected
by springs, so that the kernel of the spring stiffness $\A_{\rm s}=\A_{\rm c}+\A_{\rm bc}$ coincides
with the conforming copy fields that satisfy the essential boundary conditions,
$\ker\A_{\rm s}=\operatorname{range}\J_0$, and that the conforming stiffness $\K_0$ is positive
definite. Let $\uvec_0=\J_0\K_0^{-1}\fvec_{\rm phys}$ be the conforming solution written in copy form,
$\mathcal C_0=\fvec^{\mathsf T}\uvec_0$ its compliance and $\mathbf r=\A_{\rm el}\uvec_0-\fvec$ its
element-force residual, which collects the interelement force jumps and the support reactions
of the conforming solution, where $\K_0=\J_0^{\mathsf T}\A_{\rm el}\J_0$ is the conforming stiffness on
the free nodal displacements. For all $\alpha,\beta>0$ the compliance $\mathcal C$ of the penalty
model satisfies
\begin{equation}\label{eq:penalty-bound}
0\le\mathcal C-\mathcal C_0\le\mathbf r^{\mathsf T}\A_{\rm s}^+\mathbf r .
\end{equation}
For a nonzero load, so that $\mathcal C_0>0$, and $\alpha=\beta=\gamma k_0$ this reads
$0\le(\mathcal C-\mathcal C_0)/\mathcal C_0\le c_1/\gamma$,
where $\hat\A_{\rm s}=\A_{\rm s}/\gamma$ is independent of $\gamma$ and the coefficient
$c_1=\mathbf r^{\mathsf T}\hat\A_{\rm s}^+\mathbf r/\mathcal C_0$ is dimensionless and does not depend
on $\gamma$; $c_1/\gamma$ is also the leading term
of the penalty expansion, so the bound is asymptotically sharp.
\end{proposition}
\begin{proof}
The compliance is $\mathcal C=\max_{\mathbf v}(2\fvec^{\mathsf T}\mathbf v-\mathbf
v^{\mathsf T}\A\mathbf v)$; restricting $\mathbf v$ to the conforming fields, on which
$\A_{\rm s}$ vanishes, gives $\mathcal C_0$ and hence the lower bound. For the upper bound write
$\A=\mathbf G^{\mathsf T}\mathbf G$ with $\mathbf G=[\R;\mathbf S]$, $\R^{\mathsf T}\R=\A_{\rm el}$
and $\mathbf S^{\mathsf T}\mathbf S=\A_{\rm s}$. By the principle of complementary energy,
$\mathcal C=\fvec^{\mathsf T}\A^{-1}\fvec=\min\{\norm{\mathbf y}^2:\mathbf G^{\mathsf T}\mathbf
y=\fvec\}$. Since $\J_0^{\mathsf T}\mathbf r=\K_0\K_0^{-1}\fvec_{\rm phys}-\fvec_{\rm phys}=\mathbf 0$,
the residual is orthogonal to $\operatorname{range}\J_0$, which by assumption is the kernel of
$\mathbf S$, so
$\mathbf r$ lies in the range of $\mathbf S^{\mathsf T}$. The field $\mathbf y=(\R\uvec_0,
-(\mathbf S^{\mathsf T})^+\mathbf r)$ is therefore admissible, $\mathbf G^{\mathsf T}\mathbf
y=\A_{\rm el}\uvec_0-\mathbf r=\fvec$, and $\norm{\mathbf y}^2=\uvec_0^{\mathsf
T}\A_{\rm el}\uvec_0+\mathbf r^{\mathsf T}\A_{\rm s}^+\mathbf r=\mathcal C_0+\mathbf r^{\mathsf
T}\A_{\rm s}^+\mathbf r$.
\end{proof}

Mechanically, $\mathbf r$ collects the interelement force jumps of the conforming solution; carried
by the springs, they form a statically admissible force field whose complementary energy bounds the
additional compliance. The connectivity assumption holds on the complete Cartesian copy graphs
of Section~\ref{sec:cw}; it can fail on masked domains, e.g.\ where two elements touch only at a
corner and the face-neighbour springs do not connect their copies of the shared node. On masked
domains the unbonded void slots add further zero spring modes even when the active copy graph is
connected; like the H8 padding sector, they are decoupled from the active physical sector and carry
no load, and they are excluded from the argument. Once the conforming solution is available, the bound is evaluated independently at each physical
node, because the spring stiffness is block diagonal over the nodes. On complete Cartesian copy
graphs the blocks are diagonalized explicitly by the sum/difference transformation of
Section~\ref{sec:cw}; on other connected copy graphs the corresponding small local pseudoinverses
can be computed directly. It is therefore a computable penalty-selection rule, conditional on the conforming solution, not an inexpensive estimator that
avoids solving the elasticity problem; nor does it show that $c_1$ is bounded uniformly under mesh
refinement.

\section{Requirements of quantum algorithms on a finite element operator}\label{sec:quantum}

This section summarizes, in FE terms, the properties of an operator that quantum algorithms
require and that motivate the formulation; readers familiar with block encodings can go directly to
Section~\ref{sec:structure}. Detailed accounts are given by \citet{ChildsWiebe2012},
\citet{LowChuang2019} and \citet{Gilyen2019}. Table~\ref{tab:dictionary} lists the correspondence.

\emph{Matrix access.} Amplitude encoding represents a normalized vector of dimension $N=2^q$ in
$q$ qubits, to which reversible (unitary) operations are applied. A unitary matrix preserves lengths; permutations and
reflections are examples. A matrix that is not unitary, such as a stiffness matrix, is made available through a \emph{block encoding}: a unitary operation whose
upper-left block is $\A/\omega$, with a normalization $\omega\ge\norm{\A}_2$. (The letter
$\alpha$ is common for this normalization in the quantum literature; we use $\omega$ because
$\alpha$ is the penalty.) The most common construction writes the matrix as a weighted sum of
simple unitaries, $\A=\sum_jc_j\mathbf U_j$ (a linear combination of unitaries, LCU), which
gives $\omega=\lambda:=\sum_j|c_j|$. The implementation cost of one call depends principally on
the PREPARE and SELECT circuits, i.e.\ on the number of terms $\Un$, each of which is a branch of
the circuit, and on the cost of each branch. The normalization $\lambda$ does not make a single call
more expensive; it enters the effective inverse condition parameter below and thus the number of
calls an algorithm requires. Pauli strings, tensor products of the
$2\times2$ matrices $\Id$, $\Pa X=\bigl[\begin{smallmatrix}0&1\\1&0\end{smallmatrix}\bigr]$,
$\Pa Y=\bigl[\begin{smallmatrix}0&-i\\i&0\end{smallmatrix}\bigr]$ and
$\Pa Z=\bigl[\begin{smallmatrix}1&0\\0&-1\end{smallmatrix}\bigr]$, are the most common choice of
unitaries ($\Pa X$ is exactly the exchange of two copies in Eq.~\eqref{eq:spring-exchange}); every
matrix has a unique Pauli expansion, whose number of terms we denote by $\Th$.
Permutations and reflections are another choice; they are inexpensive to implement with reversible
integer arithmetic. Neither $\Un$ nor $\Th$ is a gate count or a run time, and neither includes the number of calls
required by a solver.

\emph{Conditioning.} The quantum singular value transformation (QSVT) applies a polynomial
approximation of $x^{-1}$ to a block-encoded matrix \citep{Gilyen2019}. The degree of the
polynomial, i.e.\ the number of calls to the block encoding, grows in proportion to
$\omega\norm{\A^{-1}}_2\log(1/\epsilon)$ for accuracy $\epsilon$; the factor
$\omega\norm{\A^{-1}}_2\ge\kappa(\A)$ plays the role that the condition number plays for the
iteration count of a classical Krylov solver. Preconditioning is therefore as important as for classical iterative solvers.

\emph{Term-wise evaluation.} Variational and energy-based algorithms instead estimate the
operator term by term, so that their cost grows with the number of terms and with their
coefficients \citep{Sato2021,BravoPrieto2023}. The load vector must in all cases be prepared as a
quantum state, and the full displacement field cannot be read out efficiently, so scalar outputs
such as the compliance are the natural results.

Different quantum algorithms therefore benefit from different operator properties: a short linear
combination for term-wise evaluation, and a small normalization together with a bounded condition
number for block-encoding-based solvers. Section~\ref{sec:structure} gives a compact description
of the operator, Section~\ref{sec:uniform-preconditioning} bounds its condition number, and
Section~\ref{sec:circuit-resources} shows that the preconditioned operator admits a factorized
block encoding with slowly growing normalization.

\begin{table}[!htbp]
\centering
\caption{Classical FE concepts and their counterparts in quantum algorithms. The number of terms
is only one component of the cost; the branch circuits, the normalization and the conditioning also
enter.}
\label{tab:dictionary}
\footnotesize
\begin{tabular}{@{}>{\raggedright\arraybackslash}p{0.30\linewidth}>{\raggedright\arraybackslash}p{0.36\linewidth}>{\raggedright\arraybackslash}p{0.28\linewidth}@{}}
\toprule
Classical FE & Quantum counterpart & In this paper\\
\midrule
Stiffness matrix in memory & block encoding with normalization $\omega$ & Sections~\ref{sec:structure}, \ref{sec:direct-block-precond}\\
Matrix-access representation & number of terms $\Un$ (unitaries) or $\Th$ (Pauli strings) & at most 13 / 81 / 23 unitaries\\
Matrix norm & normalization $\omega$, LCU 1-norm $\lambda$ & $\omega=O(\sqrt{\log m})$ for the rectangular preconditioned factor $\Hf$\\
Iteration count $\sim\sqrt\kappa$ or $\kappa$ & QSVT degree $\sim\omega\norm{\A^{-1}}$ & bounded condition number\\
Number of degrees of freedom $N$ & qubits $\log_2N$ & copies cost 2 (Q4) / 3 (H8) qubits (register conventions of Section~\ref{sec:encoding})\\
\bottomrule
\end{tabular}
\end{table}

\section{Compact operator representation on structured meshes}\label{sec:structure}

\subsection{Element-wise numbering of the degrees of freedom}\label{sec:encoding}
\begin{figure}[!htbp]
\centering
\includegraphics{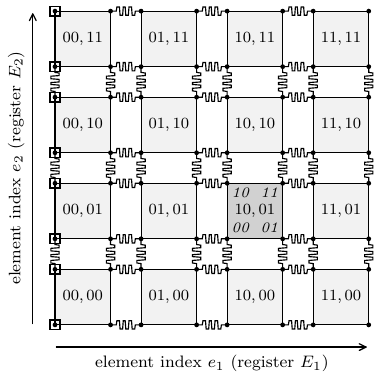}
\caption{Element-by-element numbering of the copies on a $4\times4$ Q4 grid (elements drawn apart).
Each element, labelled by its binary coordinates $e_1,e_2$, owns four copies (dots) of its corner nodes;
the italic labels in the shaded element are the local bits $\ell_2\ell_1$. Zigzag lines are the penalty
springs in directions 1 (horizontal) and 2 (vertical); squares mark the clamped copies on the face
$x_1=0$.}
\label{fig:encoding}
\end{figure}

Consider a uniform mesh with $m=2^a$ elements in each of the $d$ directions. The copies are
numbered element by element: the binary index of element $(e_1,\dots,e_d)$ occupies $a$ qubits
per direction, the local node within the element is given by one bit $\ell_i\in\{0,1\}$ per
direction (its position on the left or right face), and the displacement component occupies
$r$ further qubits. The total number of qubits is
\begin{equation}\label{eq:qubits}
q=da+d+r .
\end{equation}
For Q4, $r=1$ and every slot is a physical displacement. For H8, $r=2$ gives four slots per copy,
and the fourth is a padding degree of freedom with stiffness $k_0$ (Eq.~\eqref{eq:gamma}; one in the units
of the computations) that carries no load and is not coupled
to the physical slots; it is bonded and clamped like the physical slots (``bonded padding''),
which leaves the physical solution unchanged. For H8, $\K_e$ therefore denotes the physical
$24\times24$ element matrix embedded in this $32\times32$ padded block; where the physical matrix
alone is meant we write $\K_e^{\rm phys}$. Fig.~\ref{fig:encoding} shows the numbering on a
$4\times4$ grid. On a uniform mesh, with the same material everywhere, the element part is one
element matrix repeated for every element,
\begin{equation}\label{eq:element_tensor}
\A_{\rm el}=\Id_{\rm elements}\otimes\K_e ,
\end{equation}
which does not become more complicated as the mesh is refined.

\subsection{Exchange form of the directional couplings}\label{sec:shift}
The stiffness matrix of a single spring can be written as
\begin{equation}\label{eq:spring-exchange}
\alpha\begin{bmatrix}1&-1\\-1&1\end{bmatrix}
=\alpha\left(\begin{bmatrix}1&0\\0&1\end{bmatrix}-\begin{bmatrix}0&1\\1&0\end{bmatrix}\right),
\end{equation}
and the second matrix exchanges the two copies. Along direction $i$ the springs pair
the copies of neighbouring elements, and no copy belongs to two springs in the same direction.
All these exchanges can therefore be performed at once by a single permutation $\V_i$: it swaps
the right copy of element $e$ with the left copy of element $e+1$, for every $e$ and every
position in the other directions, and leaves the copies on the outer faces in place. Since
$\V_i^2=\Id$, it is its own inverse. With the chain index $j=2e_i+\ell_i$ along direction $i$,
\begin{equation}\label{eq:Vi-open}
\V_i:\ j=2e+1\ \leftrightarrow\ j=2e+2\quad(e=0,\ldots,m-2),\qquad j=0,\,2m-1\ \text{fixed}.
\end{equation}
Domains with openings are handled by an element activity indicator $\chi(e)\in\{0,1\}$, the
counterpart for elasticity of the indicator masks used for immersed scalar problems
\citep{Perera2026}: the
masked exchange $\V_i^\chi$ swaps a pair only if both elements are active. The supports are
described by the reflection $\mathbf Z_D$ that changes the sign of the clamped copies, so that the
projector onto clamped copies is $(\Id-\mathbf Z_D)/2$.

\begin{proposition}[Exchange form of the springs]\label{thm:shift}
\emph{(a) Uniform meshes.} For Q4 and for H8 with bonded padding,
\begin{equation}\label{eq:shift}
\A_{\rm c}^{(i)}=\alpha(\Id-\V_i),\qquad
\A_{\rm c}=\sum_{i=1}^d\alpha(\Id-\V_i).
\end{equation}
The clamped assembly-free operator is an exact linear combination of
\begin{equation}\label{eq:U-uniform}
\Un(\A)=\Th(\K_e)+d+1
\end{equation}
distinct unitaries, i.e.\ at most $13$ for Q4 and $81$ for H8 (with equality at generic values of
$\nu$; e.g.\ 11 for Q4 at $\nu=1/3$, where the element matrix has eight Pauli terms, and 57, 75
and 78 for H8 at the exceptional values of $\nu$ below), for every
$m=2^a\ge2$, with 1-norm
$\lambda=\lambda(\K_e)+2d\alpha+\beta$, where $\Th(\K_e)$ and $\lambda(\K_e)$ are the number of
Pauli terms and the 1-norm of the element matrix.\\
\emph{(b) Domains with openings.} For a masked Cartesian domain,
$\A_{\rm c}^{(i)}=\alpha(\Id-\V_i^\chi)$, and with void elements regularized by a small multiple of
the identity the masked Q4 operator admits a representation with at most $2\Th(\K_e)+d+1=23$
structured unitaries, independently of mesh refinement and of the mask, when each mask-controlled
operation is counted as one unitary. This upper bound is attained for the L-shaped and perforated
domains considered here at the benchmark material and penalty parameters; degenerate masks (e.g.\
a single active element or a single row of elements, which lack interface bonds in some
direction) need fewer.
\end{proposition}

\begin{proof}
On each interface pair, $\Id-\V_i$ restricts to
$\bigl[\begin{smallmatrix}1&-1\\-1&1\end{smallmatrix}\bigr]$, which is exactly the spring
matrix~\eqref{eq:spring-exchange}. The copies on the outer faces are fixed by $\V_i$, so
$\Id-\V_i$ vanishes there, and the pairs are disjoint; this proves~\eqref{eq:shift}, and the same
argument applies to the active pairs of $\V_i^\chi$. The counts follow by merging the identity
terms (Section~S1 of the Supplementary Material).
\end{proof}

Here $\Un$ counts the distinct unitaries of the constructed decomposition.
For $m=1$ the exchanges reduce to the identity. For $m\ge2$, the 13 terms of a generic Q4 model are the ten Pauli terms of the element matrix (listed in
Section~S1 of the Supplementary Material), two directional exchanges and one support reflection;
for H8 the cubic element matrix with unit padding stiffness has 77 terms at generic values of
$\nu$. The model grows with the mesh, but the
description of its structure does not. Each exchange $\V_i$ is an increment or decrement of the
binary chain index, with guards at the two ends, at a cost of $O(\log m)$ Toffoli gates. In the
masked case the mask-controlled exchange counts as one unitary, whose circuit must evaluate the
activity indicator of the two elements; for the L-shaped and perforated brackets below, $\chi$ is
a simple function of the leading bits of the element index, but for an arbitrary mask this predicate
may itself require a large circuit, so the count of 23 is independent of the resolution without
fixing the gate cost. Our decomposition of the assembled
operator into the same kind of unitaries requires many more terms (upper bounds for our shift-based
construction; no lower bound is proved), because the factors from
different directions multiply rather than add (Section~\ref{sec:num-shift}). This construction is
not minimal: encodings that combine the entries of the element matrix with connectivity operators
\citep{Alkadri2025} can be expected to require fewer terms (Section~\ref{sec:projection}).

\paragraph{Example: the $2\times2$ mesh}
For $m=2$ the model has four elements and 32 copies, encoded in five qubits: one bit for each
element index $e_1$, $e_2$, one for each local bit $\ell_1$, $\ell_2$ and one for the displacement
component. The conforming model has 12 free degrees of freedom. There are 16 interface springs,
eight in each direction, and eight clamped copies. The chain index $j=2e_i+\ell_i$ takes the values
$0,\dots,3$, and the exchange $\V_i$ swaps $j=1$ and $j=2$, i.e.\ the bit patterns
$(e_i,\ell_i)=(0,1)$ and $(1,0)$: on this mesh $\V_i$ is the SWAP gate acting on the two qubits
$e_i$ and $\ell_i$. The reflection $\mathbf Z_D$ changes the sign of the copies with
$(e_1,\ell_1)=(0,0)$. The complete operator is
\begin{equation}\label{eq:example22}
\A=\Id_4\otimes\K_e+\alpha\bigl(2\Id-\V_1-\V_2\bigr)+\frac\beta2\bigl(\Id-\mathbf Z_D\bigr),
\end{equation}
a combination of the ten Pauli terms of $\K_e$, two swaps and one reflection, i.e.\ 13 unitaries
after the identity terms are merged, with $\lambda=\lambda(\K_e)+4\alpha+\beta$. Refining the mesh
adds qubits to the element indices and turns the swaps into the exchanges of
Eq.~\eqref{eq:Vi-open}, but adds no terms.

\subsection{Pauli expansion}\label{sec:pauli}
In the Pauli basis an exchange $\V_i$ is not a single term, but its expansion grows only
linearly with $m$, because adding one to a binary number flips a run of trailing bits (the carry), and each run
length requires its own Pauli terms
(Section~S1 of the Supplementary Material). For Q4 plane stress with one clamped face,
\begin{equation}\label{eq:total-q4}
\Th(\A)=7m+4-(m-1)\,\mathbf 1_{\{\beta=2\alpha\}}-2\,\mathbf 1_{\{\nu=1/3\}},
\end{equation}
i.e.\ $7m+4=7\sqrt{N/8}+4$ at generic values of $\nu$ and of the penalties, $6m+5$ for
$\beta=2\alpha$, $7m+2$ for $\nu=1/3$ and $6m+3$ when both hold (the two cancellations are
established separately in the Supplementary Material, and the combined formula was checked for
$m\le8$). For H8 with bonded padding (cubic elements, unit padding stiffness, one clamped face,
$m=2^a\ge2$),
\begin{equation}\label{eq:total-h8}
\Th(\A)=\Th(\K_e)+10m-9-(m-1)\,\mathbf 1_{\{\beta=2\alpha\}},
\end{equation}
where $\Th(\K_e)=77$ for all $\nu\in[0,\frac12)$ except $\nu=\frac14$, $\frac13$ and
$\nu^*=(\sqrt{15}-1)/6\approx0.47883$, at which it is 53, 71 and 74; at $\nu^*$ the physical diagonal
entry $(2-3\nu)/[9(1+\nu)(1-2\nu)]$ equals the padding stiffness. In the generic case
$\Th(\A)=10m+68$. In $d$ dimensions, for an element with a fixed number of local slots ($2^d$ nodes and $2^r$
component slots), generic penalties and an element Pauli expansion whose only diagonal string is
the identity, the count is $\Th(\K_e)+(3d+1)m-3d$, i.e.\ proportional to $dN^{1/d}$, whereas the assembled matrix on comparable Cartesian grids has
$\Theta(N)$ Pauli strings: at least $\frac98N-O(\sqrt N)$ for Q4 and $\frac{21}{32}N-O(N^{2/3})$
for H8 with penalty supports, and at least $\frac34N-O(\sqrt N)$ and $\frac12N-O(N^{2/3})$ with
exactly eliminated supports (Proposition~S3 of the Supplementary Material). The proof uses the
tensor-product form of the assembled operator: the 1D stiffness and consistent mass matrices have
the same Pauli support, with fixed coefficient ratios that control the cancellations in the sum of
their products. In 2D the element contributions to the counted strings never cancel; in 3D they
cancel in one class of strings, and penalty supports can cancel a few further strings at isolated
values of $\beta$; the bound accounts for both. The assembly-free counts are proved in closed
form (and were confirmed numerically), and the full Pauli list can be written down without forming
any matrix. In one dimension the assembly-free operator has the same number of Pauli terms as the
assembled chain; the reduction in Pauli-count scaling arises for $d\ge2$.

\subsection{Openings and graded meshes}\label{sec:graded}
For the L-shaped bracket obtained by removing one quadrant of the square, the activity indicator acts
only on the leading bit of each element index, the element part has 40 Pauli terms at every
resolution, and the exchange form requires at most 23 unitaries (Proposition~\ref{thm:shift}b). Adding a
dyadic opening changes the indicator but not the number of unitaries. These are statements about
the operator description only: the node-local diagonalization and the preconditioner of
Section~\ref{sec:uniform-preconditioning} rely on complete Cartesian copy graphs and are not claimed
for masked domains (Section~\ref{sec:cw}).

Element sizes need not be equal. On a graded mesh with element widths $h_1(e_1),\dots,h_d(e_d)$
the topology is that of the uniform mesh. When uniform spring stiffnesses are retained, the
springs, the supports and the exchanges $\V_i$ are unchanged, and only $\A_{\rm el}$ depends on the
sizes. Because every entry of an element matrix is an
integral of a product of two shape-function derivatives, a Q4 rectangle has
\begin{equation}\label{eq:graded-q4}
\K_e=\frac{h_2}{h_1}\,\K_1+\frac{h_1}{h_2}\,\K_2+\K_3
\end{equation}
with three fixed matrices; in particular, a square plane-elasticity element has the same stiffness
matrix at every size. An H8 box has the analogous form with six fixed matrices, weighted by
$h_2h_3/h_1$, $h_1h_3/h_2$, $h_1h_2/h_3$, $h_3$, $h_2$ and $h_1$. Both identities are exact and were
checked to round-off for random sizes. The element part is therefore a sum of three (Q4) or six
(H8) structured factor terms, each a fixed matrix multiplied by a diagonal of element-size factors
that are evaluated with reversible arithmetic; these diagonal factors are not unitaries in general
and enter through their own block encodings. This assumes an efficient size oracle, e.g.\ an analytically specified
grading, in the same way as the activity predicate of a mask; a general table of element sizes is
not available to a quantum circuit at polylogarithmic cost. Under this assumption the number of terms in exchange form remains independent of the mesh
size. The size factors enter the normalization through the largest element aspect ratio (2D) or
element size (3D). In a numerical test with a uniform penalty on a strongly graded cantilever
(Supplementary Material), the penalty error remains proportional to $1/\gamma$ and increases by
factors of about 1.7 and 2.8 for largest-to-smallest element ratios of 8 and 51. This test uses
uniform springs, $\A_{\rm c}^{(i)}=\alpha(\Id-\V_i)$, in 2D. In 3D the element stiffness scales with
the element size, so locally weighted springs, $\A_{\rm c}^{(i)}=\mathbf D_{\alpha,i}(\Id-\V_i)$ with
a diagonal $\mathbf D_{\alpha,i}$ that assigns the same stiffness to both ends of each bond, would be
natural. This changes the operator although the exchange topology is unchanged: $\mathbf
D_{\alpha,i}$ is not unitary in general, compact access requires an efficient stiffness oracle and
a revised normalization, and the unweighted term counts do not carry over without further
justification. Weighted springs were not tested.

\subsection{Curved meshes}\label{sec:curved-theory}
On a logically structured mesh obtained by a smooth mapping of a square, the springs are again
unaffected, but the element matrices vary from element to element, and on the mappings and
refinements tested the exact Pauli count of $\A_{\rm el}$ grew in proportion to $N$ (a smooth
mapping does not by itself imply that all Walsh coefficients are nonzero). For smooth mappings this variation can be compressed. The
element matrices are written as $\K_e=\sum_kg_k(e)\bar\K_k$ with a few fixed matrices $\bar\K_k$,
and the smooth fields $g_k$ are approximated by a few Walsh functions, the $\pm1$ square-wave
analogues of Fourier modes. Each $\bar\K_k$ is a combination of element matrices, so the compressed
matrices still annihilate rigid translations exactly; rigid rotations are element-dependent and
are preserved only approximately. Truncating the Pauli expansion of the assembled
matrix is not analogous, because its terms carry connectivity as well as geometry; an assembled
encoding that keeps element data separate from connectivity (Section~\ref{sec:projection}) could
be compressed in the same way, which we have not examined. The compression is empirical and has three costs: classical preprocessing, consisting of a
singular value decomposition of the element matrices ($O(N)$ work), Walsh transforms of the
geometry factors ($O(N\log N)$ work) and a global positivity verification of the compressed
operator (an eigenvalue computation, Supplementary Material); a truncation tolerance
that is a threshold on the Walsh coefficients, not a bound on the displacement error, which must be
measured; and a loss of definiteness of the individual element matrices: after truncation they
are slightly indefinite (at the tolerance $10^{-4}$ used below, their smallest eigenvalue is
between $-10^{-4}$ and $-4\times10^{-3}$ times the largest), although the complete operator with
its springs remained positive definite in all cases tested, with a smallest eigenvalue within
1\% of that of the uncompressed operator. Positivity was established only for these cases; it is
not guaranteed for arbitrary geometries or truncation thresholds. Results are reported in Section~\ref{sec:curved} and the
algorithm in the Supplementary Material.

\section{Preconditioning of the assembly-free operator}\label{sec:uniform-preconditioning}

\subsection{Sources of ill-conditioning}
Without preconditioning, the assembly-free operator is ill-conditioned for two reasons
(Section~\ref{sec:num-penalty}): the well-known $h^{-2}$ growth of the elasticity operator, and the stiff
springs, which resist any separation of the copies of one node. The two effects are treated
separately. Compatible displacement fields are standard FE displacements and are handled by a
standard multilevel preconditioner (Section~\ref{sec:bpx-el}). The spring modes are artificial and
are known in closed form from the topology (Section~\ref{sec:cw}). We write
\begin{equation}\label{eq:pre-A-split-final}
\A=\A_{\rm el}+\A_{\rm s},\qquad \A_{\rm s}=\A_{\rm c}+\A_{\rm bc}=\alpha\Cj^{\mathsf T}\Cj+\beta\Cj_D^{\mathsf T}\Cj_D .
\end{equation}
The spring energy vanishes exactly on the conforming FE fields that satisfy the support
conditions.

\emph{Nondimensional form.} Throughout Sections~\ref{sec:uniform-preconditioning}
and~\ref{sec:circuit-resources}, $\A$, $\A_{\rm el}$, $\A_{\rm s}$, $\K_0$ and the spring stiffnesses
are divided once by the stiffness scale $k_0$ of the finest elements (Eq.~\eqref{eq:gamma}), so
that $\alpha=\beta=\gamma$, and lengths are measured in units of the finest element size $h$. Under
fixed-domain refinement by bisection the nondimensional element matrices of the finest level do not
depend on $h$: for Q4 because $k_0=Et$, for H8 because $\K_e^{\rm phys}$ scales with $Eh=k_0$. The
computations with $E=1$ and unit-size finest elements therefore represent fixed-domain refinement,
not domain expansion. All formulas below refer to these nondimensional operators; in dimensional
form only the physical part of the preconditioner changes (Remark~\ref{rem:units}).

\subsection{Node-local sum and difference modes}\label{sec:cw}
The copies are now grouped by physical node rather than by element. We assume a complete
Cartesian copy graph, i.e.\ a mesh without voids, so that the copies of each node form a
hypercube. At a re-entrant corner or a void boundary a node can have three active copies; the masked
exchanges then share a copy and do not commute, and Propositions~\ref{prop:copy-walsh-final}
and~\ref{prop:algebra} do not extend to such nodes without further analysis. If $k(x)$ coordinate
directions are interior at node $x$, the node has $2^{k(x)}$ copies of each displacement component, indexed by
$\mathbf r\in\{0,1\}^{k(x)}$ (four copies at an interior Q4 node). Let $b(x)=1$ on the clamped
boundary and $b(x)=0$ elsewhere, and let $\mathbf X_j$ exchange the copies that differ in the
$j$-th bit.

\begin{proposition}[Spring stiffness at a node]\label{prop:copy-walsh-final}
The spring stiffness acting on one displacement component at node $x$ is
\begin{equation}\label{eq:Px-final}
\A_{{\rm s},x}=\alpha\sum_{j=1}^{k(x)}(\Id-\mathbf X_j)+\beta b(x)\Id .
\end{equation}
In the sum/difference basis $\mathbf S_2^{\otimes k(x)}$ it is diagonal,
\begin{equation}\label{eq:Px-diag-final}
\mathbf S_2^{\otimes k(x)}\A_{{\rm s},x}\mathbf S_2^{\otimes k(x)}
=\operatorname{diag}_{\mathbf z\in\{0,1\}^{k(x)}}\bigl(\beta b(x)+2\alpha|\mathbf z|\bigr),
\end{equation}
where $|\mathbf z|$ is the number of directions in which mode $\mathbf z$ is a difference. At a
free node the only zero-stiffness mode is $\mathbf z=0$, the normalized sum mode associated with
the physical nodal displacement: for $n_x$ identical copies $u_x$ its coefficient is
$u_s=\sqrt{n_x}\,u_x$.
\end{proposition}

The proof extends the two-copy calculation~\eqref{eq:two-copy} direction by direction
(\ref{app:precond-proofs}). Let $\W_c$ collect all modes with positive spring stiffness
(the difference modes, and all modes at clamped nodes) and set
\begin{equation}\label{eq:Dc-final}
\D_c=\operatorname{diag}\bigl(\beta b(x)+2\alpha|\mathbf z|\bigr)^{-1/2}
\end{equation}
on those modes. Then $(\W_c\D_c)^{\mathsf T}\A_{\rm s}(\W_c\D_c)=\Id$: the spring
stiffness becomes exactly the identity on the scaled spring modes. Because the transformation
acts on the few copies of one node at a time, no global system has to be solved.

The sum/difference transformation is not an additional ingredient: it is the common eigenbasis of
the operations that already describe the springs in Section~\ref{sec:shift}.

\begin{proposition}[One set of commuting operations]\label{prop:algebra}
On a uniform Q4 mesh, or an H8 mesh with bonded padding (the padding slot counted as a fourth
displacement component), the exchanges $\V_1,\ldots,\V_d$ and the support reflection $\mathbf Z_D$
are commuting involutions, and:
\begin{enumerate}
\item[(a)] the spring stiffness is their linear combination,
$\A_{\rm s}=\alpha\sum_{i=1}^d(\Id-\V_i)+\tfrac\beta2(\Id-\mathbf Z_D)$;
\item[(b)] on the physical displacement sector, the orthogonal projector onto the conforming
displacements written in copy form is the normalized product of $\Id+\V_i$ and $\Id+\mathbf Z_D$
(for H8 the product also contains the compatible padding modes, which are excluded here, as in
Theorem~\ref{thm:full-pre-final}),
\begin{equation}\label{eq:projector}
\boldsymbol\Pi_0=\mathbf Q\mathbf Q^{\mathsf T}
=\frac1{2^{d+1}}\Bigl[\prod_{i=1}^d(\Id+\V_i)\Bigr](\Id+\mathbf Z_D),\qquad
\mathbf Q=\J_0\mathbf M_{\rm copy}^{-1/2},
\end{equation}
so that the conforming stiffness in copy form is $\boldsymbol\Pi_0\A_{\rm el}\boldsymbol\Pi_0$,
which on the compatible subspace is congruent to the assembled matrix through a diagonal scaling,
$\mathbf Q^{\mathsf T}\A_{\rm el}\mathbf Q=\mathbf M_{\rm copy}^{-1/2}\K_0\mathbf M_{\rm copy}^{-1/2}$ (Eq.~\eqref{eq:projected});
\item[(c)] the pseudoinverse of the spring stiffness (its inverse on the spring modes, zero on the
conforming fields) is a function of the same operations,
\begin{equation}\label{eq:spring-inverse}
\W_c\D_c^2\W_c^{\mathsf T}=\A_{\rm s}^+
=\sum_{\mathbf s:\,\lambda_{\mathbf s}>0}\frac{1}{\lambda_{\mathbf s}}\,
\frac1{2^{d+1}}\Bigl[\prod_{i=1}^d(\Id+s_i\V_i)\Bigr](\Id+s_{d+1}\mathbf Z_D),\qquad
\lambda_{\mathbf s}=\alpha\sum_{i=1}^d(1-s_i)+\tfrac\beta2(1-s_{d+1}),
\end{equation}
with $\mathbf s\in\{1,-1\}^{d+1}$.
\end{enumerate}
\end{proposition}

The proof is given in \ref{app:precond-proofs}; all three identities were also verified to
round-off for Q4 ($m\le8$) and H8 ($m=2$), and identity (b) for Q4 up to $m=32$. The same $d+1$ operations thus give the penalty
operator as a sum, the conforming operator as a product with the element stiffness, and the spring
part of the preconditioner through their common eigenbasis. Section~\ref{sec:projection} compares
the penalty and projected forms.

\subsection{Multilevel preconditioning of the physical displacements}\label{sec:bpx-el}
The compatible displacements carry the standard elasticity stiffness, whose condition number
grows as $h^{-2}$. For them we use the BPX multilevel preconditioner
\citep{BramblePasciakXu1990}, which can be viewed as an additive multigrid method: a displacement
on the finest mesh is represented as a sum of contributions from all coarser nested meshes, so
that long- and short-wavelength components are each treated on the mesh that resolves them. Let
$\G_{\rm BPX}$ be this componentwise multilevel representation, with the level-$\ell$ hat
functions weighted by $(h_\ell/h)^{(2-d)/2}$, where $h_\ell$ is the element size of level $\ell$
and $h$ that of the finest level; it is a redundant set of basis vectors (a frame). Let
$\K_0=\J_0^{\mathsf T}\A_{\rm el}\J_0$ be the conforming stiffness matrix on the free nodal
displacements, i.e.\ the assembled matrix of Eq.~\eqref{eq:standard} with the supports
eliminated. For the scalar Laplacian the BPX-preconditioned matrix has a positive spectrum in an
interval $[c_B,C_B]$ independent of the number of levels
\citep{Oswald1990,DahmenKunoth1992,Xu1992}. Two standard inequalities carry this over to clamped
elasticity. The strain energy is bounded by the strains,
$m_D\norm{\boldsymbol\varepsilon(u)}^2\le a(u,u)\le M_D\norm{\boldsymbol\varepsilon(u)}^2$, with
$m_D=1/(1+\nu)$ and $M_D=1/(1-\nu)$ in plane stress ($M_D=1/(1-2\nu)$ in 3D), and Korn's
inequality $\norm{\nabla u}\le C_K\norm{\boldsymbol\varepsilon(u)}$ bounds the displacement
gradient by the strains, because the body is clamped on a part $\Gamma_D$ of its boundary
\citep{Horgan1995}. The positive spectrum of the BPX-preconditioned conforming elasticity matrix
therefore lies in $[c_{\rm BPX},C_{\rm BPX}]$ with
\begin{equation}\label{eq:c0C0-final}
c_{\rm BPX}=\frac{m_D}{C_K^2}c_B,\qquad C_{\rm BPX}=M_DC_B,\qquad
\kappa_+\bigl(\G_{\rm BPX}^{\mathsf T}\K_0\G_{\rm BPX}\bigr)\le\frac{C_B}{c_B}\,C_K^2\,\frac{M_D}{m_D},
\end{equation}
independently of the mesh size. Here and below, because the multilevel frame is redundant and the
preconditioned matrices are singular, $\kappa_+$ denotes the positive-spectrum condition number, the
ratio of the largest to the smallest positive eigenvalue; all condition numbers of preconditioned
operators reported in this paper are of this kind. The material factor $M_D/m_D$ grows without bound as
$\nu\to1/2$ in 3D; no robustness is claimed in the nearly incompressible limit.

\subsection{The combined preconditioner}\label{sec:combined}
The complete transformation combines the two parts,
\begin{equation}\label{eq:F-final}
\F=\bigl[\F_{\rm p},\F_{\rm c}\bigr]=\bigl[\J_0\G_{\rm BPX},\ \W_c\D_c\bigr],
\end{equation}
where $\J_0$ maps nodal displacements to all copies. The transformation $\F$ is constructed from the
mesh hierarchy and the spring stiffness, without the element matrices, but it is applied to the \emph{complete}
operator $\A=\A_{\rm el}+\A_{\rm s}$. Since the springs do no work on compatible fields
and are scaled to the identity on the spring modes,
\begin{equation}\label{eq:block-structure}
\F^{\mathsf T}\A\F=
\begin{bmatrix}
\G_{\rm BPX}^{\mathsf T}\K_0\G_{\rm BPX} & \F_{\rm p}^{\mathsf T}\A_{\rm el}\F_{\rm c}\\[2pt]
\F_{\rm c}^{\mathsf T}\A_{\rm el}\F_{\rm p} & \Id+\F_{\rm c}^{\mathsf T}\A_{\rm el}\F_{\rm c}
\end{bmatrix}.
\end{equation}
The upper-left block is the BPX-preconditioned elasticity matrix. The element stiffness that
remains in the spring sector and its coupling to the physical sector are not removed; their
relative size is $\eta$ and $\sqrt\eta$, respectively, the latter measured relative to the physical
energy norm, which is small when the springs are much
stiffer than the elements. With
\begin{equation}\label{eq:eta-final}
\bar k=\norm{\K_e}_2,\qquad p_{\min}=\min\{\beta,2\alpha\},\qquad \eta=\frac{\bar k}{p_{\min}},
\end{equation}
one has $\norm{\F_{\rm c}^{\mathsf T}\A_{\rm el}\F_{\rm c}}_2\le\eta$, and the following bound
holds (proof in \ref{app:precond-proofs}).

\begin{theorem}[Mesh- and penalty-robust preconditioning]\label{thm:full-pre-final}
For uniform Cartesian Q4 or H8 mesh hierarchies (complete copy graphs) with a clamped boundary of
positive measure and fixed isotropic material with $0\le\nu<1/2$, in the nondimensional form of
Section~\ref{sec:uniform-preconditioning}, assume $\eta<1$.
For H8 the padding component is decoupled from the physical components,
$\A=\A_{\rm phys}\oplus\A_{\rm pad}$; its operator $\A_{\rm pad}$ (identity plus springs) is diagonal
in the sum/difference basis with the known entries $1+\beta b+2\alpha|\mathbf z|$, and is scaled
exactly to the identity, $\F=\F_{\rm phys}\oplus\F_{\rm pad}$ with $\F_{\rm pad}^{\mathsf T}\A_{\rm
pad}\F_{\rm pad}=\Id_{\rm pad}$; the quantities below then refer to the physical components, with
$\bar k=\norm{\K_e^{\rm phys}}_2$. Then the positive spectrum of $\F^{\mathsf
T}\A\F$ lies in
$\bigl[(1-\sqrt\eta)\min\{c_{\rm BPX},1\},\ (1+\sqrt\eta+\eta)\max\{C_{\rm BPX},1\}\bigr]$, and
\begin{equation}\label{eq:conditioning_bound}
\kappa_+\bigl(\F^{\mathsf T}\A\F\bigr)\le
\frac{1+\sqrt\eta+\eta}{1-\sqrt\eta}\,\frac{\max\{C_{\rm BPX},1\}}{\min\{c_{\rm BPX},1\}} ,
\end{equation}
which is independent of the mesh size and, for any fixed $\eta_0<1$, uniform over all penalties with
$\eta\le\eta_0$. Since the BPX interval may always be enlarged so that
$c_{\rm BPX}\le1\le C_{\rm BPX}$, the bound can be read as the conforming bound
$C_{\rm BPX}/c_{\rm BPX}$ multiplied by the penalty factor; without this enlargement the factors
$\max\{C_{\rm BPX},1\}$ and $\min\{c_{\rm BPX},1\}$ must be kept. The penalty-dependent factor of this upper bound decreases as
the springs are made stiffer and tends to one as $\alpha,\beta\to\infty$.
\end{theorem}

\begin{remark}[Units and range of the penalty]\label{rem:units}
In dimensional form the physical part of the preconditioner is
$\F_{\rm p}=k_0^{-1/2}\J_0\G_{\rm BPX}$, which is equivalent to the nondimensional form used above;
$\F_{\rm c}$ is unchanged because $\D_c$ is built from the dimensional spring stiffness. Without the
factor $k_0^{-1/2}$ the bound depends on the system of units: for the Q4 cantilever with $m=8$ and
$\alpha=\beta=10E$ the condition number is $59.3$ for $E=1$ but $6.3\times10^{3}$ for $E=10^{3}$
without this factor. The bound is pessimistic as $\eta$ approaches one: for $\gamma=2$
($\eta=0.71$) the computed condition number exceeds its large-penalty limit only by a factor of 1.3.
\end{remark}

In mechanical terms, the condition $\eta<1$ requires the springs to be stiffer than the elements
they connect; for $E=1$, $\nu=0.3$ and $\alpha=\beta=10$, $\eta=1/7$ for Q4 and $0.125$ for H8. The
theorem is the reason why, for the ideal preconditioner and sufficiently stiff penalties
($\eta\le\eta_0<1$), the penalty can be chosen for accuracy alone: stiffer springs reduce the
penalty error \emph{and} improve the upper bound (that the computed condition number also decreases
is a numerical observation, Section~\ref{sec:numerics}), whereas the condition number of the unpreconditioned
matrix grows in proportion to $\gamma$. Its scope should be stated precisely. The
mesh-independent bound is inherited from the conforming model through
Eq.~\eqref{eq:c0C0-final}; the contribution of the sum/difference scaling is that the springs add
no more than the factor $(1+\sqrt\eta+\eta)/(1-\sqrt\eta)$ to it. The theorem is an instance of
auxiliary-space preconditioning \citep{Xu1996} and of the principle behind robust subspace
correction for nearly singular systems \citep{LeeWuXuZikatanov2007}: a large penalty term does not
destroy mesh-independent conditioning if the decomposition contains the kernel of that term exactly. Its
ingredients are standard: the BPX bound for the scalar Laplacian, the material bounds and Korn's
inequality give Eq.~\eqref{eq:c0C0-final}, and the springs enter through a two-by-two block
estimate (\ref{app:precond-proofs}). What is specific to the assembly-free formulation is that both
the kernel of the springs (the conforming displacements) and its complement (the spring modes,
Proposition~\ref{prop:algebra}) are known in closed form and node-local, so that the complement is
scaled exactly rather than smoothed, and the penalty-dependent factor of the bound is explicit and
improves as $\gamma$ increases. Because the multilevel basis is
redundant, $\F^{\mathsf T}\A\F$ is singular, like the stiffness matrix of an unsupported
structure, and only its positive eigenvalues matter; Section~\ref{sec:direct-block-precond} shows
how it is block-encoded without forming the product.

\section{Numerical assessment}\label{sec:numerics}

\subsection{Models and verification protocol}
\begin{figure}[!htbp]
\centering\includegraphics{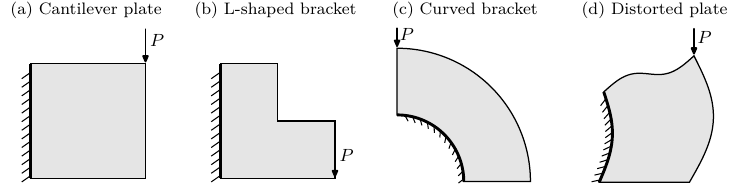}
\caption{Two-dimensional benchmark problems: (a) cantilever plate, (b) L-shaped bracket, (c)
curved bracket, (d) smoothly distorted plate. Hatched edges are clamped; arrows mark the point
loads used for the displacement comparisons.}
\label{fig:problems}
\end{figure}

\begin{figure}[!htbp]
\centering\includegraphics[width=\textwidth]{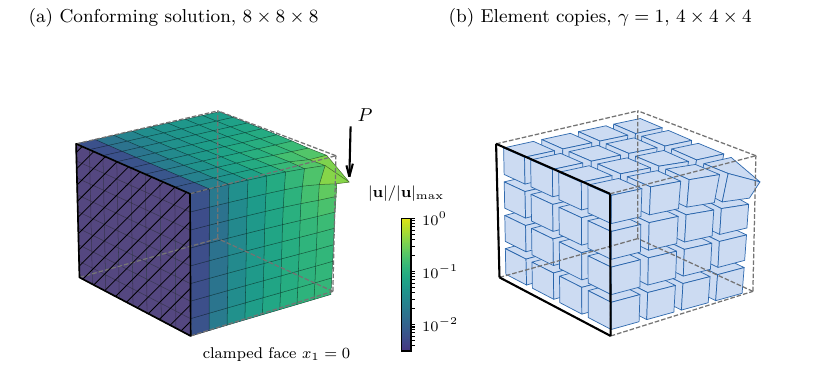}
\caption{Three-dimensional H8 cantilever: unit cube clamped on $x_1=0$ (hatched in (a)), unit
downward point load $P$ at a corner of the free face. (a) Conforming solution on an
$8\times8\times8$ mesh, with displacements magnified so that the loaded-corner deflection is 20\% of
the edge length and the undeformed outline dashed; the shading is the displacement magnitude
normalized by its maximum on a logarithmic scale (colour bar), and the deformation concentrates at
the loaded corner because the point load is singular. (b) Assembly-free solution with
the dimensionless penalty $\gamma=1$ on  a $4\times4\times4$ mesh, same
magnification; every element is drawn shrunk by 22\% towards its centroid to show that it carries
its own copies of the nodal displacements. At finite spring stiffness the copies also separate
physically; the imposed centroid shrinkage makes the copies visually distinguishable and should not
be interpreted as the magnitude of the physical compatibility error.}
\label{fig:h8}
\end{figure}

Unless stated otherwise $E=1$, $\nu=0.3$ and the thickness is one; Q4 elements use $2\times2$ and
H8 elements $2\times2\times2$ Gauss integration. The two-dimensional benchmarks are shown in
Fig.~\ref{fig:problems}: (a) a clamped square plate (cantilever) with a corner load, (b) an
L-shaped bracket, (c) a curved bracket on a polar mesh and (d) a smoothly distorted plate; a
further, randomly perturbed version of the plate mesh (not shown in Fig.~\ref{fig:problems}) tests
the limits of compression. A three-dimensional H8 cantilever, a unit cube clamped on the face
$x_1=0$ under a unit downward point load at a corner of the free face (Fig.~\ref{fig:h8}), and the
two-dimensional L-shaped
bracket with an opening of Section~\ref{sec:num-hole} complete the set. Point loads serve for
operator verification and discrete displacement comparisons; continuum-accuracy statements use an
edge traction (Section~\ref{sec:num-penalty}), because the compliance
under a point force does not converge under refinement. Every decomposition into
unitaries was checked entry by entry against the sparse operator (largest deviation
$7\times10^{-15}$), and with very large penalties the assembly-free and assembled solutions agreed
to at least six significant digits in every case. Displacement errors are measured against the
conforming FE solution on the same mesh. Details of the computations are given in the
Supplementary Material.

\subsection{Size of the operator}\label{sec:num-shift}
Table~\ref{tab:counts} compares the Pauli counts with those of the assembled operator at equal
numbers of qubits. Equal qubits do not mean equal resolution: at $q=11$ the assembly-free Q4 mesh
has $16\times16$ elements and 116 terms, whereas the assembled mesh has $31\times31$ elements and
5,426 terms; at approximately matched resolution, the assembly-free $32\times32$ mesh has 228
terms. In three dimensions the assembly-free H8 counts grow from 88 to 148, against 1,222 to
118,822 for the assembled operator.

\begin{table}[!htbp]
\centering
\begin{threeparttable}
\caption{Exact Pauli term counts $\Th$ on uniform meshes at equal numbers of qubits $q$.
Assembly-free meshes have $m=2^a$ elements per direction, assembled meshes $m'=2m-1$ at the same
$q$; $\alpha=7.3$, $\beta=11.1$. Assembly-free counts are proved in closed form; assembled counts
are computed and are $\Theta(N)$ by Proposition~S3 of the Supplementary Material.}
\label{tab:counts}
\footnotesize\setlength{\tabcolsep}{5pt}
\begin{tabular}{@{}llrrrr@{}}\toprule
Element & Qubits $q$ & Assembly-free $m$ & Assembly-free $\Th$ & Assembled $m'$ & Assembled $\Th$\\\midrule
Q4 & 5  & 2  & 18  & 3  & 64\\
Q4 & 7  & 4  & 32  & 7  & 302\\
Q4 & 9  & 8  & 60  & 15 & 1{,}306\\
Q4 & 11 & 16 & 116 & 31 & 5{,}426\\
Q4 & 13 & 32 & 228 & -- & --\\\addlinespace
H8 & 8  & 2  & 88  & 3  & 1{,}222\\
H8 & 11 & 4  & 108 & 7  & 12{,}990\\
H8 & 14 & 8  & 148 & 15 & 118{,}822\\\bottomrule
\end{tabular}
\begin{tablenotes}[flushleft]\footnotesize
\item H8 with bonded padding. Term counts measure the size of the description, not gate counts,
query counts or run times.
\end{tablenotes}
\end{threeparttable}
\end{table}

The exchange form is even more compact (Table~\ref{tab:shift}). Q4 requires 13 unitaries at every
tested size, against 109 for the assembled operator with penalty supports and 196 with exact Dirichlet
elimination with identity padding (the constrained rows and columns are replaced by decoupled
identity rows, so that the dimension remains a power of two; this is not the free-node matrix
$\K_0$ of Section~\ref{sec:bpx-el}), decomposed with the same kind of structured unitaries; for H8 the numbers are
81, 2,776 and 10,072. Both formulations have mesh-independent counts, but the assembled one is
larger by a factor of 8 to 124, because the factors of different directions multiply. These
assembled counts are upper bounds for the particular shift-based decomposition implemented here,
which expands the element matrix entry by entry over shifted node pairs; they do not establish a
lower bound on the number of terms required by other assembled-matrix encodings, and the entry-wise assembled encoding of
\citet{Alkadri2025}, which we have not implemented for elasticity, can be expected to require fewer
terms (Section~\ref{sec:projection}).
Moreover, the number of branches is only one cost component: the 1-norm, the branch circuits, the
conditioning and the number of calls also enter. The 1-norm $\lambda$ shows the opposing trade-off: it is larger for the assembly-free
operator (42.0 against 16.9 for Q4 at the parameters of the table), because the springs are
stiffer than the elements. The L-shaped bracket, with or without an opening, requires 23 unitaries.

\paragraph{Penalty-error-controlled cost proxy}\label{sec:num-cost}
A fair comparison must include the penalty error. In the Supplementary Material its
leading term is removed by extrapolation from $\gamma=2$ and 8, which keeps the worst tested penalty
error near $3.3\times10^{-3}$ for the four two-dimensional benchmarks and at $2.0\times10^{-4}$ for the
H8 cantilever at $m=4$ and 8 (the exactly constrained assembled operator has none), and a cost proxy, the number of
terms multiplied by $\kappa_{\rm eff}=\lambda\norm{\A^{-1}}_2$, is compared with that of the
assembled operator with exact Dirichlet elimination with identity padding. In Pauli form the
assembly-free operator has a smaller value of this cost proxy from $m=32$ in 2D and from $m=8$ in 3D.
In exchange form it has a smaller value in 3D (proxy ratio 0.40--0.50) but not in 2D (3.6--3.9),
where the smaller 1-norm and larger smallest
eigenvalue of the assembled operator outweigh its 15 times larger number of terms. The proxy
counts every unitary equally, ignores discretization, inversion and readout errors, and applies to
the unpreconditioned operators; it does not establish actual quantum execution cost, and for the
preconditioned factor constructed in Section~\ref{sec:circuit-resources} the relevant parameter is
Eq.~\eqref{eq:factor-qsvt} instead.

\begin{table}[!htbp]
\centering
\caption{Number of distinct unitaries $\Un$ and 1-norm $\lambda$ of the exchange-form
decompositions ($\alpha=7.3$, $\beta=11.1$). Assembled operators are decomposed with the same
kind of structured unitaries, with penalty supports or exact Dirichlet elimination with identity
padding (``exact supports''); the assembled counts are
upper bounds for this shift-based decomposition, not lower bounds for assembled encodings in
general. All counts were verified for every listed $N$ and are independent of it. $\Un$ counts
branches of the linear combination; it is not a gate or query count.}
\label{tab:shift}
\footnotesize\setlength{\tabcolsep}{4pt}
\begin{tabular}{@{}llrrr@{}}
\toprule
Problem & Formulation & $N$ verified & $\Un$ & $\lambda$\\
\midrule
Q4 cantilever & assembly-free & $32$--$131{,}072$ & $13$ & $42.0$\\
 & assembled, penalty supports & $32$--$32{,}768$ & $109$ & $16.9$\\
 & assembled, exact supports & $32$--$8{,}192$ & $196$ & $6.6$\\
H8 cantilever & assembly-free & $256$--$131{,}072$ & $81$ & $57.4$\\
 & assembled, penalty supports & $256$--$16{,}384$ & $2{,}776$ & $28.2$\\
 & assembled, exact supports & $256$--$2{,}048$ & $10{,}072$ & $13.6$\\
L-shaped bracket & assembly-free & $128$--$32{,}768$ & $23$ & $42.5$\\
L-bracket with opening & assembly-free & $512$--$32{,}768$ & $23$ & $42.5$\\
\bottomrule
\end{tabular}
\end{table}

\subsection{Penalty accuracy}\label{sec:num-penalty}
\begin{figure}[!htbp]
\centering\includegraphics{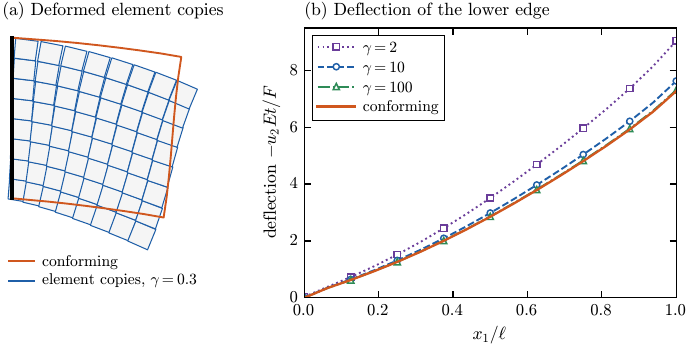}
\caption{Q4 cantilever under a uniform edge traction of total force $F$ (length $\ell$, thickness
$t$). (a) Deformed element copies for a deliberately soft penalty, $\gamma=0.3$, on an $8\times8$
mesh (magnified), with the outline of the conforming solution: the springs between the elements
open visibly. (b) Deflection of the lower edge on a $16\times16$ mesh for three penalties; the
difference from the conforming solution is approximately proportional to $1/\gamma$
(asymptotically, for large $\gamma$).}
\label{fig:physical}
\end{figure}

Fig.~\ref{fig:physical} shows the effect of the penalty in physical terms for the Q4 cantilever
under a uniform edge traction of total force $F$.
With deliberately soft springs the element copies separate visibly (Fig.~\ref{fig:physical}a);
with $\gamma=10$ and $100$ the deflection at the loaded end exceeds the conforming one by $4.9\%$
and $0.49\%$ (Fig.~\ref{fig:physical}b). Table~\ref{tab:traction} quantifies the penalty error for
this load, with $\alpha=\beta=\gamma$, in three measures that are relevant in design: the
compliance, the strain-energy seminorm and the largest error of the element-centre stresses. All three exhibit the expected $O(1/\gamma)$ decay in the large-penalty regime; the coefficients for $\gamma=100$ and $1000$ agree to within one
percent. The coefficient of the compliance error decreases from $0.51$ to $0.46$ under refinement;
it agrees, to the reported precision, with the bound coefficient $c_1$ of
Proposition~\ref{prop:penalty-bound} ($0.512$, $0.488$, $0.475$, $0.468$, $0.465$ for $m=8$--$128$,
and $0.462$ at $m=512$). For the tested equal penalties $0.5\le\gamma\le10$ and the additional
unequal-penalty combinations, the computed compliance error is between $99.1\%$ and $99.95\%$ of the
upper bound~\eqref{eq:penalty-bound} (Table~\ref{tab:penbound-main} and Supplementary Material); this
tightness is empirical, not a general property of the bound. The corresponding coefficient for the
strain-energy error decreases from $0.069$ to $0.043$. The penalty error in these global measures
therefore remains bounded over the tested meshes. The largest stress error is smaller than the compliance error, between
$0.11/\gamma$ and $0.17/\gamma$ of the peak stress, but its coefficient grows with $m$; the
maximum occurs next to the ends of the clamped edge, where the stresses are singular. The
observation that the penalty error is bounded therefore applies to the tested global outputs, not to
pointwise stresses; local stress accuracy near the clamped--free corners requires separate care.
Mechanically, the springs act as a thin compliant layer between the elements and soften the
structure by a relative amount of about $1/(2\gamma)$. Without extrapolation, a sufficient
condition for the relative compliance penalty error to remain below $\tau_{\rm pen}$ is
$\gamma\ge c_1/\tau_{\rm pen}$, i.e.\ about $\gamma\approx0.5/\tau_{\rm pen}$ on the tested meshes; by
Proposition~\ref{prop:penalty-bound} this choice is guaranteed on any mesh for which $c_1$ has been
computed from the conforming solution.

\begin{table}[!htbp]
\centering
\caption{Test of the compliance bound~\eqref{eq:penalty-bound} for the Q4 cantilever under the edge
traction, $m=32$: relative penalty error $e=(\mathcal C-\mathcal C_0)/\mathcal C_0$, bound
$b=\mathbf r^{\mathsf T}\A_{\rm s}^+\mathbf r/\mathcal C_0$ and their ratio, for soft, stiff and
unequal springs. Results for $m=8$ and 16 are in the Supplementary Material.}
\label{tab:penbound-main}
\footnotesize
\begin{tabular}{@{}rrccc@{}}
\toprule
$\alpha$ & $\beta$ & $e$ & $b$ & $e/b$\\
\midrule
0.5 & 0.5  & $9.453\times10^{-1}$ & $9.502\times10^{-1}$ & 0.9949\\
2   & 2    & $2.371\times10^{-1}$ & $2.375\times10^{-1}$ & 0.9980\\
10  & 10   & $4.749\times10^{-2}$ & $4.751\times10^{-2}$ & 0.9995\\
7.3 & 11.1 & $6.356\times10^{-2}$ & $6.360\times10^{-2}$ & 0.9995\\
100 & 3    & $1.478\times10^{-2}$ & $1.484\times10^{-2}$ & 0.9958\\
\bottomrule
\end{tabular}
\end{table}

\begin{table}[!htbp]
\centering
\caption{Q4 cantilever under a uniform edge traction, $\alpha=\beta=\gamma$: penalty error
multiplied by $\gamma$, measured against the conforming solution on the same mesh
($\gamma=100$). Compliance: relative error. Energy: $\norm{\uvec-\uvec_0}_{\A_{\rm el}}$
divided by $\norm{\uvec_0}_{\A_{\rm el}}$, where $\norm{\mathbf v}_{\A_{\rm el}}=(\mathbf
v^{\mathsf T}\A_{\rm el}\mathbf v)^{1/2}$ is the strain-energy seminorm (it vanishes on
element-wise rigid motions and is a norm on conforming fields). Stress: largest error of any element-centre stress
component divided by the largest stress component.}
\label{tab:traction}
\footnotesize
\begin{tabular}{@{}rccc@{}}
\toprule
$m$ & $\gamma\times$ compliance error & $\gamma\times$ energy error & $\gamma\times$ max.\ stress error\\
\midrule
8   & 0.512 & 0.069 & 0.107\\
16  & 0.488 & 0.057 & 0.111\\
32  & 0.475 & 0.049 & 0.118\\
64  & 0.468 & 0.045 & 0.126\\
128 & 0.465 & 0.043 & 0.167\\
\bottomrule
\end{tabular}
\end{table}

For a point load the error of the loaded-corner displacement is likewise proportional to
$1/\gamma$ in two and three dimensions. To isolate the interface springs, this test uses $\alpha=\gamma$
and effectively exact support enforcement, $\beta=10^8$: on the $16\times16$ Q4 mesh the error is
$3.95\%$, $0.396\%$ and $0.0396\%$ for $\gamma=10$, 100 and 1000 (Table~S4 of the Supplementary
Material); with $\alpha=\beta=\gamma$ it is $4.30\%$ at $\gamma=10$ (Table~S11). Without preconditioning, the
condition number of the assembly-free operator is larger than that of the assembled operator with
eliminated supports by a factor of about $(2.4$--$3.4)\gamma$ in 2D, e.g.\ $1.0\times10^{5}$
against $3.0\times10^{3}$ at $m=16$ and $\gamma=10$; both grow as $h^{-2}$.

\subsection{Preconditioned condition numbers}
\begin{figure}[!htbp]
\centering\includegraphics{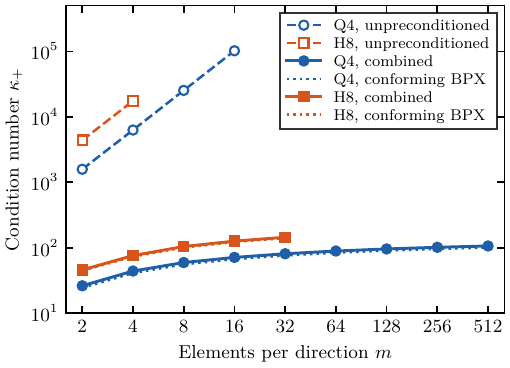}
\caption{Condition numbers under mesh refinement before (open symbols, $\gamma=10$) and after
(filled symbols, $\alpha=\beta=10$) sum/difference scaling and BPX preconditioning, for Q4 and H8
cantilevers; the dotted lines show conforming BPX alone, which the combined values exceed by only
2--8\%. Data from Table~\ref{tab:bpx-q4h8} (including the finest Q4 mesh, $m=512$) and the
Supplementary Material. Extreme eigenvalues were
computed by Lanczos iterations; they are numerical diagnostics, not certified eigenvalue
enclosures, and uniform boundedness rests on Theorem~\ref{thm:full-pre-final}.}
\label{fig:cond}
\end{figure}

Table~\ref{tab:bpx-q4h8} and Fig.~\ref{fig:cond} report the condition numbers after the combined
transformation of Section~\ref{sec:combined}. For Q4 the value rises from 26.2 at $m=2$ to 106.2
at $m=512$; for H8 from 45.7 at $m=2$ to 144.4 at $m=32$. The smallest positive eigenvalue is
essentially independent of the mesh, consistent with the lower bound of
Theorem~\ref{thm:full-pre-final} (0.102--0.108 for Q4, 0.061--0.071 for H8). The growth comes from the largest eigenvalue
and slows down from level to level, which is typical of BPX and compatible with a
level-independent bound. The computed values have not yet reached a plateau, however: for Q4 the
increments shrink by a factor of about 0.84 per level; if they continued to do so, the values would
level off near 130, but this is an extrapolation, not a result; and
for H8 they are still large at $m=32$ (30.0, 28.4, 21.8, 18.5). On these benchmarks the combined values exceed those
of conforming BPX alone by only 2--8\% (a measured overhead, not a bound), which is the cost of the
duplicated degrees of freedom after preconditioning. The finite-$m$ data are consistent with, but
cannot by themselves establish, mesh independence; that rests on Theorem~\ref{thm:full-pre-final}. In contrast, the unpreconditioned condition number grows by a factor of four per
refinement (Fig.~\ref{fig:cond}).

\begin{table}[!htbp]
\centering
\caption{Condition numbers of the BPX-preconditioned scalar, conforming elasticity and complete
assembly-free (sum/difference scaling + BPX) operators; $E=1$, $\nu=0.3$, $\alpha=\beta=10$, one
clamped face. For the complete operator the smallest positive eigenvalue is also given. ``--'':
not computed.}
\label{tab:bpx-q4h8}
\footnotesize\setlength{\tabcolsep}{4pt}
\begin{tabular}{@{}r|rrrr|rrrr@{}}
\toprule
& \multicolumn{4}{c|}{Q4 (2D)} & \multicolumn{4}{c}{H8, bonded padding (3D)}\\
$m$ & scalar Q1 & conforming & combined & $\lambda_{\min}^+$ & scalar Q1 & conforming & combined & $\lambda_{\min}^+$\\
\midrule
2   & 4.19  & 24.36 & 26.16 & 0.102 & 8.63  & 44.63  & 45.73  & 0.061\\
4   & 6.24  & 41.05 & 43.96 & 0.105 & 12.81 & 72.98  & 75.76  & 0.067\\
8   & 7.79  & 55.59 & 59.26 & 0.106 & 16.34 & 100.64 & 104.14 & 0.070\\
16  & 9.07  & 66.85 & 71.04 & 0.107 & 19.15 & 121.92 & 125.92 & 0.071\\
32  & 10.14 & 76.16 & 80.79 & 0.108 & 21.42 & 139.97 & 144.39 & 0.071\\
64  & 11.03 & 83.89 & 88.88 & 0.108 & --    & --     & --     & --\\
128 & 11.78 & 90.35 & 95.67 & 0.108 & --    & --     & --     & --\\
256 & 12.42 & 95.78 & 101.39& 0.108 & --    & --     & --     & --\\
512 & --    & 100.37& 106.22& 0.108 & --    & --     & --     & --\\
\bottomrule
\end{tabular}
\end{table}

A penalty sweep at fixed resolution is consistent with the second half of the theorem. At $m=32$ the Q4
condition number after preconditioning decreases from 99.3 at $\gamma=2$ to 76.6 at $\gamma=100$
and 76.2 at $\gamma=1000$ (Supplementary Material), consistent with the decreasing upper bound:
on this mesh stiffer springs improve rather than damage the transformed system.

All values above are for $\nu=0.3$. In plane stress the dependence on Poisson's ratio is mild: at
$m=16$ and $\gamma=10$ the Q4 condition number rises from 61.6 at $\nu=0$ to 88.4 at $\nu=0.499$
(Supplementary Material),
as expected because plane stress does not lock. No such test was made in 3D, where the
bound~\eqref{eq:c0C0-final} grows as $1/(1-2\nu)$. The masked (Section~\ref{sec:num-hole}), graded (Section~\ref{sec:graded}) and curved
(Section~\ref{sec:curved}) meshes were not preconditioned; for them neither a bound nor computed condition
numbers are available.

\subsection{L-shaped bracket with an opening}\label{sec:num-hole}
\begin{figure}[!htbp]
\centering
\includegraphics[width=\textwidth]{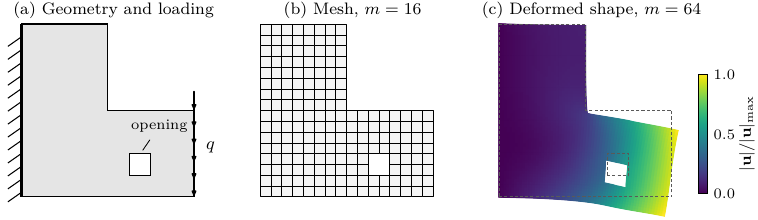}
\caption{L-shaped bracket with an opening: (a) geometry, clamped edge and traction $q$; (b)
assembly-free mesh with void elements removed, $m=16$; (c) deformed shape (magnified) from the
extrapolated assembly-free solution, $m=64$.}
\label{fig:lhole}
\end{figure}

\begin{figure}[!htbp]
\centering
\includegraphics{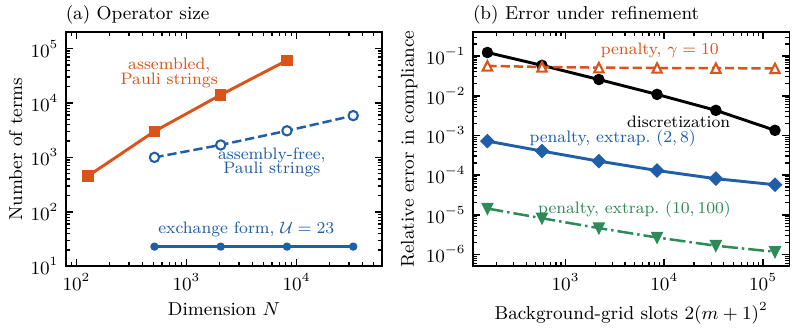}
\caption{L-shaped bracket with an opening: (a) number of terms against the dimension $N$: Pauli
strings of the assembled ($m'=m-1$) and assembly-free operators ($\alpha=7.3$, $\beta=11.1$), and
unitaries of the exchange form; (b) relative error in the compliance due to discretization (assembled
solution, relative to the $m=512$ reference solution, not to the exact continuum solution) and due to the penalty (assembly-free against assembled on the
same mesh), for a single penalty ($\gamma=10$) and for two-penalty extrapolation from $\gamma=2$ and
8 and from $\gamma=10$ and 100. The horizontal axis of (b) counts the displacement slots
$2(m+1)^2$ of the background grid, including nodes in the removed regions and constrained nodes.}
\label{fig:lhole-conv}
\end{figure}

The next example combines a re-entrant corner, an interior opening, a clamped edge
and a distributed load. The L-shaped bracket receives a square opening
$[5/8,3/4]\times[1/8,1/4]$ in its lower arm; the edge $x_1=0$ is clamped, and a uniform downward
traction of total magnitude one acts on the end $x_1=1$ of the lower arm (Fig.~\ref{fig:lhole}). The void quadrant and the opening
are unions of dyadic blocks, so the mask is exact, and the exchange form requires 23 unitaries at
every resolution; the mask predicate depends only on a few leading bits of the element index. The Pauli count is 1,003 at $m=8$, larger than the 458 of the assembled operator,
but smaller from $m=16$ on: 5,861 against 60,658 at $m=64$ (Fig.~\ref{fig:lhole-conv}a). With a
single penalty $\gamma=10$ the
penalty error stays near 5\% and exceeds the discretization error from $m=32$ on, so a single
moderate penalty is not sufficient for fine meshes; with extrapolation from $\gamma=2$ and 8 it
stays 24--170 times below the discretization error up to $m=256$, and with extrapolation from
$\gamma=10$ and 100 it lies between $1.2\times10^{-6}$ and $1.4\times10^{-5}$, three to four orders
of magnitude below the discretization error (Fig.~\ref{fig:lhole-conv}b). The
penalty error must therefore be checked alongside the operator size.

\subsection{Curved meshes and compression}\label{sec:curved}
For the curved bracket and the distorted plate, the exact assembly-free counts grew in proportion
to $N$ over the tested refinements but remain smaller by factors of approximately 3.3 and 1.9 than the assembled counts at comparable resolution
(6,485 against 21,168 and 30,668 against 57,824 at $m=32$; Supplementary Material). Compression of the element part reduces them further, to 2,126 and
10,913 terms at relative displacement errors of $0.5\%$ and $0.1\%$ (2-norm of the displacement
vector, measured against the uncompressed penalty solution on the same mesh; these compression
errors are separate from the penalty and discretization errors), and over two successive refinements that each quadruple $N$ the compressed counts grow by
factors of 1.6--1.8 and then 1.3--1.35. The corresponding direct truncation of the assembled Pauli expansion, a different coefficient
representation from the geometry-factor compression, produces large displacement errors unless
almost all terms are kept. On the randomly perturbed mesh
compression is ineffective, because the tested randomly perturbed geometry does not exhibit
sufficiently rapid Walsh-coefficient decay for effective compression; this is a genuine limit of the
approach.

\section{Quantum compatibility of the preconditioned formulation}\label{sec:circuit-resources}

This section shows that the preconditioned formulation can be provided to quantum algorithms that
act on block encodings, without committing it to a particular solution algorithm. It constructs a
factor of the preconditioned operator whose normalization grows only as $\sqrt{\log m}$, and it
expresses loads and the compliance through this factor.

\subsection{Direct access to the preconditioned operator}\label{sec:direct-block-precond}
Block-encoding $\F$, $\A$ and $\F^{\mathsf T}$ separately and multiplying them would multiply their
normalizations and forfeit much of the benefit of preconditioning. Instead we follow the
full-operator approach of \citet{Deiml2026} and factorize the preconditioned operator. Let $\R$
be an element-level energy factor with $\R^{\mathsf T}\R=\A_{\rm el}$, for example the product of
the square root of the constitutive matrix and the strain--displacement matrix at the
integration points, weighted by the square roots of the quadrature weights and Jacobian
determinants, and define the rectangular matrix
\begin{equation}\label{eq:Hhat-def}
\Hf=\begin{bmatrix}\R\F_{\rm p} & \R\F_{\rm c}\\ \mathbf 0 & \Id_{\rm c}\end{bmatrix}.
\end{equation}
Because the springs vanish on $\F_{\rm p}$ and equal the identity on $\F_{\rm c}$,
\begin{equation}\label{eq:Hhat-gram}
\Hf^{\mathsf T}\Hf=\F^{\mathsf T}\A_{\rm el}\F+\begin{bmatrix}\mathbf 0&\mathbf 0\\ \mathbf 0&\Id_{\rm c}\end{bmatrix}
=\F^{\mathsf T}\A_{\rm el}\F+\F^{\mathsf T}\A_{\rm s}\F=\F^{\mathsf T}\A\F ,
\end{equation}
i.e.\ the preconditioned matrix is exactly the Gram matrix of $\Hf$, in the same way as a
stiffness matrix is $\int\mathbf B^{\mathsf T}\mathbf D\mathbf B$. The nonzero singular values of
$\Hf$ are the square roots of the positive eigenvalues of $\F^{\mathsf T}\A\F$, so
$\kappa_+(\Hf)=\sqrt{\kappa_+(\F^{\mathsf T}\A\F)}$, with $\kappa_+(\Hf)$ the ratio of the
largest to the smallest nonzero singular value, and a quantum algorithm that acts on block
encodings, for example a pseudoinverse by QSVT \citep{Gilyen2019}, can act on $\Hf$ directly.
For H8, Eqs.~\eqref{eq:Hhat-def} and~\eqref{eq:Hhat-gram} are applied to the physical components
only, with $\R^{\mathsf T}\R=\A_{\rm el}^{\rm phys}$ and $\F_{\rm c}^{\mathsf T}\A_{\rm s}^{\rm
phys}\F_{\rm c}=\Id_{\rm c}$, which gives $\Hf_{\rm phys}$. The padding sector, scaled exactly to the
identity as in Theorem~\ref{thm:full-pre-final}, contributes a separate identity block,
$\Hf=\Hf_{\rm phys}\oplus\Id_{\rm pad}$, so that $\Hf^{\mathsf T}\Hf=\F^{\mathsf T}\A\F$ without
counting the padding stiffness twice.

\begin{theorem}[Normalization of the preconditioned factor]\label{thm:direct-precond-block}
Under the assumptions of Theorem~\ref{thm:full-pre-final}, consider a uniform Cartesian hierarchy
clamped on the face $x_1=0$, with $m=2^L$ elements per direction ($L=a$ refinements, i.e.\ $L+1$ levels) and with the dimensionless
penalty fixed under refinement, so that $\eta<1$ does not depend on $m$, $\Hf$ admits a block
encoding, constructed at the level of the specified circuit model in Section~S1.2 of the
Supplementary Material, with
\begin{equation}\label{eq:Hhat-final-bound}
\omega(\Hf)^2\le1+\eta+C_{\rm el}^2(L+1),\qquad C_{\rm el}=2\sqrt d\,\norm{\M_{\rm el}}_2,
\end{equation}
where, in the nondimensional form of Section~\ref{sec:uniform-preconditioning},
$\M_{\rm el}=(\mathbf D/E)^{1/2}\mathbf S$ maps displacement gradients (in units of $1/h$) to
$(\mathbf D/E)^{1/2}\boldsymbol\varepsilon$, with $\mathbf S$ the symmetrization to Voigt strains,
and $\norm{\M_{\rm el}}_2^2=1/(1-\nu)$ in plane stress ($1/(1-2\nu)$ in 3D). (In dimensional
variables the energy factor $\R$ carries $k_0^{1/2}=(Eh^{d-2})^{1/2}$, which cancels against the factor
$k_0^{-1/2}$ of $\F_{\rm p}$ in Remark~\ref{rem:units}.) For H8 the bound applies to $\Hf_{\rm phys}$; the padding sector enters through a sector-controlled
direct sum, with $\omega(\Hf_{\rm phys}\oplus\Id_{\rm pad})=\max\{\omega(\Hf_{\rm phys}),1\}$, and
therefore does not increase the bound. Hence the rectangular factor has
$\omega(\Hf)=O(\sqrt{\log m})$, and, at the level of the specified circuit model and for a fixed
precision of the arithmetic and rotation angles, the index arithmetic, level transfers,
sum/difference transformation and material maps use a number of gates polylogarithmic in $N$.
\end{theorem}

The construction composes isometries, validity projectors, coordinate conversions and
controlled dilations at the level of a specified circuit model; it has not been compiled to a
gate-level network, and only the Gram identity, not an encoded circuit, has been verified
numerically. In practical terms, the rectangular factor of the preconditioned operator admits a
block encoding whose normalization grows only as $O(\sqrt{\log m})$ and whose arithmetic is
polylogarithmic in $N$ in this circuit model; by the standard product construction the
preconditioned matrix itself, $\Hf^{\mathsf T}\Hf=\F^{\mathsf T}\A\F$, then has normalization
$\omega(\F^{\mathsf T}\A\F)\le\omega(\Hf)^2=O(\log m)$. The proof (Section~S1 of the Supplementary Material) combines one block per multilevel level
with one block for the spring modes. The level blocks are the gradient construction of
\citet{Deiml2026}, applied to each displacement component, adapted to the one-face-clamped
boundary conditions through partial-permutation endpoint gathers of normalization one, and
followed by the joint strain/material map $\M_{\rm el}$, encoded by a single dilation normalized
by $\norm{\M_{\rm el}}_2$. Both branches use a common output basis: the orthonormal
discontinuous-Q1 gradient coefficients of the multilevel construction are converted to
quadrature-weighted Gauss-point values by a fixed element-local orthogonal transformation (the
$d$-fold tensor product of a $2\times2$ orthogonal matrix), so that they represent the same element
energy factor $\R$ as the spring branch, including the mixed physical--spring blocks of the Gram
identity, without increasing the normalization. The new ingredient is the spring block, which by Proposition~\ref{prop:algebra}(c) is a
function of the exchanges and the support reflection, evaluated in their common eigenbasis. Combined with
Theorem~\ref{thm:full-pre-final}, the parameter that governs the cost of a block-encoding-based
pseudoinverse of $\Hf$ is
\begin{equation}\label{eq:factor-qsvt}
\omega(\Hf)\,\norm{\Hf^+}_2=O(\sqrt{\log m}) .
\end{equation}
For Q4 the Gram identity~\eqref{eq:Hhat-gram} is reproduced to round-off (below $10^{-13}$) for
$m\le8$, and the bound~\eqref{eq:Hhat-final-bound} exceeds the computed $\norm{\Hf}_2$ by a factor
between 2.7 and 3.0 for $m=2,\dots,64$.

One call of the encoding has the following logical structure (Supplementary Material). Because
$\Hf$ is a horizontal concatenation of blocks acting on different coefficient spaces, the input
itself carries a selector that labels a multilevel level or the spring sector; no preparation on
the input side is needed. Conditioned on this selector, the physical branch applies the level
transfer and gradient network of \citet{Deiml2026}, the conversion to the common Gauss-point output
basis and the joint strain/material map, and the spring branch applies the node-local scaling $\D_c$, undoes the sum/difference transformation,
converts to element-local nodal displacements and applies the complete element energy factor
(quadrature-weighted gradient and constitutive map), and it prepares an output flag that separates the energy rows from the identity rows, with amplitudes
proportional to $\omega_{\rm c}$ and 1. Finally the selector is projected onto a fixed state whose
amplitudes are proportional to the normalizations of the blocks, which yields $\Hf/\Omega$ with
$\Omega^2=1+\omega_{\rm c}^2+\sum_\ell\omega_\ell^2$. Apart from $O(\log m)$-bit index arithmetic, the assembly-free parts of the circuit do not depend on
the mesh size.

\subsection{Loads and the compliance}\label{sec:readout}
Here $\A$, $\F$, $\fvec$ and $\mathcal C$ are taken in the same nondimensional units (Remark~\ref{rem:units}).
The shared load~\eqref{eq:compatible-load} has no component on the spring modes, so
$\F^{\mathsf T}\fvec=[\G_{\rm BPX}^{\mathsf T}\fvec_{\rm phys};\,\mathbf 0]$: the springs are needed
to precondition the operator, but they do not complicate the load. For a corner point load in the
two-dimensional Q4 cantilever the transformed load is a uniform superposition over the $L+1$ levels,
and for a uniform edge traction on the same benchmark each level carries the consistent traction of
that level; both are prepared with $O(L)$ rotations and no $O(N)$ data loading. In three dimensions
the level weights $(h_\ell/h)^{-1/2}$ make the level amplitudes nonuniform. With $\tilde\fvec=\F^{\mathsf T}\fvec$, Eq.~\eqref{eq:Hhat-gram} gives
\begin{equation}\label{eq:compliance-factor}
\mathcal C=\fvec^{\mathsf T}\uvec=\bigl\|(\Hf^{+})^{\mathsf T}\tilde\fvec\bigr\|_2^2 ,
\end{equation}
so the compliance is the squared length of a vector produced by a pseudoinverse of $\Hf$, and no
displacement field has to be reconstructed. Other linear outputs
$\gvec^{\mathsf T}\uvec$ follow as overlaps of two such vectors \citep{Deiml2026}. Dense checks for
$m=2$ and 4 reproduce the compliance from~\eqref{eq:compliance-factor} to below
$3\times10^{-14}$.

\subsection{Scope}\label{sec:qc-scope}
A complete quantum solver additionally requires a procedure that applies the pseudoinverse (for
example a polynomial approximation, as in methods based on QSVT) and an estimation procedure for
the output. Their costs depend on the algorithm chosen, on a numerical
lower bound for the positive spectrum and on the size of the output amplitude, and they are not
analysed here. The formulation-dependent inputs to such an analysis are the bounded condition
number of Theorem~\ref{thm:full-pre-final}, the normalization of
Theorem~\ref{thm:direct-precond-block} and the load representation above; the penalty
contribution to the error of the compliance is bounded by Proposition~\ref{prop:penalty-bound}.

\section{Discussion}\label{sec:discussion}

\subsection{Implications for finite element practice}
The broader implication is methodological: a formulation that is optimal for classical solvers
need not be optimal for quantum ones, and choices that classical practice avoids, such as
duplicated unknowns, can become advantageous. Three aspects are of particular relevance to
practitioners. The formulation retains the standard
element stiffness matrices, shape functions and boundary conditions, and changes how
compatibility and essential boundary conditions are enforced. The more compact operator comes at a cost: additional copies of the
displacements and an $O(1/\gamma)$ penalty error, which must be controlled by the choice of the
penalty or by extrapolation. Finally, the exchange structure is exact for Cartesian meshes, graded
meshes and dyadic masks, whereas the preconditioning theorem is proved only for uniform Cartesian
meshes; neither should be extrapolated to arbitrary unstructured industrial meshes.

\subsection{Benefits and limits of retaining element-local degrees of freedom}\label{sec:broken-role}
By retaining element-local displacements, the assembly-free form confines, for uniform meshes, the
constitutive physics to a fixed-size matrix and isolates the mesh topology in simple exchange operators. This
yields encodings whose number of terms is independent of the mesh, exposes spring modes that can
be scaled exactly, and allows multilevel preconditioning to act only on the physical
displacements. It reverses a familiar principle: for classical solvers fewer degrees of freedom
are preferable and assembly is the natural first step, whereas for quantum algorithms, where the data-register size grows logarithmically with the
encoded dimension, a larger number of degrees of freedom with
more structure can be advantageous. Domain decomposition methods of the FETI and
BDDC type \citep{FarhatRoux1991,Dohrmann2003} and discontinuous Galerkin schemes
\citep{Arnold2002} also duplicate interface degrees of freedom, but in order to solve subdomains in parallel
or to allow discontinuous approximations; here the duplicates are retained because their structure
is inexpensive to express as a quantum circuit. Element-wise representations also appear inside
quantum FE encodings: the factorization of \citet{Deiml2026} evaluates gradients in auxiliary
spaces of discontinuous Q1 functions on every level, and \citet{Alkadri2025} and
\citet{SatoKondo2026} use connectivity or element restriction operators. In these constructions
the element-wise values are auxiliary quantities of the encoding and the unknowns are nodal. Here
they are the unknowns of the discretization, as in interior-penalty methods; this is what turns the
connectivity into commuting exchanges and the supports into a reflection, and it is also the source
of the penalty error.

Table~\ref{tab:related} compares the formulation with related quantum FE constructions. The
assembly-free form is most attractive for non-periodic supports, three-dimensional elements and
masked domains; for periodic plane-stress problems the encoding of \citet{Suresh2026} is very
compact, and a conforming formulation combined with the construction of \citet{Deiml2026} and
Korn's inequality would reach the same asymptotic parameter~\eqref{eq:factor-qsvt} as ours. Two
encodings should be distinguished. The exchange form of
Section~\ref{sec:structure} describes the unpreconditioned penalty operator, whose condition number
grows as $\gamma h^{-2}$. The construction of Section~\ref{sec:circuit-resources} block-encodes the
factor $\Hf$: its physical branch is the multilevel network of \citet{Deiml2026}, applied to each
displacement component, and its spring branch is the function of the exchanges and the support
reflection given by Proposition~\ref{prop:algebra}(c). The two encodings therefore share the
operations $\V_i$ and $\mathbf Z_D$, the element factor and the data layout, in which element data,
loads and masks are attached to element copies without gathering at the nodes; what the solver adds
is the multilevel prolongation. A
table in Section~S1 of the Supplementary Material lists which results are proved, which are verified numerically and which
rest on modelling assumptions.

\begin{table}[!htbp]
\centering
\caption{Position of the present formulation among related quantum FE operator constructions, as
described in the cited works and in Sections~\ref{sec:intro} and~\ref{sec:broken-role}.}
\label{tab:related}
\footnotesize
\begin{tabular}{@{}>{\raggedright\arraybackslash}p{0.22\linewidth}>{\raggedright\arraybackslash}p{0.21\linewidth}>{\raggedright\arraybackslash}p{0.25\linewidth}>{\raggedright\arraybackslash}p{0.21\linewidth}@{}}
\toprule
Approach & Problem class & Size of the description & Preconditioning\\
\midrule
Pauli expansion of the assembled matrix \citep{Hantzko2024,Koska2024,Georges2025,Philo2026} & any assembled FE matrix & $\Theta(N)$ Pauli terms on uniform grids (proved here); $\propto N$ on the curved meshes (observed) & none\\\addlinespace
\citet{Suresh2026} & periodic plane-stress Q4 & 17 terms, mesh-independent normalization & none\\\addlinespace
\citet{Wu2026} & voxel meshes, variational solver; supported plane stress and heat conduction shown & mesh-independent bound; 18 terms in the examples & none\\\addlinespace
\citet{Holscher2026} & supported plane-stress Q4, compliance output & one term per element & none; named as future work\\\addlinespace
\citet{Alkadri2025} & scalar problems, tensor-product Lagrange elements & element-matrix entries combined with connectivity operators; mesh-independent & none\\\addlinespace
\citet{Deiml2026} & scalar elliptic problems, conforming & multilevel factor circuits & BPX, built into the algorithm\\\addlinespace
\citet{Fressart2026} & Poisson problem, conforming & bounds on the block-encoding parameters of the preconditioned system & two-level additive Schwarz, BPX as local solver\\\addlinespace
\citet{SatoKondo2026} & two-dimensional isotropic elasticity, conforming; strain-energy output & factorized assembly with analytically diagonalized element stiffness; gate complexity logarithmic in $N$ & none reported\\\addlinespace
This work & clamped vector elasticity, Q4 and H8, dyadic openings & at most 13 / 81 / 23 unitaries (penalty operator); conforming operator as a product of the same operations; Pauli terms $\propto dN^{1/d}$ & spring modes diagonalized exactly by the same operations and scaled to unit stiffness, BPX on the physical part; bound proved on uniform Cartesian hierarchies\\
\bottomrule
\end{tabular}
\end{table}

\subsection{Penalty or projection}\label{sec:projection}
Keeping element copies is not the only way to separate element physics from connectivity. The
direct stiffness method itself, Eq.~\eqref{eq:standard}, is a product of a gather operator, the
element matrices and a scatter operator, and this structure has been used for quantum encodings
of the assembled operator: \citet{Alkadri2025} write the global matrix of scalar tensor-product
elements as a linear combination over the entries of the element matrix, each paired with a
connectivity operator, with a number of terms that does not depend on the mesh, and the
factorization of \citet{Deiml2026} places element-wise coefficients between two gradient
operators. In the copy layout the same operator follows from Proposition~\ref{prop:algebra}(b):
projecting the difference modes out instead of penalizing them gives
\begin{equation}\label{eq:projected}
\mathbf Q^{\mathsf T}\A_{\rm el}\mathbf Q=\mathbf M_{\rm copy}^{-1/2}\K_0\mathbf M_{\rm copy}^{-1/2},
\end{equation}
the assembled stiffness matrix up to a diagonal scaling, or in copy form
$\boldsymbol\Pi_0\A_{\rm el}\boldsymbol\Pi_0$, a product of the element part with the $d$ exchanges
and the support reflection. The penalty and projected forms thus use the same operations, combined
as a sum and as a product, and they trade accuracy, normalization and structure differently
(Table~\ref{tab:penalty-projection}).

\begin{table}[!htbp]
\centering
\caption{Penalty and projected forms of the clamped Q4 cantilever ($E=1$, $\nu=0.3$,
$\alpha=\beta=\gamma=10$). The unitary counts were computed for $m=2$ and 4; $\kappa_{\rm eff}$ of
the projected form uses its smallest positive eigenvalue.}
\label{tab:penalty-projection}
\footnotesize
\begin{tabular}{@{}>{\raggedright\arraybackslash}p{0.36\linewidth}>{\raggedright\arraybackslash}p{0.27\linewidth}>{\raggedright\arraybackslash}p{0.29\linewidth}@{}}
\toprule
& Penalty form (sum) & Projected form (product)\\
\midrule
Unitaries when written as one linear combination & 13 & 360 (chosen expanded representation; the product itself need not be implemented this way)\\
1-norm $\lambda$ (normalization) & 51.7 & 1.73\\
Penalty error & $O(1/\gamma)$ & none\\
Operator on the copy space & nonsingular & singular; kernel = spring modes (72\% of the copies at $m=8$)\\
$\kappa_{\rm eff}=\lambda\norm{\A^{+}}_2$, $m=2$ / 4 / 8 & 2{,}022 / 7{,}945 / 31{,}903 & 64 / 251 / 1{,}014\\
Preconditioned operator representation & factor $\Hf$: multilevel and spring branches & physical branch of $\Hf$ \citep{Deiml2026}\\
\bottomrule
\end{tabular}
\end{table}

The compactness of the penalty representation does not necessarily translate into an advantage
for every quantum solution algorithm. Quantum linear-system methods based on block encodings may
benefit from projecting onto the conforming subspace: the projected form has no penalty error, in
the tabulated $\gamma=10$ representations its normalization is smaller by a factor of approximately thirty, and
$\kappa_{\rm eff}$ is smaller by a factor of about $3\gamma$, because the penalty adds about $5\gamma$
to the 1-norm but leaves the smallest eigenvalue almost unchanged; with multilevel preconditioning it
leads to the conforming construction of \citet{Deiml2026}. Its drawbacks are structural: it is
singular on the copy space and must be inverted as a pseudoinverse on the range of
$\boldsymbol\Pi_0$, and it is a product rather than a sum, which as one linear combination requires
360 distinct unitaries instead of 13, with a Pauli expansion that grows in proportion to $N$ (112
and 782 strings for $m=2$ and 4). The penalty form is therefore the shorter description of a
nonsingular operator, which may be beneficial for algorithms that evaluate the operator term by
term \citep{Sato2021,BravoPrieto2023}; this potential advantage, and the measurement variance and
optimization costs that such algorithms incur, have not been tested here. After
preconditioning, the springs contribute robustness rather than speed:
Theorem~\ref{thm:full-pre-final} shows that they do not destroy mesh-independent conditioning, not
that they make a solve cheaper. The principal contribution of the present work is thus a finite element formulation
that exposes operator structure and permits an exact treatment of the artificial compatibility
modes, rather than the optimization of a particular quantum solution algorithm.

The comparison of Table~\ref{tab:shift} is with our own decomposition of the assembled operator and
not with the best available one. An entry-wise construction in the manner of \citet{Alkadri2025},
extended to vector elasticity, would require a number of terms of the order of the number of
entries of the element matrix (64 for Q4) instead of 109 or 196; its terms are block encodings of
connectivity operators rather than single signed permutations, its supports are imposed by
Lagrange multipliers, and we have not implemented it.

\subsection{Limitations}\label{sec:limits}
The mesh-independent conditioning bound is a property of BPX applied to the conforming
displacements; the contribution of the assembly-free form is that its stiff springs do not destroy
mesh-independent conditioning, and they introduce a modest conditioning overhead in the reported
experiments: the measured increase in the
condition number is 2--8\%. The preconditioning theorem covers uniform Cartesian Q4/H8 hierarchies (complete copy graphs) with
a clamped boundary of positive measure and fixed isotropic material away from the incompressible
limit; the node-local diagonalization does not extend automatically to re-entrant or void-adjacent
nodes of masked domains. Its numerical support is limited to $\nu=0.3$ in 3D and to meshes on which the condition number is still
rising slowly, and no condition numbers were computed for masked, graded or curved meshes; in 3D,
springs scaled with the local element size may be beneficial on graded meshes, although a varying
stiffness is not mathematically required. Curved-mesh compression is empirical
and fails for rough geometry, and fully unstructured meshes are not covered. The discretization is
a penalty method: extrapolation removes the leading $O(1/\gamma)$ error, but local stress errors
next to singular points decrease more slowly than global measures. The Pauli counts of the assembled operators are $\Theta(N)$ only on uniform grids (proved) and
were computed on curved meshes; the numbers of unitaries of the assembled operators in the
shift-based form are upper bounds and were not compared with entry-wise assembled encodings
\citep{Alkadri2025}. For quantum solvers based on block encodings the projected form of
Section~\ref{sec:projection} may be preferable to the penalty form; the advantage of the penalty
form, a short linear combination of unitaries of a nonsingular operator, concerns term-wise
algorithms that were not tested. Loads and outputs are treated for point loads, edge
tractions and compatible linear functionals; general loads and full displacement fields need
efficient state preparation and readout. Finally, no quantum solution algorithm is implemented,
simulated or costed: the paper establishes operator representations, conditioning and a factorized
block encoding, but neither a solver-level validation nor gate counts, error budgets or
fault-tolerant resource estimates; the comparisons are with other quantum FE formulations, and no
experiment on quantum hardware and no speed-up is reported.

\section{Conclusions}\label{sec:conclusions}

This work has approached quantum finite element analysis from the formulation rather than from
the encoding. In the assembly-free formulation each element retains its local displacement degrees
of freedom, and compatibility and clamped supports are enforced by penalty springs; under the
register conventions used here this costs two additional qubits for Q4 and three for H8 and
separates the fixed element stiffness from the mesh topology. On uniform meshes each directional
coupling is a single exchange and the supports a single reflection, so that clamped Q4 and H8
penalty operators are sums of at most 13 and 81 structured unitaries, independent of mesh
refinement (at most 23 for dyadically masked Q4 domains, given the mask predicate), and on uniform
grids their Pauli expansions contain $\Theta(dN^{1/d})$ strings, against $\Theta(N)$ for the assembled operators in the
standard binary global-node encoding on comparable grids.

On complete Cartesian copy graphs the exchanges and the support reflection commute. Their weighted
linear combination is the spring stiffness, their product
gives the conforming stiffness on the compatible copy subspace, and their common eigenbasis, a
node-local sum/difference transformation, diagonalizes the spring stiffness with closed-form mode
stiffnesses. Scaling the spring modes exactly and applying BPX preconditioning to the physical
displacements gives a positive-spectrum condition number bounded independently of mesh refinement
on complete Cartesian hierarchies, uniformly for sufficiently stiff penalties; on the reported
benchmarks the springs add 2--8\% to the condition numbers of conforming BPX. For one-face-clamped Cartesian hierarchies, the preconditioned operator admits a rectangular factor
$\Hf$, with $\Hf^{\mathsf T}\Hf=\F^{\mathsf T}\A\F$, that can be block-encoded with normalization
$O(\sqrt{\log m})$.

A complementary-energy argument bounds the penalty error in the compliance by a computable,
asymptotically sharp quantity; this bound controls the unextrapolated error, while the analytic
penalty expansion justifies Richardson extrapolation. Numerical investigations of two- and
three-dimensional cantilevers verify the uniform-grid operator representations and assess the
predicted conditioning, while the Q4 traction benchmark demonstrates the tightness of the
compliance bound. The masked, graded and curved-mesh examples illustrate the wider applicability
and the limitations of the operator representation: compression is
empirical on curved meshes and ineffective on the tested randomly perturbed geometry, and the conditioning theorem is
proved only for complete Cartesian hierarchies.

The benefits are algorithm dependent. Among the decompositions considered here, the penalty form provides the shorter
linear-combination representation, of a nonsingular operator, and is potentially beneficial for algorithms that evaluate it term by term, whereas the
projected conforming form has no penalty error and a smaller normalization and may be preferable
for solvers based on block encodings; no solver-level advantage has been established. Future work
will address solver-level validation, compilation and resource estimation, general loads and
outputs, nearly incompressible and heterogeneous elasticity, the extension of the preconditioning
theorem to masked, graded and multi-block domains, and consistent Nitsche or symmetric
interior-penalty couplings \citep{Arnold2002}, which would remove the $O(1/\gamma)$ penalty error if
the exchange structure is preserved. More broadly, choosing the unknowns so that physics, topology
and supports separate into a few commuting operations is a route towards quantum-compatible
preconditioned solvers for structured three-dimensional linear elasticity.

\appendix
\section{Principal notation}\label{app:notation}
\begin{center}\footnotesize
\begin{tabular}{@{}l>{\raggedright\arraybackslash}p{0.66\linewidth}@{}}
\toprule
Symbol & Meaning\\
\midrule
$d$, $m=2^a$, $h$ & dimension, elements per direction, element size\\
$m'$ & elements per direction of an assembled comparison mesh\\
$N$, $q$ & number of encoded degrees of freedom (copies), number of qubits ($N=2^q$)\\
$\K_e$, $\bar k$, $k_0$ & element stiffness matrix (for H8 the physical $\K_e^{\rm phys}$ embedded in a padded $32\times32$ block), its spectral norm, stiffness scale $Eh^{d-2}$\\
$\K_0$ & conforming stiffness on the free nodal displacements, $\J_0^{\mathsf T}\A_{\rm el}\J_0$\\
$\mathcal B$, $\mathcal D$, $\Cj$, $\Cj_D$ & bonds; clamped copies; jump matrix; selection of clamped copies\\
$\A=\A_{\rm el}+\A_{\rm c}+\A_{\rm bc}$ & assembly-free operator: element part, interface springs, support springs\\
$\alpha$, $\beta$, $\gamma$ & interface and support spring stiffness; dimensionless penalty ($\alpha=\beta=\gamma k_0$)\\
$\V_i$, $\mathbf Z_D$, $\chi$ & exchange in direction $i$; support reflection; element activity indicator\\
$\J_0$, $\mathbf M_{\rm copy}$ & map from free nodal displacements to copies; number of copies per node\\
$\mathbf Q$, $\boldsymbol\Pi_0$ & normalized sum modes $\J_0\mathbf M_{\rm copy}^{-1/2}$; projector $\mathbf Q\mathbf Q^{\mathsf T}$ onto conforming displacements\\
$\W_c$, $\D_c$ & spring modes with positive stiffness; their scaling\\
$\G_{\rm BPX}$, $\F=[\F_{\rm p},\F_{\rm c}]$ & multilevel (BPX) frame; complete preconditioning transformation\\
$p_{\min}$, $\eta=\bar k/p_{\min}$ & smallest spring-mode stiffness $\min\{\beta,2\alpha\}$; ratio of element stiffness to it\\
$\M_{\rm el}$ & nondimensional material map $(\mathbf D/E)^{1/2}\mathbf S$ from displacement gradients to $(\mathbf D/E)^{1/2}\boldsymbol\varepsilon$\\
$\kappa$, $\kappa_+$, $\kappa_{\rm eff}$ & condition number; positive-spectrum condition number (singular preconditioned matrices); $\lambda\norm{\A^{+}}_2$, with $\A^{+}=\A^{-1}$ for nonsingular $\A$\\
$c_{\rm BPX}$, $C_{\rm BPX}$ & bounds of the positive spectrum of the BPX-preconditioned conforming matrix\\
$\mathbf D$ & constitutive matrix (the support set of clamped copies is $\mathcal D$)\\
$\Un$, $\Th$, $\lambda$ & number of unitaries, number of Pauli strings, 1-norm of the coefficients\\
$\omega$ & normalization of a block encoding\\
$\Hf$, $\R$, $L$ & energy factor with $\Hf^{\mathsf T}\Hf=\F^{\mathsf T}\A\F$; element energy factor; number of refinements ($L+1$ levels)\\
$\mathcal C$, $\mathcal C_0$, $c_1$, $\tau$ & compliance; conforming compliance; penalty coefficient of Proposition~\ref{prop:penalty-bound}; target relative accuracy\\
\bottomrule
\end{tabular}
\end{center}

\section{Proofs of Propositions~\ref{prop:copy-walsh-final} and~\ref{prop:algebra} and Theorem~\ref{thm:full-pre-final}}\label{app:precond-proofs}

\paragraph{Proposition~\ref{prop:copy-walsh-final}}
In one interior direction, the two copies of a node that lie in neighbouring elements are joined by
one spring, with matrix $\alpha(\Id-\mathbf X)$. Different directions act on different bits of the
copy index and therefore add. The support spring acts equally on every copy of a clamped node and
contributes $\beta\Id$. Since $\mathbf S_2\mathbf X\mathbf S_2=\operatorname{diag}(1,-1)$, the
sum/difference basis gives $1-(-1)^{z_j}\in\{0,2\}$ for each direction, which
proves~\eqref{eq:Px-diag-final}. \qed

\paragraph{Proposition~\ref{prop:algebra}}
Each $\V_i$ is a permutation that swaps disjoint pairs of copies along direction $i$ and leaves all
other indices unchanged; exchanges in different directions act on different index registers and
commute. $\mathbf Z_D$ changes the sign of the copies with chain index $j=0$ in direction 1, which
$\V_1$ never moves, and it acts on the registers of direction 1 only, so it commutes with every
$\V_i$. Commuting involutions have a common
eigenbasis with eigenvalues $\pm1$, and the products in~\eqref{eq:projector}
and~\eqref{eq:spring-inverse} are the orthogonal projectors onto their joint eigenspaces. (a) is
Proposition~\ref{thm:shift}. (b) $(\Id+\V_i)/2$ replaces the two copies of every bond in direction
$i$ by their average and leaves the copies on outer faces unchanged, and $(\Id+\mathbf Z_D)/2$
removes the clamped copies; the product therefore projects onto vectors whose copies agree at
every node and vanish on $\Gamma_D$, i.e.\ onto the range of $\J_0$, with orthonormal basis
$\mathbf Q$. Then $\mathbf Q^{\mathsf T}\A_{\rm el}\mathbf Q=\mathbf M_{\rm copy}^{-1/2}\J_0^{\mathsf T}\A_{\rm
el}\J_0\mathbf M_{\rm copy}^{-1/2}=\mathbf M_{\rm copy}^{-1/2}\K_0\mathbf M_{\rm copy}^{-1/2}$. (c) On the joint eigenspace with signs
$\mathbf s$, (a) gives $\A_{\rm s}=\lambda_{\mathbf s}\Id$; this is the diagonalization of
Proposition~\ref{prop:copy-walsh-final}, with $|\mathbf z|$ the number of $s_i=-1$ ($i\le d$) and
$b=1$ for $s_{d+1}=-1$, and $\W_c\D_c^2\W_c^{\mathsf T}$ collects exactly the reciprocals of the
positive eigenvalues. For H8 the padding slot is bonded and clamped like the displacement
components, so (a) and (c) apply unchanged on the padded space; in (b) the product projects onto the
compatible fields of all four components, and $\boldsymbol\Pi_0=\J_0\mathbf M_{\rm copy}^{-1}\J_0^{\mathsf T}$
holds on the physical sector (equivalently, on the padded space once $\J_0$ is extended by the
compatible padding component). \qed

\paragraph{Theorem~\ref{thm:full-pre-final}}
On a uniform grid $\A_{\rm el}$ is block diagonal over elements, so $\norm{\A_{\rm el}}_2=\bar k$.
Since $\W_c$ has orthonormal columns and $\norm{\D_c}_2\le p_{\min}^{-1/2}$,
$\norm{\F_{\rm c}^{\mathsf T}\A_{\rm el}\F_{\rm c}}_2\le\eta$. Let $\R^{\mathsf T}\R=\A_{\rm el}$,
and for coefficient vectors $\mathbf x$, $\mathbf y$ of the physical and spring sectors set
$\mathbf a=\R\F_{\rm p}\mathbf x$ and $\mathbf b=\R\F_{\rm c}\mathbf y$. The springs vanish on
$\F_{\rm p}$, have no cross term with it, and equal the identity on $\F_{\rm c}$; hence
\begin{equation}\label{eq:energy-ab-final}
\begin{bmatrix}\mathbf x\\ \mathbf y\end{bmatrix}^{\mathsf T}\F^{\mathsf T}\A\F
\begin{bmatrix}\mathbf x\\ \mathbf y\end{bmatrix}=\norm{\mathbf a+\mathbf b}^2+\norm{\mathbf y}^2 ,
\end{equation}
with $\norm{\mathbf b}^2\le\eta\norm{\mathbf y}^2$. Young's inequality
$2|\mathbf a^{\mathsf T}\mathbf b|\le t\norm{\mathbf a}^2+t^{-1}\norm{\mathbf b}^2$ with
$t=\sqrt\eta$ yields
\begin{equation}
(1-\sqrt\eta)\bigl(\norm{\mathbf a}^2+\norm{\mathbf y}^2\bigr)
\le\norm{\mathbf a+\mathbf b}^2+\norm{\mathbf y}^2
\le(1+\sqrt\eta+\eta)\bigl(\norm{\mathbf a}^2+\norm{\mathbf y}^2\bigr).
\end{equation}
The kernel of $\F^{\mathsf T}\A\F$ consists of the vectors $(\mathbf x,\mathbf 0)$ with
$\G_{\rm BPX}\mathbf x=\mathbf 0$, because $\A$ is positive definite and $\J_0$ is injective; all
estimates below are taken on its orthogonal complement (the positive-spectrum subspace), where
$\mathbf x\perp\ker\G_{\rm BPX}$ and $\mathbf y$ is arbitrary. Finally
$\norm{\mathbf a}^2=\mathbf x^{\mathsf T}\G_{\rm BPX}^{\mathsf T}\K_0\G_{\rm BPX}\mathbf x$ lies
between $c_{\rm BPX}\norm{\mathbf x}^2$ and $C_{\rm BPX}\norm{\mathbf x}^2$ on the complement of the kernel of the
physical frame, by~\eqref{eq:c0C0-final}, which gives the spectral interval and the
bound~\eqref{eq:conditioning_bound}. The prefactor decreases monotonically in $p_{\min}$ and tends to
one as $\alpha,\beta\to\infty$. For H8 the padding component is decoupled from the physical
components and its operator, the identity plus the springs, is diagonal in the sum/difference
basis with entries $1+\beta b+2\alpha|\mathbf z|\ge1$; scaled by the inverse square roots of these
entries it becomes exactly the identity, which lies in the spectral interval above, and the
argument applies to the physical components with $\bar k=\norm{\K_e^{\rm phys}}_2$. \qed


\section*{Declaration of competing interest}
The authors declare that they have no known competing financial interests or personal relationships
that could have appeared to influence the work reported in this paper.

\section*{Funding}
A.D.acknowledges support from DST INSPIRE(DST/INSPIRE/04/2020/000928).

\section*{Data availability}
The numerical results supporting the findings of this study, including operator term counts,
condition numbers, penalty errors and the verification of the compliance bound, are reported in
the article and its Supplementary Material. The computational procedures, model parameters and
details of the independent numerical verification are documented in the Supplementary Material.
The computational implementation is not publicly distributed.

\bibliographystyle{elsarticle-harv}
\bibliography{refs}

\end{document}


\maketitle

\noindent
This document contains the derivations and data behind the results of the main paper.
Section~\ref{sec:mainproofs} gives the register form of the operator, the proofs of the Pauli and
exchange-form counts and of Theorem~\ref{M-thm:direct-precond-block} of the main paper, and the status
of all results; the proofs of Propositions~\ref{M-prop:copy-walsh-final} and~\ref{M-prop:algebra} and
Theorem~\ref{M-thm:full-pre-final} are in Appendix~B of the main paper.
The remaining sections are grouped as follows.
\begin{itemize}\itemsep1pt
\item Operator counts: Section~\ref{sec:tools} collects notation and the Pauli toolkit;
Sections~\ref{sec:2el}--\ref{sec:1dgeneral} treat one dimension, starting from a two-element bar
that can be solved in closed form; Sections~\ref{sec:q4} and~\ref{sec:2d} treat plane stress and
Section~\ref{sec:3d} the extension to $d$ dimensions; Section~\ref{sec:assembled} gives the
assembled operators, and Section~\ref{sec:perm} shows that the counts are invariant under
permutations of the qubit registers.
\item Penalty and geometry: Section~\ref{sec:penalty} treats the penalty error and conditioning,
Sections~\ref{sec:lshape}--\ref{sec:curved} masked, graded and curved meshes, and
Section~\ref{sec:opextra} collects further figures and tables of the operator study.
\item Exchange form and accuracy: Sections~\ref{sec:shiftsupp}--\ref{sec:lholesupp} support the
exchange form, the penalty extrapolation, the penalty-error-controlled cost proxy and the L-shaped
bracket with an opening.
\item Preconditioning and block encoding: Section~\ref{sec:bpxsupp} gives the Q4 and H8
preconditioning data, Section~\ref{sec:circuitsupp} the derivations behind the block encoding of the
preconditioned factor and the transformed point load, and Section~\ref{sec:tractionsupp} the
discretization and penalty accuracy of the traction benchmark, including a test of the compliance
bound of the main paper.

\end{itemize}
Sections, equations, tables,
figures and theorems of this document carry the prefix~S (e.g.\ Eq.~\eqref{eq:Ael},
Table~\ref{tab:status}); numbers without the prefix refer to the main paper.

\paragraph{Units and penalty convention.}
Unless stated otherwise, stiffnesses are given in the nondimensional form of the main paper: they
are divided by the element stiffness scale $k_0=Eh^{d-2}$ of the (finest) elements, i.e.\ $k_0=Et$
for Q4 plane stress and $k_0=Eh$ for H8 (Eq.~\eqref{M-eq:gamma} of the main paper). In this form
the Q4 element matrix is independent of the element size, the H8 element matrix is that of a unit
cube on every mesh, and the H8 padding stiffness is one. Penalties are specified in the same
nondimensional form: $\alpha=\beta=\gamma$ in a table or caption means $\alpha=\beta=\gamma k_0$
in dimensional variables, and fixed values such as $\alpha=7.3$, $\beta=11.1$ are likewise in units
of $k_0$. Relative errors and condition numbers do not depend on this choice of units.

\tableofcontents

\section{Proofs and status of the results stated in the main paper}\label{sec:mainproofs}
This section gives the register form of the operator, the Pauli and exchange-form counts, the proof
of Theorem~\ref{M-thm:direct-precond-block} of the main paper and the status of all main results,
in the notation of the main paper (bold matrices). The proofs of
Propositions~\ref{M-prop:copy-walsh-final} and~\ref{M-prop:algebra} and
Theorem~\ref{M-thm:full-pre-final} are given in Appendix~B of the main paper and are not repeated
here.

\begingroup
\newcommand{\A}{\mathbf{A}}\newcommand{\K}{\mathbf{K}}\newcommand{\F}{\mathbf{F}}\newcommand{\Id}{\mathbf{I}}
\newcommand{\V}{\mathbf{V}}\newcommand{\J}{\mathbf{J}}\newcommand{\W}{\mathbf{W}}\newcommand{\G}{\mathbf{G}}
\newcommand{\D}{\mathbf{D}}\newcommand{\R}{\mathbf{R}}\newcommand{\M}{\mathbf{M}}\newcommand{\Hf}{\mathcal{H}}
\newcommand{\Th}{\mathcal{T}}\newcommand{\Un}{\mathcal{U}}\newcommand{\norm}[1]{\lVert #1\rVert}
\subsection{Register form, Pauli counts and the exchange-form count}\label{app:proofs}

\paragraph{Operator in register form}
We use the quantum notation $|j\rangle$ for the $j$-th unit vector and $|j\rangle\langle k|$ for
the matrix with a single one in row $j$ and column $k$. Let $S_a=\sum_{j=0}^{2^a-2}|j+1\rangle\langle j|$ be the non-cyclic increment on $a$ qubits, let
$P_0=|0\rangle\langle0|$, $P_1=|1\rangle\langle1|$, $L_{01}=|0\rangle\langle1|$,
$L_{10}=|1\rangle\langle0|$, and let $D^+=I-P_1^{\otimes a}$ and $D^-=I-P_0^{\otimes a}$ mark the
elements that have an upper and a lower neighbour. A spring in direction $i$ joins the copy with
$\ell_i=1$ in element $e$ to the copy with $\ell_i=0$ in element $e+\mathbf e_i$. With homogeneous
material, the three parts of Eq.~\eqref{M-eq:split} are
\begin{align}
A_{\rm el}&=I_E\otimes K_e,\label{eq:Ael}\\
A_{\rm c}&=\alpha\sum_{i=1}^d\Big[D^+\otimes P_1+D^-\otimes P_0-S_a\otimes L_{01}-S_a^\dagger\otimes L_{10}\Big]_{(E_i,\ell_i)}\otimes\Pi_{\mathsf D},\label{eq:Ac}\\
A_{\rm bc}&=\beta\,\big[P_0^{\otimes a}\big]_{E_1}\otimes\big[P_0\big]_{\ell_1}\otimes\Pi_{\mathsf D}
\quad\text{(clamped face $x_1=0$)},\label{eq:Abc}
\end{align}
where $[\cdot]_{(E_i,\ell_i)}$ acts on the element register $E_i$ and the local bit $\ell_i$,
identities act elsewhere, and $\Pi_{\mathsf D}$ is the projector onto bonded displacement slots
($\Pi_{\mathsf D}=I$ for Q4 and for H8 with bonded padding). Let $s_\Pi$ and $\pi_\Pi$ be the
number of Pauli strings and the 1-norm of $\Pi_{\mathsf D}$ ($s_\Pi=\pi_\Pi=1$ with bonded padding).

\begin{proposition}[Pauli expansion of the springs]\label{thm:main}
For $m=2^a\ge2$, the spring operator~\eqref{eq:Ac} has
\begin{equation}\label{eq:main}
\Th(A_{\rm c})=s_\Pi(3dm-3d+1),\qquad
\lambda(A_{\rm c})=d\,\alpha\,\pi_\Pi\Bigl(\log_2m+2-\frac2m\Bigr).
\end{equation}
\end{proposition}

Each direction contributes $s_\Pi(3m-2)$ strings, $s_\Pi(2m-2)$ off the diagonal and $s_\Pi m$ on it,
and directions share only the strings that are the identity on all element and local registers.
The count follows from writing the increment as a sum of ladder-operator products: adding one to a
binary number flips a trailing block $0\,1^{t-1}$ into $1\,0^{t-1}$, so a spring that crosses a
carry of length $t$ becomes a $(t+1)$-body term, which together with its transpose is a sum of
$2^t$ Pauli strings. The step-by-step derivation is given in Section~\ref{sec:1dgeneral}, and
Algorithm~\ref{alg:closed} (Section~\ref{sec:opextra}) lists all strings without forming a matrix.

\paragraph{Q4 element matrix}
For a square Q4 element of unit thickness, with local qubits ordered $(\ell_2,\ell_1,\mathsf D)$,
\begin{multline}\label{eq:q4main}
\frac{1-\nu^2}{E}K_e=\frac{3-\nu}{24}\big(4\Pa{III}-\Pa{IXI}-\Pa{XII}-2\Pa{XXI}\big)
+\frac{1+\nu}{8}\big(\Pa{XIZ}-\Pa{IXZ}+\Pa{YYX}+\Pa{ZZX}\big)\\
+\frac{3\nu-1}{8}\big(\Pa{YZY}-\Pa{ZYY}\big),
\end{multline}
independently of the element size, so that $\Th(K_e)=10$ for $\nu\neq\frac13$. The first group acts
identically on both displacement components and carries the Laplacian-like part of the energy; the
others couple the components. Combining the element, spring and support parts and counting shared
strings gives, for Q4 with one clamped face and $m\ge2$,
\begin{equation}\label{eq:q4-complete}
\Th(A)=7m+4-(m-1)\,\mathbf 1_{\{\beta=2\alpha\}}-2\,\mathbf 1_{\{\nu=1/3\}},
\end{equation}
i.e.\ $7m+4$ for generic parameters, $6m+5$ if $\beta=2\alpha$, $7m+2$ if $\nu=\frac13$ and $6m+3$
if both hold (the two cancellations involve disjoint sets of strings; verified by direct Pauli
expansions for $m=2,4,8$). For H8 with bonded padding (cubic elements, unit padding stiffness, one clamped face),
\begin{equation}\label{eq:h8-complete}
\Th(A)=\Th(K_e)+10m-9-(m-1)\,\mathbf 1_{\{\beta=2\alpha\}},
\end{equation}
with $\Th(K_e)=77$ except at the Poisson ratios $\nu=\frac14$, $\frac13$ and
$\nu^*=(\sqrt{15}-1)/6\approx0.4788305577$ in $[0,\frac12)$, where it is 53, 71 and 74. The value
$\nu^*$ is exact: it is the root of $18\nu^2+6\nu-7=0$, at which the physical diagonal entry
$(K_e)_{ii}=(2-3\nu)/[9(1+\nu)(1-2\nu)]$ equals the padding stiffness 1. Unlike the
cancellations at $\nu=\frac14$ and $\frac13$, which are properties of the physical element matrix,
this third exceptional value depends on the chosen padding stiffness and on the normalization of
the physical element matrix; it is not a material property of the H8 element. Hence
\begin{center}
\begin{tabular}{@{}lccc@{}}
\toprule
Poisson ratio & $\Th(K_e)$ & $\Th(A)$, $\beta\neq2\alpha$ & unitaries\\
\midrule
generic & 77 & $10m+68$ & 81\\
$1/4$ & 53 & $10m+44$ & 57\\
$1/3$ & 71 & $10m+62$ & 75\\
$\nu^*$ & 74 & $10m+65$ & 78\\
\bottomrule
\end{tabular}
\end{center}
Symbolically, with the Lam\'e constants $\lambda_{\rm L}$, $\mu_{\rm L}$ of $E=1$, the coefficient of
each element string is $a_P\lambda_{\rm L}+b_P\mu_{\rm L}+p_P$: 24 strings have coefficients
proportional to $\lambda_{\rm L}-\mu_{\rm L}$, which vanish at $\nu=\frac14$; 6 strings are
proportional to $\lambda_{\rm L}-2\mu_{\rm L}$, which vanish at $\nu=\frac13$; and 3 strings combine
the physical diagonal with the padding stiffness, with coefficients
$\pm\frac14\bigl[(\lambda_{\rm L}+4\mu_{\rm L})/9-1\bigr]=\pm\frac14[(K_e)_{ii}-1]$, which vanish at
$\nu^*$; the remaining 44 coefficients do not vanish for $0\le\nu<\frac12$ (symbolic computation
of the coefficients). The formula~\eqref{eq:h8-complete} was verified by direct Pauli expansions for
$m=2,4$, both cases of $\beta$ and all four values of $\nu$, using an absolute coefficient threshold
of $10^{-10}$. The exceptional values themselves follow from the exact rational dependence of the
coefficients on $\nu$. The assembled counts are invariant under any permutation of the qubit registers, which includes
the lexicographic and Morton node orderings (Section~\ref{sec:perm}), so the comparisons are not an
artefact of these orderings; arbitrary permutations of the node indices are not covered.

\paragraph{Exchange-form count, Proposition~\ref{M-thm:shift}(a)}
The element part contributes the $\Th(K_e)$ local Pauli unitaries, including the identity. Each
direction adds one nontrivial exchange $\V_i$; all directional identity terms merge with the element
identity. The support projector $(\Id-\mathbf Z_D)/2$ adds one reflection $\mathbf Z_D$, with its identity branch
again merged. Hence $\Un=\Th(K_e)+d+1$, and the 1-norm is $\lambda(K_e)+2d\alpha+\beta$. Direct
construction gives 13 terms for Q4 and 81 for H8 with bonded padding at every tested resolution
(generic $\nu$; fewer when the element matrix has fewer strings, e.g.\ 11 for Q4 at $\nu=1/3$),
with entrywise errors below $8\times10^{-15}$.

\paragraph{Exchange-form count, Proposition~\ref{M-thm:shift}(b)}
With $\Pi_\chi=(\Id+\mathbf Z_\chi)/2$ the projector onto active elements, the element part
$\Pi_\chi\otimes K_e+c(\Id-\Pi_\chi)\otimes\Id$ gives two unitaries for every local term of $K_e$,
while the void regularization changes only the coefficients of the identity and of $\mathbf Z_\chi$. After
merging identities, the support reflection and the $d$ masked exchanges give at most
$2\Th(K_e)+d+1$ unitaries, i.e.\ at most 23 for Q4 (attained for the L-shaped and perforated masks
at the benchmark parameters; degenerate masks without interface bonds in some direction need
fewer); these masks duce the sparse
matrices to below $2\times10^{-15}$.

\subsection{Construction for Theorem~\ref{M-thm:direct-precond-block} of the main paper}\label{app:normalization}
Write the BPX part by levels, $\F_{\rm p}=[\F_0,\ldots,\F_L]$ with
$\F_\ell=2^{(L-\ell)(2-d)/2}\J_0\mathbf I_{\ell,L}$ (the weight $(h_\ell/h)^{(2-d)/2}$ of Section~\ref{M-sec:bpx-el}), where $\mathbf I_{\ell,L}$ interpolates from level
$\ell$ to the finest level, and set $\Hf_\ell=\R\F_\ell$ and $\Hf_{\rm c}=\R\F_{\rm c}$. For H8
everything in this subsection refers to the physical components. The independently scaled padding
sector is incorporated through a sector-controlled direct-sum block encoding,
$\Hf=\Hf_{\rm phys}\oplus\Id_{\rm pad}$ (Section~\ref{M-sec:direct-block-precond} of the main paper):
controlled on a sector qubit, apply the encoding of $\Hf_{\rm phys}$ or the identity scaled to the
same normalization by an ancilla rotation, which gives the common normalization
$\max\{\omega(\Hf_{\rm phys}),1\}$. It is not an additional horizontal branch, which would add one to
the squared normalization; the padding sector therefore does not increase the bound below.

\emph{Rectangular concatenation.} $\Hf$ is a horizontal concatenation of blocks that act on
different coefficient spaces, so an LCU in which the same preparation is applied before and after
the selection would produce products of preparation amplitudes rather than the required weights.
We therefore use the following construction. Suppose that $\Hf_\ell$ and $\Hf_{\rm c}$ have block
encodings $U_\ell$ and $U_{\rm c}$ with normalizations $\omega_\ell$ and $\omega_{\rm c}$, i.e.\
$\Hf_\ell=\omega_\ell(\bra0\otimes\Id)U_\ell(\ket0\otimes\Id)$, with compatible embeddings of the
inputs and outputs in common registers (invalid codes are removed by validity projectors).
\begin{enumerate}\itemsep1pt
\item \emph{Spring branch.} Combine the two spring rows into the rectangular block
$\mathbf B_{\rm c}=\bigl[\begin{smallmatrix}\Hf_{\rm c}\\ \Id_{\rm c}\end{smallmatrix}\bigr]$ with
$\Omega_{\rm c}=(\omega_{\rm c}^2+1)^{1/2}$: prepare an output flag in
$(\omega_{\rm c}\ket0+\ket1)/\Omega_{\rm c}$ and apply $U_{\rm c}$ if the flag is 0 and the
identity if it is 1. Projecting the ancillas of $U_{\rm c}$ onto $\ket0$ gives
$\Omega_{\rm c}^{-1}(\ket0\otimes\Hf_{\rm c}\mathbf y+\ket1\otimes\mathbf y)$, i.e.\ a block
encoding of $\mathbf B_{\rm c}$ with normalization $\Omega_{\rm c}$; the flag is part of the output
and distinguishes energy and identity rows. In the level branches the flag stays $\ket0$, so they
encode $[\Hf_\ell;\mathbf 0]$.
\item \emph{Selection by the input.} The input register contains a selector
$\ket j$, $j\in\{0,\dots,L,{\rm c}\}$, and the coefficient vector of that block. Apply
$\sum_\ell\ket\ell\bra\ell\otimes U_\ell+\ket{\rm c}\bra{\rm c}\otimes U_{\mathbf B_{\rm c}}$; no
preparation is applied on the input side.
\item \emph{Output projection.} Project the selector, now an ancilla, onto
\begin{equation}\label{eq:selector-state}
\ket P=\frac1\Omega\Bigl(\sum_{\ell=0}^L\omega_\ell\ket\ell+\Omega_{\rm c}\ket{\rm c}\Bigr),
\qquad \Omega^2=\sum_{\ell=0}^L\omega_\ell^2+\Omega_{\rm c}^2 ,
\end{equation}
by applying the inverse of a unitary that prepares $\ket P$ and projecting onto $\ket0$.
\end{enumerate}
For an input $\sum_j\ket j\otimes\mathbf x_j$ the projected output is
$\sum_\ell(\omega_\ell/\Omega)(\Hf_\ell\mathbf x_\ell/\omega_\ell)+(\Omega_{\rm c}/\Omega)(\mathbf
B_{\rm c}\mathbf x_{\rm c}/\Omega_{\rm c})=\Hf\mathbf x/\Omega$: each block's normalization cancels
against its amplitude in $\ket P$. Hence
\begin{equation}\label{eq:Hhat-normalization-abstract}
\omega(\Hf)^2\le\Omega^2=1+\omega_{\rm c}^2+\sum_{\ell=0}^L\omega_\ell^2 .
\end{equation}
This is a construction at the algebraic level of block encodings, not a compiled circuit.
\emph{Spring block.} A norm bound alone does not give a normalization, so we state the
construction. $\Hf_{\rm c}=\R\,\W_c\D_c$ is a product of three factors. (i) $\W_c$ is an isometry,
implemented by at most $d$ Hadamard gates controlled by the boundary flags of the node. (ii) The
diagonal contraction $\sqrt{p_{\min}}\D_c$ takes only a few values, since $b\in\{0,1\}$ and
$|\mathbf z|\le d$, and is block-encoded with normalization one by a fixed set of rotations
controlled by the mode label. (iii) On a uniform mesh $\R=\Id\otimes\R_e$ with one fixed
element matrix $\R_e$ of norm $\sqrt{\bar k}$, and $\R_e/\sqrt{\bar k}$ is a contraction of
constant size, which has an exact block encoding with normalization one by unitary dilation on a
constant number of qubits. The normalizations of a product multiply, so
$\omega_{\rm c}=p_{\min}^{-1/2}\sqrt{\bar k}=\sqrt\eta$.
\emph{Level blocks.} We use the multilevel energy construction of \citet{Deiml2026}, extended from
the scalar gradient to the symmetric gradient of elasticity in two steps: their scalar
construction is applied to each displacement component, i.e.\ tensored with the identity on the
component register, which leaves its normalization unchanged, and the resulting displacement
gradient is multiplied by the fixed map $\M_{\rm el}$ of constant size, block-encoded by unitary
dilation with normalization $\norm{\M_{\rm el}}_2$. A level block is
$\Hf_\ell=2^{(L-\ell)(2-d)/2}(\M_{\rm el}\otimes\Id)\widetilde{\mathbf T}_{\ell,L}\mathbf C_\ell^{\nabla}$,
where $\mathbf C_\ell^\nabla$ is the Cartesian gradient on level $\ell$, $\widetilde{\mathbf
T}_{\ell,L}$ transfers it to the finest level with orthonormal columns, and $\M_{\rm el}=(\mathbf
D/E)^{1/2}\mathbf S$ maps the displacement gradient (in units of $1/h$) to the Voigt strain and applies
the square root of the nondimensional constitutive matrix. In dimensional variables $R$ carries the
factor $k_0^{1/2}=(Eh^{d-2})^{1/2}$, and $F_{\rm p}$ the factor $k_0^{-1/2}$ (Remark~\ref{M-rem:units}
of the main paper); the two cancel in $RF_{\rm p}$, so only $(\mathbf D/E)^{1/2}$ remains. \citet{Deiml2026} show that the weighted level gradient has normalization
$h_\ell^{(2-d)/2}\omega(\mathbf C_\ell^\nabla)\le2\sqrt d$ on every level of the unit domain, which
in the units of Section~\ref{M-sec:uniform-preconditioning} reads $2^{(L-\ell)(2-d)/2}\omega(\mathbf
C_\ell^\nabla)\le2\sqrt d$, so
\begin{equation}\label{eq:level-gamma-bound}
\omega_\ell\le C_{\rm el}:=2\sqrt d\,\norm{\M_{\rm el}}_2,\qquad
\norm{\M_{\rm el}}_2^2=\begin{cases}1/(1-\nu),&\text{plane stress},\\ 1/(1-2\nu),&\text{3D},\end{cases}
\end{equation}
for $0\le\nu<1/2$. The strain mixer $\mathbf S$ and the constitutive square root are encoded
jointly as the constant-size map $\M_{\rm el}$, by one unitary dilation normalized by
$\norm{\M_{\rm el}}_2$; encoding them separately would give the larger product
$\norm{(\mathbf D/E)^{1/2}}_2\norm{\mathbf S}_2$ (e.g.\ $\norm{\mathbf S}_2=\sqrt2$ in plane stress).

\emph{Common output basis.} The Gram identity~\eqref{M-eq:Hhat-gram} holds for one common energy
factor $\R$ with $\R^{\mathsf T}\R=\A_{\rm el}$; correct diagonal blocks alone would not establish
the mixed block $\F_{\rm p}^{\mathsf T}\A_{\rm el}\F_{\rm c}$. The physical and spring branches
therefore use a common output basis. The multilevel construction of \citet{Deiml2026} produces
gradient coefficients in an orthonormal discontinuous-Q1 basis, whereas the spring branch
(Section~\ref{sec:circuitsupp}) produces quadrature-weighted values at the Gauss points. On the
reference interval with the normalized inner product the orthonormal basis is $\psi_0=1$,
$\psi_1=\sqrt3\,\xi$, and its weighted values at the two Gauss points $\xi=\mp1/\sqrt3$ form the
orthogonal matrix
\begin{equation}\label{eq:gauss-basis}
U_G=\frac1{\sqrt2}\begin{bmatrix}1&-1\\1&1\end{bmatrix}.
\end{equation}
With consistent element scaling, its $d$-fold tensor product converts the coefficients to weighted
Gauss-point values. The conversion is element-local, of constant size and orthogonal, so it does
not change the normalization, and both branches represent the same element energy factor $\R$,
including the mixed physical--spring blocks of the Gram identity. (Alternatively, both branches
can be written directly in the discontinuous polynomial basis.)

\emph{One-face-clamped boundary conditions: the level-gradient normalization.} The scalar
construction of \citet{Deiml2026} assumes homogeneous Dirichlet conditions at both ends of each
one-dimensional interval, whereas the cantilever hierarchy is clamped only on $x_1=0$ and has free
nodes on the other faces, on every level including the coarsest. We therefore establish the level
bound directly. For one-dimensional nodal coefficients $v$ on level $\ell$, let
\begin{equation}\label{eq:endpoint-gathers}
(E_0v)_j=v_j,\qquad (E_1v)_j=v_{j+1}
\end{equation}
gather the left and right endpoints of element $j$, with the clamped nodal entries (direction 1,
node 0) set to zero and the free boundary nodes kept. Each gather is a partial permutation, so it
has a block encoding with normalization one, and their vertical concatenation with a retained
output flag, $\mathcal E=[E_0;E_1]$, has $\omega(\mathcal E)\le\sqrt2$. In the local orthonormal
polynomial basis $\psi_0=1$, $\psi_1=\sqrt3\,\xi$ (normalized inner product on the reference
interval) the linear interpolant of the endpoint values has the coefficients given by the
interpolation mixer
\begin{equation}
M_v=\frac12\begin{bmatrix}1&1\\-1/\sqrt3&1/\sqrt3\end{bmatrix},\qquad \norm{M_v}_2=\frac1{\sqrt2}.
\end{equation}
With the level element size $h_\ell$, the $L^2$-normalized interpolation and derivative maps are
\begin{equation}
C_v=\sqrt{h_\ell}\,M_v\,\mathcal E,\quad \omega(C_v)\le\sqrt{h_\ell},\qquad
C_\partial=\frac1{\sqrt{h_\ell}}\,[-1\ \ 1]\,\mathcal E,\quad \omega(C_\partial)\le\frac2{\sqrt{h_\ell}} .
\end{equation}
A directional gradient in $d$ dimensions is the tensor product of one derivative map and $d-1$
interpolation maps, so its normalization is at most $2h_\ell^{(d-2)/2}$, and stacking the $d$
directional gradients gives
\begin{equation}
\omega(\mathbf C_\ell^\nabla)\le2\sqrt d\,h_\ell^{(d-2)/2}.
\end{equation}
The BPX weight cancels the mesh-size dependence, $h_\ell^{(2-d)/2}\omega(\mathbf
C_\ell^\nabla)\le2\sqrt d$, and with the orthonormal level transfers and the joint strain/material map
$\omega_\ell\le2\sqrt d\,\norm{\M_{\rm el}}_2$, as in~\eqref{eq:level-gamma-bound}. This
accommodates the free boundary nodes without changing the stated bound. It is an algebraic
normalization argument: the efficient implementation of the level transfers, of the input and
output validity embeddings and of the controlled dilations is specified at the level of the circuit
model but has not been compiled. Consistently, the numerically evaluated level-block norms of the
clamped cantilever satisfy the bound for $m\le64$ ($\Gamma_{\rm ideal}\le\Gamma_{\rm explicit}$ in
Table~\ref{tab:direct-q4}).

Inserting both bounds for the $L+1$ levels into
\eqref{eq:Hhat-normalization-abstract} gives~\eqref{M-eq:Hhat-final-bound}. The asymptotic
(polylogarithmic) gate-complexity statement of the theorem follows from the Cartesian transfer and
gradient circuits of \citet{Deiml2026}, reversible $O(\log m)$-bit index arithmetic and
constant-size copy and material registers. \qed

The construction is specified at the level of isometries, validity projectors, coordinate
conversions and controlled dilations, with the normalizations stated above; it has not been
compiled into a reversible gate-level network, and only the Gram identity~\eqref{M-eq:Hhat-gram},
not an encoded circuit, was verified numerically (Section~\ref{sec:bpxsupp}). No compiled
gate-level counts or explicit resource estimates are given.

\subsection{Status of the results}
Table~\ref{tab:status} summarizes which results of the main paper are proved, which are verified
numerically and which rest on modelling assumptions.

\begin{table}[!htbp]
\centering
\caption{Status of the main results. Proved: theorem or exact identity. Verified: duced
numerically to round-off or computed for the stated meshes. Model: rests on stated assumptions.}
\label{tab:status}
\footnotesize
\begin{tabular}{@{}>{\raggedright\arraybackslash}p{0.42\linewidth}>{\raggedright\arraybackslash}p{0.17\linewidth}>{\raggedright\arraybackslash}p{0.35\linewidth}@{}}
\toprule
Result & Status & Evidence\\
\midrule
Exchange form: at most 13 / 81 / 23 unitaries (13 / 81 attained at generic $\nu$; 23 attained for the tested masks) & proved & entrywise check up to $N=131{,}072$\\
Pauli counts $\propto dN^{1/d}$ on uniform meshes & proved (closed form) & numerical confirmation up to $N=2^{15}$ \\
Assembled Pauli counts $\Theta(N)$ on uniform grids & proved (Proposition~\ref{prop:assembled-lb}) & full numerical expansions, Q4 $M\le64$, H8 $M\le8$; curved meshes: observed\\
Graded-mesh element identities & exact (scaling of the element integrals) & round-off check, random sizes\\
Exact diagonalization of the springs (complete Cartesian copy graphs) & proved & check to $10^{-14}$\\
Penalty expansion and extrapolation & proved (fixed-mesh asymptotics in $1/\gamma$, not a uniform-in-mesh error theorem) & four 2D benchmarks ($m\le64$), perforated bracket ($m\le256$), H8 cantilever ($m\le8$)\\
Curved-mesh compression & empirical & two smooth meshes, one rough mesh\\
Mesh- and penalty-robust conditioning (uniform Cartesian hierarchies) & proved (BPX, Korn) & Q4 up to $m=512$, H8 up to $m=32$, $\nu=0.3$; plateau not reached\\
Assembled counts 109 / 196 / 2,776 / 10,072 & upper bounds & entrywise check; minimality not proved\\
Projected (assembled) form of the copies, Proposition~\ref{M-prop:algebra}(b) & known identity & verified for Q4, $m\le32$\\
Commuting exchanges and reflection; projector product and spring pseudoinverse, i.e.\ exact diagonalization of the spring modes (Proposition~\ref{M-prop:algebra}) & proved & round-off check of all three identities, Q4 $m\le8$, H8 $m=2$\\
Penalty vs.\ projected form (Table~\ref{M-tab:penalty-projection}) & computed & Q4, $m=2$--8; unitary count $m=2$, 4\\
Factor normalization $O(\sqrt{\log m})$ & proved for the specified construction & diagnostics up to $m=64$; circuit not compiled\\
Two-sided compliance bound for the penalty error (Proposition~\ref{M-prop:penalty-bound}) & proved & Q4 cantilever, $m\le32$, 15 spring pairs; error $99.1$--$99.95\%$ of the bound (Table~\ref{tab:penbound})\\
\bottomrule
\end{tabular}
\end{table}

\endgroup

\section{Notation and Pauli toolkit}\label{sec:tools}

\paragraph{Pauli strings.}
$\Pa I,\Pa X,\Pa Y,\Pa Z$ are the Pauli matrices. Registers are big-endian: in
$\ket{b_1b_2\cdots b_q}$ the first qubit is the most significant bit. A string
$\Pa{P_1P_2\cdots P_q}$ means $P_1\otimes\cdots\otimes P_q$. Every real or complex
$N\times N$ matrix, $N=2^q$, has the unique expansion
\begin{equation}\label{eq:pexp}
A=\sum_{P}c_P\,P,\qquad c_P=2^{-q}\tr(PA),
\end{equation}
and $A$ is Hermitian if and only if all $c_P$ are real.

\paragraph{Ladder operators and projectors.}
\begin{equation}
\Lup=\ket1\!\bra0=\tfrac12(\Pa X-i\Pa Y),\quad
\Ldn=\ket0\!\bra1=\tfrac12(\Pa X+i\Pa Y),\quad
P_0=\tfrac12(\Pa I+\Pa Z),\quad P_1=\tfrac12(\Pa I-\Pa Z).
\end{equation}
For the all-zero and all-one projectors on $a$ qubits, expanding the tensor products
gives
\begin{equation}\label{eq:allproj}
P_0^{\otimes a}=2^{-a}\sum_{\mathbf z\in\{0,1\}^a}Z^{\mathbf z},\qquad
P_1^{\otimes a}=2^{-a}\sum_{\mathbf z\in\{0,1\}^a}(-1)^{|\mathbf z|}Z^{\mathbf z},
\end{equation}
where $Z^{\mathbf z}=\bigotimes_k\Pa Z^{z_k}$ and $|\mathbf z|$ is the Hamming weight.

\paragraph{Fast exact decomposition.}
Every Pauli string is, up to a phase in $\{\pm1,\pm i\}$, of the form
$X^{\mathbf x}Z^{\mathbf z}$; the phase is fixed by the number of $\Pa Y$ factors,
which equals $|\mathbf x\wedge\mathbf z|$. We call $\mathbf x$ the \emph{X-pattern}.
Since $\bra j X^{\mathbf x}=\bra{j\oplus\mathbf x}$,
\begin{equation}
\tr\!\big(X^{\mathbf x}Z^{\mathbf z}A\big)
=\sum_j(-1)^{\mathbf z\cdot(j\oplus\mathbf x)}A_{j\oplus\mathbf x,\,j}
=(-1)^{\mathbf z\cdot\mathbf x}\sum_j(-1)^{\mathbf z\cdot j}\,v_{\mathbf x}(j),
\qquad v_{\mathbf x}(j)=A_{j\oplus\mathbf x,\,j}.
\end{equation}
Thus, for fixed $\mathbf x$, all $2^q$ coefficients are a Walsh--Hadamard transform of
the generalized diagonal $v_{\mathbf x}$, computable in $O(N\log N)$ operations. Only
X-patterns for which $v_{\mathbf x}\neq0$ need be processed; for a sparse matrix these
are the values $i\oplus j$ over its nonzeros $(i,j)$. The decomposition is exact in exact
arithmetic. Computed numerically, the coefficients $c_P$, and hence the 1-norms, are obtained to
floating-point precision, and coefficients below a threshold are treated as zero, so numerically
evaluated term counts depend on that threshold.
That the threshold lies far below the smallest genuine coefficient was verified for the
configurations reported here; it is not guaranteed for other parameters,
which may produce small but nonzero coefficients.

\paragraph{Cost measures.}
$\T(A)$ is the number of nonzero $c_P$; $\mathcal X(A)$ the number of distinct
X-patterns among them; $\lambda(A)=\sum_P|c_P|$ the LCU 1-norm; and
$\kappa_{\rm eff}(A)=\lambda(A)\|A^{-1}\|_2$. Two facts are used repeatedly.
\begin{enumerate}
\item For operators on disjoint registers,
$\T(A\otimes B)=\T(A)\,\T(B)$ and $\lambda(A\otimes B)=\lambda(A)\lambda(B)$, because the
strings of $A\otimes B$ are the products of the strings of $A$ and $B$, all distinct.
\item For a sum, $\T(A+B)=\T(A)+\T(B)-\#(\text{shared strings})-\#(\text{strings that cancel})$.
\end{enumerate}

\section{The two-element bar in closed form}\label{sec:2el}

\subsection{Model}
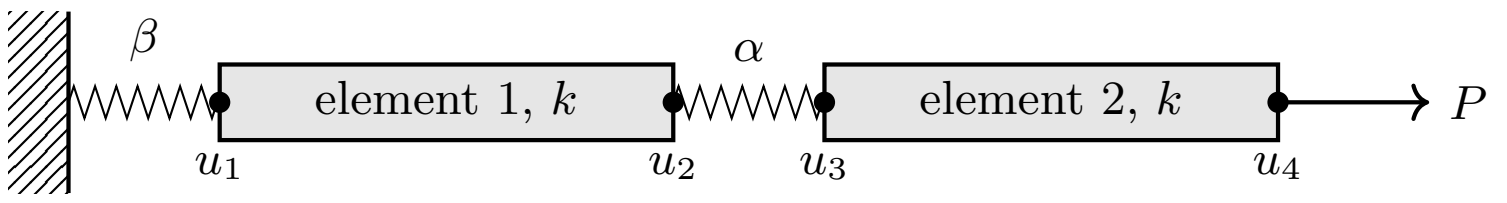
\begin{figure}[ht]
\centering
\begin{tikzpicture}[every node/.style={font=\small}]
  \fill[pattern=north east lines] (-0.4,-0.6) rectangle (0,0.6);
  \draw[thick] (0,-0.6) -- (0,0.6);
  \draw[decorate,decoration={zigzag,segment length=4,amplitude=3}] (0,0) -- (1,0);
  \node[above] at (0.5,0.15) {$\beta$};
  \draw[thick,fill=gray!20] (1,-0.25) rectangle (4,0.25); \node at (2.5,0) {element 1, $k$};
  \fill (1,0) circle (2pt) node[below=6pt] {$u_1$};
  \fill (4,0) circle (2pt) node[below=6pt] {$u_2$};
  \draw[decorate,decoration={zigzag,segment length=4,amplitude=3}] (4,0) -- (5,0);
  \node[above] at (4.5,0.15) {$\alpha$};
  \draw[thick,fill=gray!20] (5,-0.25) rectangle (8,0.25); \node at (6.5,0) {element 2, $k$};
  \fill (5,0) circle (2pt) node[below=6pt] {$u_3$};
  \fill (8,0) circle (2pt) node[below=6pt] {$u_4$};
  \draw[->,thick] (8,0) -- (9,0) node[right] {$P$};
\end{tikzpicture}
\caption{The assembly-free two-element bar with penalty springs for compatibility ($\alpha$)
and the support ($\beta$).}
\label{fig:bar2}
\end{figure}

A bar of length $L$, axial rigidity $EA$, fixed at $x=0$ and loaded by $P$ at $x=L$ is
discretized by two linear elements of length $h=L/2$ and stiffness $k=EA/h$
(Fig.~\ref{fig:bar2}). Each element owns its end displacements:
$\mathbf u=(u_1,u_2,u_3,u_4)=(u_1^{L},u_1^{R},u_2^{L},u_2^{R})$. Compatibility
$u_2=u_3$ is enforced by a penalty $\alpha$ and the support condition $u_1=0$ by a
penalty $\beta$:
\begin{equation}\label{eq:pi2}
\Pi=\tfrac{k}{2}(u_2-u_1)^2+\tfrac{k}{2}(u_4-u_3)^2+\tfrac{\alpha}{2}(u_2-u_3)^2
+\tfrac{\beta}{2}u_1^2-Pu_4 .
\end{equation}

With $\Pi=\frac12\mathbf u^{\mathsf T}A\mathbf u-\mathbf f^{\mathsf T}\mathbf u$,
\begin{equation}\label{eq:A2}
A=\begin{bmatrix}k+\beta&-k&0&0\\-k&k+\alpha&-\alpha&0\\0&-\alpha&k+\alpha&-k\\0&0&-k&k\end{bmatrix},
\qquad\mathbf f=(0,0,0,P)^{\mathsf T}.
\end{equation}

\subsection{Exact solution}
The model is four springs in series (support $\beta$, element $k$, interface $\alpha$,
element $k$), each carrying $P$. Hence $\beta u_1=P$, $k(u_2-u_1)=P$,
$\alpha(u_3-u_2)=P$, $k(u_4-u_3)=P$:
\begin{equation}\label{eq:u2}
\mathbf u=P\Big(\tfrac1\beta,\ \tfrac1\beta+\tfrac1k,\ \tfrac1\beta+\tfrac1k+\tfrac1\alpha,\
\tfrac1\beta+\tfrac2k+\tfrac1\alpha\Big)\ \xrightarrow{\ \alpha,\beta\to\infty\ }\ \tfrac Pk(0,1,1,2),
\end{equation}
which matches the exact solution $u(x)=Px/EA$ at the nodes. The interface gap is
$P/\alpha$ and the tip error is $P(1/\alpha+1/\beta)$.

\subsection{Encoding and Pauli decomposition}
Encode on two qubits $\ket{e}\ket{s}$ (element, end):
$u_1\mapsto\ket{00}$, $u_2\mapsto\ket{01}$, $u_3\mapsto\ket{10}$, $u_4\mapsto\ket{11}$.

\emph{Elements.} Each element block is $k\begin{bmatrix}1&-1\\-1&1\end{bmatrix}=k(\Pa I-\Pa X)$
on the end qubit; together, $A_{\rm el}=k(\Pa{II}-\Pa{IX})$.

\emph{Compatibility.} With $\mathbf v=\ket{01}-\ket{10}$, the penalty is
$\alpha\mathbf v\mathbf v^{\mathsf T}$. Using
$\ket{01}\!\bra{01}+\ket{10}\!\bra{10}=\frac12(\Pa{II}-\Pa{ZZ})$ and
$\ket{01}\!\bra{10}+\ket{10}\!\bra{01}=\frac12(\Pa{XX}+\Pa{YY})$,
\begin{equation}\label{eq:singlet}
A_{\rm c}=\frac\alpha2\big(\Pa{II}-\Pa{XX}-\Pa{YY}-\Pa{ZZ}\big)=2\alpha\ket{\Psi^-}\!\bra{\Psi^-},
\qquad \ket{\Psi^-}=\tfrac{1}{\sqrt2}(\ket{01}-\ket{10}).
\end{equation}
The compatibility penalty is an isotropic Heisenberg exchange interaction between the
element and end qubits: it penalizes the singlet component of the state.

\emph{Support.} $A_{\rm bc}=\beta\ket{00}\!\bra{00}=\frac\beta4(\Pa{II}+\Pa{IZ}+\Pa{ZI}+\Pa{ZZ})$.

Summing,
\begin{equation}\label{eq:A2pauli}
A=\big(k+\tfrac\alpha2+\tfrac\beta4\big)\Pa{II}-k\,\Pa{IX}-\tfrac\alpha2\Pa{XX}
-\tfrac\alpha2\Pa{YY}+\big(\tfrac\beta4-\tfrac\alpha2\big)\Pa{ZZ}+\tfrac\beta4\Pa{IZ}+\tfrac\beta4\Pa{ZI}.
\end{equation}
Check of one diagonal entry, $\ket{00}$:
$(k+\frac\alpha2+\frac\beta4)+(\frac\beta4-\frac\alpha2)+\frac\beta4+\frac\beta4=k+\beta$.
For $\beta\neq2\alpha$ there are seven strings (six if $\beta=2\alpha$, when the $\Pa{ZZ}$
coefficient vanishes), in three qubit-wise commuting groups,
$\{\Pa{II},\Pa{ZZ},\Pa{IZ},\Pa{ZI}\}$, $\{\Pa{IX},\Pa{XX}\}$, $\{\Pa{YY}\}$. Because
$\mathbf f=P\ket{11}$, the squared magnitude of the load overlap is $|\braket{f|\psi}|^2=P^2p_{11}$,
where $p_{11}$ is the probability of outcome $11$ in the computational basis; such a measurement
does not recover the sign or phase of the overlap.

\section{The four-element bar and general $n$}\label{sec:4el}

\subsection{Solution and penalty error}
For $n$ elements the series-spring argument gives, for copy $j=0,\dots,2n-1$
(element $e=\lfloor j/2\rfloor$, end $s=j\bmod2$),
\begin{equation}\label{eq:ugen}
u_j=P\Big(\frac1\beta+\frac{e+s}{k}+\frac{e}{\alpha}\Big),\qquad
u_{\rm tip}=P\Big(\frac1\beta+\frac nk+\frac{n-1}{\alpha}\Big).
\end{equation}
Penalty errors accumulate over the $n-1$ interfaces. Relative to the exact tip
displacement $nP/k$ the error is $\frac{k}{\alpha}\frac{n-1}{n}+\frac{k}{n\beta}$. The
interface term is bounded by $k/\alpha$ and the support term by $k/(n\beta)\le k/\beta$, so the
relative error is $O(k/\alpha+k/\beta)$ uniformly in $n$ when both $\alpha$ and $\beta$ are scaled
with the element stiffness $k=EA/h$; the support term alone decreases as $1/n$.

\subsection{Pauli decomposition for $n=4$}
Three qubits $\ket{e_1e_0s}$. The element part is $k(\Pa{III}-\Pa{IIX})$. The interfaces
join copies $(1,2)=(001,010)$, $(3,4)=(011,100)$ and $(5,6)=(101,110)$.

\emph{In-block interfaces.} Copies $(1,2)$ and $(5,6)$ differ only in the last two bits
and together give $\Pa I\otimes\frac\alpha2(\Pa{II}-\Pa{XX}-\Pa{YY}-\Pa{ZZ})$: exchange
interactions as in~\eqref{eq:singlet}.

\emph{Carry interface.} Going from $3=011$ to $4=100$ flips all three bits. With
$\mathbf v=\ket{011}-\ket{100}$,
\begin{align}
\ket{011}\!\bra{011}+\ket{100}\!\bra{100}&=\tfrac14\big(\Pa{III}+\Pa{IZZ}-\Pa{ZIZ}-\Pa{ZZI}\big),\\
\ket{011}\!\bra{100}+\ket{100}\!\bra{011}&=\tfrac14\big(\Pa{XXX}-\Pa{XYY}+\Pa{YXY}+\Pa{YYX}\big),
\end{align}
the second following from $\ket{011}\!\bra{100}=\Ldn\otimes\Lup\otimes\Lup$ and
keeping real terms. This is a three-body flip term. The support penalty
$\beta\ket{000}\!\bra{000}$ contributes all eight $\Pa Z$-type strings with coefficient
$\beta/8$. In total (verified symbolically):

\begin{table}[ht]
\centering\small
\caption{Pauli coefficients of $A$ for the four-element bar.}
\begin{tabular}{@{}ll|ll@{}}
\toprule
String & Coefficient & String & Coefficient\\
\midrule
$\Pa{III}$ & $k+\frac{3\alpha}{4}+\frac\beta8$ & $\Pa{IZZ},\Pa{ZIZ},\Pa{ZZI}$ & $\frac{\beta-2\alpha}{8}$\\
$\Pa{IIX}$ & $-k$ & $\Pa{IIZ},\Pa{IZI},\Pa{ZII},\Pa{ZZZ}$ & $\frac\beta8$\\
$\Pa{IXX},\Pa{IYY}$ & $-\frac\alpha2$ & $\Pa{XXX},\Pa{YXY},\Pa{YYX}$ & $-\frac\alpha4$\\
& & $\Pa{XYY}$ & $+\frac\alpha4$\\
\bottomrule
\end{tabular}
\end{table}

\noindent
For $\beta\neq2\alpha$ there are $15=2N-1$ strings (12 if $\beta=2\alpha$, when the three
coefficients $(\beta-2\alpha)/8$ vanish). The exchange structure survives within dyadic blocks, but
bonds that cross a carry become many-body terms.

\section{General one-dimensional analysis}\label{sec:1dgeneral}

Let $n=2^a$ elements be encoded as $\ket{e}\ket{s}$, with $e$ on register $E$ of $a$
qubits and $s$ on one local qubit $\ell$. Then $A_{\rm el}=k\,I_E\otimes(\Pa I-\Pa X)$,
and the compatibility bonds join $\ket{e,1}$ with $\ket{e+1,0}$ for $e=0,\dots,n-2$. The counts
below refer to generic penalties ($\beta\neq2\alpha$); the cancellation at $\beta=2\alpha$ is noted
after the total count.

\begin{lemma}[Increment]\label{lem:inc}
$S_a=\sum_{j=0}^{2^a-2}\ket{j+1}\!\bra j=\sum_{t=1}^{a}I^{\otimes(a-t)}\otimes\Lup\otimes\Ldn^{\otimes(t-1)}$.
\end{lemma}
\begin{proof}
Every $j<2^a-1$ has a binary representation ending in a block $0\,1^{t-1}$ for a unique
$t\in\{1,\dots,a\}$ (the position of the lowest zero). Adding one turns this block into
$1\,0^{t-1}$ and leaves the leading $a-t$ bits unchanged. The operator
$I^{\otimes(a-t)}\otimes\Lup\otimes\Ldn^{\otimes(t-1)}$ maps exactly the states ending in
$0\,1^{t-1}$ to the corresponding states ending in $1\,0^{t-1}$, for every value of the
leading bits, and annihilates all other states. Summing over $t$ gives $S_a$. The
all-ones state has no zero and is annihilated, so the increment is non-cyclic.
\end{proof}

The bond operator is
\begin{equation}
\sum_{e=0}^{n-2}(\ket{e,1}-\ket{e+1,0})(\bra{e,1}-\bra{e+1,0})
=\underbrace{D^+\otimes P_1+D^-\otimes P_0}_{\rm diagonal}
-\underbrace{\big(S_a\otimes\Ldn+S_a^\dagger\otimes\Lup\big)}_{\rm off\text{-}diagonal},
\end{equation}
with $D^+=I-P_1^{\otimes a}$ (elements with a right neighbour) and
$D^-=I-P_0^{\otimes a}$ (elements with a left neighbour). The off-diagonal term uses
$\ket{e+1,0}\!\bra{e,1}=(\ket{e+1}\!\bra e)\otimes\Ldn$.

\begin{lemma}[Hermitized ladder strings]\label{lem:herm}
Let $H_t=\Lup\otimes\Ldn^{\otimes t}+\Ldn\otimes\Lup^{\otimes t}$ on $t+1$ qubits. Then
$H_t$ consists of the $2^t$ strings of $\{\Pa X,\Pa Y\}^{\otimes(t+1)}$ having an even
number of $\Pa Y$ factors, each with coefficient $\pm2^{-t}$. Hence $\T(H_t)=2^t$,
$\lambda(H_t)=1$, and all its strings share one X-pattern.
\end{lemma}
\begin{proof}
Expanding $\Lup\otimes\Ldn^{\otimes t}=2^{-(t+1)}(\Pa X-i\Pa Y)\otimes(\Pa X+i\Pa Y)^{\otimes t}$,
a string whose set of $\Pa Y$ positions is $Q$ has coefficient
$2^{-(t+1)}(-1)^{[1\in Q]}\,i^{|Q|}$. The second term is the Hermitian conjugate, whose
coefficients are the complex conjugates. The sum therefore has coefficient
$2^{-t}\operatorname{Re}\big((-1)^{[1\in Q]}i^{|Q|}\big)$, which is $\pm2^{-t}$ for even
$|Q|$ and zero for odd $|Q|$. Exactly half of the $2^{t+1}$ subsets $Q$ are even. All
strings have $\Pa X$ or $\Pa Y$ in every position and so share the X-pattern
$1^{t+1}$.
\end{proof}

By Lemma~\ref{lem:inc},
$S_a\otimes\Ldn+S_a^\dagger\otimes\Lup=\sum_{t=1}^aI^{\otimes(a-t)}\otimes H_t$, where
$H_t$ acts on the last $t$ qubits of $E$ and on $\ell$. Different $t$ have different
identity patterns, so the strings are all distinct:
\begin{equation}
\T_{\rm off}=\sum_{t=1}^a2^t=2^{a+1}-2=2n-2,\qquad\lambda_{\rm off}=a,\qquad
\mathcal X_{\rm off}=a .
\end{equation}

\begin{lemma}[Boundary projectors]\label{lem:proj}
$D^+\otimes P_1+D^-\otimes P_0
=I-2^{-a}\Big[\sum_{|\mathbf z|\,{\rm even}}Z^{\mathbf z}\otimes\Pa I
+\sum_{|\mathbf z|\,{\rm odd}}Z^{\mathbf z}\otimes\Pa Z\Big]$.
\end{lemma}
\begin{proof}
$D^+\otimes P_1+D^-\otimes P_0=I-P_1^{\otimes a}\otimes P_1-P_0^{\otimes a}\otimes P_0$.
By~\eqref{eq:allproj},
$P_1^{\otimes a}\otimes P_1+P_0^{\otimes a}\otimes P_0
=2^{-a-1}\sum_{\mathbf z}Z^{\mathbf z}\otimes\big[(-1)^{|\mathbf z|}(\Pa I-\Pa Z)+(\Pa I+\Pa Z)\big]$,
and the bracket equals $2\Pa I$ for even $|\mathbf z|$ and $2\Pa Z$ for odd.
\end{proof}
Each $\mathbf z$ contributes one string; the $\mathbf z=\mathbf 0$ string merges with
$I$ to give coefficient $1-2^{-a}$. Hence
\begin{equation}
\T_{\rm diag}=2^a=n,\qquad\lambda_{\rm diag}=(1-2^{-a})+(2^a-1)2^{-a}=2-\frac2n .
\end{equation}
Diagonal and off-diagonal strings are distinct (different X-patterns), so
\begin{equation}\label{eq:1dcount}
\T(A_{\rm c})=3n-2,\qquad\lambda(A_{\rm c})=\alpha\Big(\log_2n+2-\frac2n\Big),\qquad
\mathcal X(A_{\rm c})=\log_2n+1 .
\end{equation}

\paragraph{Total count with a point support.}
$A_{\rm bc}=\beta P_0^{\otimes(a+1)}$ contributes all $2^{a+1}=N$ strings of $\Pa Z$-type
with coefficient $\beta/N$. The overlaps are: the identity (common to all three
parts) and the $n$ diagonal compatibility strings, which are $\Pa Z$-type and so also lie
in $A_{\rm bc}$. The element strings $\Pa{I\cdots I}$ and $\Pa{I\cdots IX}$ meet the
compatibility part only in the identity. Inclusion--exclusion gives
\begin{equation}
\T(A)=2+(3n-2)+2n-1-1-n+1=4n-1=2N-1 .
\end{equation}
The non-identity overlapping strings have coefficient $(\beta/2-\alpha)2^{-a}$ and cancel
when $\beta=2\alpha$, leaving $3n$ strings (verified numerically).

\begin{proposition}[Equivalence in 1D]
The assembly-free operator for $n$ elements is the stiffness matrix of a chain of
$N=2n$ nodes with springs alternating $k,\alpha,k,\dots$, plus the support penalty.
\end{proposition}
\begin{proof}
In the left-to-right order of copies, element bonds join $(2e,2e+1)$ and compatibility
bonds join $(2e+1,2e+2)$: these are consecutive springs of a chain.
\end{proof}
An assembled bar with $N$ nodes and uniform springs $k$ is the case $\alpha=k$, and the
count above is independent of the value of $\alpha$ for generic parameters ($\beta\neq2\alpha$,
i.e.\ $\beta\neq2k$ for the assembled bar). Hence, for generic parameters in both formulations,
both have $2N-1$ strings at equal $N$: in one dimension, avoiding assembly
gives no reduction.

\section{The Q4 plane-stress element}\label{sec:q4}

\subsection{Element matrix}
Consider a square element $[0,h]^2$ of unit thickness. Local node $\ell$ has offsets
$(\delta_1,\delta_2)\in\{0,1\}^2$ with index $\ell=\delta_1+2\delta_2$, i.e.\ nodes
$(0,0),(h,0),(0,h),(h,h)$; DOF $2\ell+c$ is displacement component $c\in\{0,1\}$. The
local index thus reads $\ket{\delta_2}\ket{\delta_1}\ket{c}$ on three qubits
$(\ell_2,\ell_1,\mathsf D)$. With bilinear shape functions
$N_\ell=\frac14(1+\sigma_1\xi)(1+\sigma_2\eta)$, $\sigma_i=2\delta_i-1$, the strain
matrix $B$ and plane-stress matrix
$D=\frac{E}{1-\nu^2}\begin{bmatrix}1&\nu&0\\\nu&1&0\\0&0&\frac{1-\nu}2\end{bmatrix}$,
\begin{equation}
K_e=\int_{-1}^1\!\!\int_{-1}^1B^{\mathsf T}DB\,\frac{h^2}{4}\,d\xi\,d\eta .
\end{equation}
Because $\partial/\partial x=(2/h)\partial/\partial\xi$, the factors of $h$ cancel and
$K_e$ is independent of $h$. The integrand is polynomial of degree at most two in each
variable, so $2\times2$ Gauss quadrature is exact.

\subsection{Pauli expansion}
Computing $c_P=\frac18\tr(PK_e)$ for all 64 strings (symbolically in $\nu$) gives
\begin{equation}\label{eq:q4}
\frac{1-\nu^2}{E}K_e=\frac{3-\nu}{24}\big(4\Pa{III}-\Pa{IXI}-\Pa{XII}-2\Pa{XXI}\big)
+\frac{1+\nu}{8}\big(\Pa{XIZ}-\Pa{IXZ}+\Pa{YYX}+\Pa{ZZX}\big)
+\frac{3\nu-1}{8}\big(\Pa{YZY}-\Pa{ZYY}\big).
\end{equation}
\emph{Checks.} (i) The only $\Pa Z$-type string is $\Pa{III}$, so all diagonal entries
equal $\frac{E}{1-\nu^2}\frac{3-\nu}{6}=\frac{E}{1-\nu^2}\big(\frac12-\frac\nu6\big)$, the
known diagonal of the square Q4 element. (ii) The first group is $\Pa I$ on the
component qubit: it acts identically on both components and contains the
Laplacian-like part. (iii) Rigid translations $\ket{++}\ket{c}$, with
$\ket+=(\ket0+\ket1)/\sqrt2$, have zero energy, and since $K_e$ is positive semidefinite they are
therefore annihilated: for them
$\braket{\Pa{IXI}}=\braket{\Pa{XII}}=\braket{\Pa{XXI}}=1$, so the first group gives
$4-1-1-2=0$. In the second group $\braket{\Pa{XIZ}}=\braket{\Pa{IXZ}}$, so these two cancel,
and every remaining string contains $\Pa Y$ or $\Pa Z$ on a local-node qubit, whose
expectation in $\ket+$ vanishes. (iv) The expansion was checked
numerically against Gauss integration at $\nu=0.3$.

Hence $\T(K_e)=10$ for generic $\nu$ and $\T(K_e)=8$ for $\nu=1/3$, where the last group
vanishes. For generic $\nu$ the X-patterns are $000,010,100,110,001,111,101,011$, i.e.\ all
eight, so $\mathcal X(K_e)=8$; for $\nu=1/3$ the two strings of the last group carry X-patterns
that occur nowhere else, and $\mathcal X(K_e)=6$. The 1-norm is
$\lambda(K_e)=\frac{E}{1-\nu^2}\big[\frac{3-\nu}{3}+\frac{1+\nu}{2}+\frac{|3\nu-1|}{4}\big]$.

\section{Two-dimensional compatibility operator}\label{sec:2d}

\subsection{Encoding and bonds}
With $m=2^a$ elements per direction, the state space is
$E_1\otimes E_2\otimes\ell_2\otimes\ell_1\otimes\mathsf D$ ($q=2a+3$), element
$(e_1,e_2)$ having index $e_1m+e_2$. Two elements sharing an edge normal to
direction $i$ share two nodes, each with two components. A bond along direction $i$
joins the copy with $\delta_i=1$ in element $e$ to the copy with $\delta_i=0$ in
element $e+\mathbf e_i$, for both values of the other offset and both components.
Hence, as in Section~\ref{sec:1dgeneral},
\begin{equation}\label{eq:Ac2}
A_{\rm c}=\alpha\sum_{i=1}^{2}\Big[\big(D^+\otimes P_1+D^-\otimes P_0\big)
-\big(S_a\otimes\Ldn+S_a^\dagger\otimes\Lup\big)\Big]_{(E_i,\ell_i)} ,
\end{equation}
with identity on all other registers.

\subsection{The $2\times2$ mesh in explicit form}
For $m=2$ ($a=1$), $S_1=\Lup$, and by Lemma~\ref{lem:herm} with $t=1$ and
Lemma~\ref{lem:proj}, each bracket in~\eqref{eq:Ac2} equals
$\frac12(\Pa{II}-\Pa{XX}-\Pa{YY}-\Pa{ZZ})_{(E_i,\ell_i)}$. Thus
\begin{equation}
A_{\rm c}=2\alpha\Big(\ket{\Psi^-}\!\bra{\Psi^-}_{(E_1,\ell_1)}+\ket{\Psi^-}\!\bra{\Psi^-}_{(E_2,\ell_2)}\Big),
\end{equation}
two exchange interactions, one per direction, sharing only the identity: $4+4-1=7$
strings, as found numerically. Physically, $\ket{\Psi^-}_{(E_i,\ell_i)}$ is the
antisymmetric combination of ``element $0$, right copy'' and ``element $1$, left
copy'' along direction $i$, i.e.\ the displacement jump across the interface.

\subsection{General $m$}
Per direction, Section~\ref{sec:1dgeneral} gives $3m-2$ strings with 1-norm
$\log_2m+2-2/m$ and $\log_2m+1$ X-patterns. Strings from the two directions are
distinct unless they are the identity on $E_1,E_2,\ell_1,\ell_2$: off-diagonal
strings carry $\Pa X$ or $\Pa Y$ on their own register; diagonal strings for
$\mathbf z\neq\mathbf 0$ carry $\Pa Z$ on their own register. The identity appears in both
with positive coefficient $\alpha(1-1/m)$. Therefore
\begin{equation}
\T(A_{\rm c})=2(3m-2)-1=6m-5,\qquad
\lambda(A_{\rm c})=2\alpha\Big(\log_2m+2-\frac2m\Big),\qquad
\mathcal X(A_{\rm c})=2\log_2m+1 .
\end{equation}
For $m=4$, each direction contains one in-block term ($t=1$, an exchange interaction on
the last qubit of $E_i$ and $\ell_i$) and one carry term ($t=2$, a three-body term on
both qubits of $E_i$ and $\ell_i$), exactly as in the four-element bar.

\subsection{Support and total count}
The face $x_1=0$ consists of the copies with $\delta_1=0$ in elements with $e_1=0$:
\begin{equation}
A_{\rm bc}=\beta\,P_0^{\otimes a}{}^{(E_1)}\otimes I_{E_2}\otimes I_{\ell_2}\otimes P_0^{(\ell_1)}\otimes I_{\mathsf D},
\end{equation}
which has $2^a\cdot2=2m$ strings, each $\Pa Z^{\mathbf z}\otimes(\Pa I\text{ or }\Pa Z)$
on $(E_1,\ell_1)$ with coefficient $\beta/(2m)$, and $\lambda(A_{\rm bc})=\beta$.

Inclusion--exclusion over $A_{\rm el}=I\otimes K_e$, $A_{\rm c}$ and $A_{\rm bc}$:
\begin{itemize}
\item $A_{\rm el}\cap A_{\rm c}$: element strings act as the identity on $E_1,E_2$,
whereas every non-identity compatibility string acts non-trivially on $E_1$ or $E_2$
(with $\Pa X$ or $\Pa Y$ off the diagonal, and with $Z^{\mathbf z}$, $\mathbf z\neq\mathbf 0$,
on the diagonal). The overlap is $\{\Pa I\}$.
\item $A_{\rm el}\cap A_{\rm bc}$: bc strings with $\mathbf z=\mathbf 0$ are $\Pa I$ and
$\Pa Z_{\ell_1}$; $\Pa{IZI}\notin$~\eqref{eq:q4}, so the overlap is $\{\Pa I\}$.
\item $A_{\rm c}\cap A_{\rm bc}$: the $m$ diagonal strings of direction 1 (including
$\Pa I$) all lie in $A_{\rm bc}$; no other compatibility string does.
\item Triple overlap: $\{\Pa I\}$.
\end{itemize}
Hence, for $m\ge2$ and generic parameters,
\begin{equation}\label{eq:q4total}
\T(A)=10+(6m-5)+2m-1-1-m+1=7m+4 .
\end{equation}
\emph{Cancellations.} On the $m-1$ shared non-identity strings, the coefficients are
$-\alpha/m$ from $A_{\rm c}$ and $+\beta/(2m)$ from $A_{\rm bc}$, which cancel when
$\beta=2\alpha$, giving $6m+5$. If $\nu=1/3$, two element strings vanish, giving $7m+2$. The two
cancellations involve disjoint sets of strings, so if both hold the count is $6m+3$, which gives
the complete formula~\eqref{eq:q4-complete}. All three special cases were confirmed numerically for
$m=2,4,8$.

\section{Extension to $d$ dimensions and to 3D}\label{sec:3d}

For general $d$, the compatibility operator is the sum over $i=1,\dots,d$ of the
one-direction operator on $(E_i,\ell_i)$, tensored with the DOF projector
$\Pi_{\mathsf D}$ onto the physically coupled components. Since $\Pi_{\mathsf D}$ acts on a
disjoint register, each direction contributes $s_\Pi(3m-2)$ strings with 1-norm
$\pi_\Pi(\log_2m+2-2/m)$. The strings shared between directions are those that are the
identity on all element and local registers, i.e.\ $I\otimes P$ for the $s_\Pi$ strings
$P$ of $\Pi_{\mathsf D}$; each is shared by all $d$ directions, with coefficients of the
same sign. Therefore
\begin{align}
\T(A_{\rm c})&=d\,s_\Pi(3m-2)-(d-1)s_\Pi=s_\Pi(3dm-3d+1),\\
\lambda(A_{\rm c})&=d\,\alpha\,\pi_\Pi\Big(\log_2m+2-\frac2m\Big).
\end{align}
With $N=n_{\rm loc}m^d$, where $n_{\rm loc}$ is the number of local DOF slots per
element, this gives
\[
\T(A_{\rm c})=\Theta\big(d\,s_\Pi\,(N/n_{\rm loc})^{1/d}\big).
\]

\paragraph{Hexahedral elements: unbonded padding (alternative convention).}
For $d=3$ the three displacement components occupy a two-qubit register whose fourth
slot is a padding DOF, decoupled and carrying unit stiffness in each element. The main paper bonds
and clamps the padding slot like the physical slots (bonded padding, next paragraph). This
paragraph treats the alternative in which only the three physical slots are bonded. Then
$\Pi_{\mathsf D}=I-\ket{11}\!\bra{11}=\frac14(3\Pa{II}+\Pa{IZ}+\Pa{ZI}-\Pa{ZZ})$, with
$s_\Pi=4$ and $\pi_\Pi=3/2$. Hence $\T(A_{\rm c})=36m-32$ and
$\lambda(A_{\rm c})=\frac92\alpha(\log_2m+2-2/m)$; both were confirmed for $m=2,4,8$.
For the nondimensional cubic element with unit padding stiffness, the element matrix has
$\T(K_e)=77$ strings at $\nu=0.3$ (and at every $\nu\in[0,\frac12)$ except the three values of
Section~\ref{app:proofs}), and $A_{\rm bc}$ has
$8m$ strings. The unbonded count below was computed at $\nu=0.3$ with $\beta\neq2\alpha$; other
parameter values may cancel further strings. The computed overlaps are
$|K\cap C|=|K\cap B|=|K\cap C\cap B|=4$ and $|C\cap B|=4m$ (the diagonal strings of
direction 1 times the four strings of $\Pi_{\mathsf D}$), so
\begin{equation}
\T(A)=77+(36m-32)+8m-4-4-4m+4=40m+41,
\end{equation}
as observed for $m=2,4,8$ ($121$, $201$, $361$).

\paragraph{Bonded padding (convention of the main paper).}
If the padding slot is bonded and clamped like the physical slots, then $\Pi_{\mathsf D}=I$ and
$s_\Pi=\pi_\Pi=1$, so $\T(A_{\rm c})=9m-8$, $\lambda(A_{\rm c})=3\alpha(\log_2m+2-2/m)$ and
$A_{\rm bc}$ has $2m$ strings. The overlaps become $|K\cap C|=|K\cap B|=|K\cap C\cap B|=1$ (the
identity) and $|C\cap B|=m$, so
\begin{equation}\label{eq:3dtotal-bonded}
\T(A)=77+(9m-8)+2m-1-1-m+1=10m+68,
\end{equation}
observed for $m=2,4,8,16$ ($88$, $108$, $148$, $228$) at $\nu=0.3$ and $\beta\neq2\alpha$. This is the
generic 77-term case of the unified count~\eqref{eq:h8-complete}: in general $77$ is replaced by
$\T(K_e)$, and for $\beta=2\alpha$ the $m-1$ shared support strings cancel. The padding block of the operator is
decoupled from the physical block and unloaded, so the physical solution is unchanged. The
condition number changed by about one percent: $4.39\times10^3$ against $4.34\times10^3$ at
$m=2$ and $1.750\times10^4$ against $1.736\times10^4$ at $m=4$ ($\gamma=10$, $\beta=\alpha$).

\section{Assembled operators}\label{sec:assembled}

\paragraph{One dimension.}
The free assembled bar stiffness on $M=2^{a'}$ nodes, and the consistent mass matrix,
each have $\frac32M-1$ strings (computed for $M=4,\dots,32$). With a point support penalty
the stiffness has $2M-1$ strings (Section~\ref{sec:1dgeneral}).

\paragraph{Tensor-product structure.}
For bilinear elements on a uniform grid of square elements, the scalar Laplacian stiffness matrix
is exactly $K=K^{(1)}\otimes M^{(1)}+M^{(1)}\otimes K^{(1)}$, where $K^{(1)}$ and $M^{(1)}$ are
the 1D stiffness and consistent mass matrices on the node registers ($2\times2$ Gauss quadrature
is exact for these integrands). The plane-stress operator has the same structure block by block,
with additional terms $G^{(1)}\otimes G^{(1)\mathsf T}$ built from the 1D first-derivative matrix
in the off-diagonal component blocks, and the H8 operator is the analogous sum of threefold
products. Common element-size factors of the normalized 1D stiffness and mass factors (e.g.\
$h^{-1}$ and $h$, and for H8 the corresponding products of three such factors) are omitted
throughout; they scale all coefficients of a tensor product equally and leave the counts
unchanged. Each 1D factor has at most $\frac32M$ strings, so each tensor product has $O(M^d)=O(N)$
strings and the assembled operator has $O(N)$ strings. The matching lower bound is proved below.

\paragraph{Lower bound.}
Let $M=2^b$ ($b\ge1$) be the number of nodes per direction and index the nodes of each direction in
binary on a register of $b$ qubits, as in Section~\ref{sec:perm} (any permutation of the qubits
gives the same counts). On one register let
\[
\mathcal D=\{Z^{\mathbf z}:|\mathbf z|\ \text{even}\},\qquad
\mathcal O=\{I^{\otimes(b-t)}\otimes W:\ 1\le t\le b,\ W\in\{\Pa X,\Pa Y\}^{\otimes t}\ \text{with an
even number of}\ \Pa Y\},
\]
so that $|\mathcal D|=M/2$ (including $I$), $|\mathcal O|=\sum_{t=1}^b2^{t-1}=M-1$, and the strings
of $\mathcal O$ with a given $t$ share the X-pattern $0^{b-t}1^t$. Write $\mathcal S=\mathcal
D\cup\mathcal O$, $|\mathcal S|=\frac32M-1$, and $c_P(\cdot)$ for Pauli coefficients.

\begin{lemma}[1D stiffness and mass]\label{lem:1dKM}
Let $L=K^{(1)}$ be the free 1D stiffness matrix (diagonal $1,2,\dots,2,1$, off-diagonal $-1$) and
$M^{(1)}$ the consistent mass matrix (diagonal $\frac13,\frac23,\dots,\frac23,\frac13$, off-diagonal
$\frac16$) on $M=2^b$ nodes. Both are supported exactly on $\mathcal S$, $c_P(L)\ne0$ for every
$P\in\mathcal S$, and
\[
c_P(M^{(1)})=r_P\,c_P(L),\qquad r_P=\tfrac13\ (P\in\mathcal D),\qquad r_P=-\tfrac16\ (P\in\mathcal O).
\]
\end{lemma}
\begin{proof}
With $S_b$ the increment and $E=P_0^{\otimes b}+P_1^{\otimes b}$ the projector onto the two end nodes,
$L=2I-E-(S_b+S_b^\dagger)$ and $M^{(1)}=\frac13(2I-E)+\frac16(S_b+S_b^\dagger)$. By
Lemma~\ref{lem:proj}, $E=2^{1-b}\sum_{|\mathbf z|\,\rm even}Z^{\mathbf z}$, so the diagonal part
$2I-E$ has the strings of $\mathcal D$ with coefficients $2-2^{1-b}\ge1$ (identity) and $-2^{1-b}$
(all others). By Lemma~\ref{lem:inc}, $S_b+S_b^\dagger=\sum_{t=1}^bI^{\otimes(b-t)}\otimes H_{t-1}$,
and by Lemma~\ref{lem:herm} each $H_{t-1}$ consists of the $2^{t-1}$ strings of $\mathcal O$ with
X-pattern $0^{b-t}1^t$, with coefficients $\pm2^{-(t-1)}$; different $t$ have different
X-patterns, so nothing cancels. The diagonal parts of $M^{(1)}$ and $L$ are in the ratio
$\frac13$ and the off-diagonal parts in the ratio $-\frac16$.
\end{proof}

\begin{proposition}[Assembled Pauli counts are $\Theta(N)$]\label{prop:assembled-lb}
Consider the assembled stiffness matrix of a uniform Cartesian grid of square Q4 (plane stress) or
cubic H8 elements with $M=2^b$ nodes per direction, any $E>0$ and $0\le\nu<\frac12$, the components
on one (Q4) or two (H8, padding slot with a stiffness proportional to the identity, possibly zero)
qubits, and a clamped face $x_1=0$ imposed either by a penalty $\beta>0$ on its degrees of freedom
or by exact Dirichlet elimination with identity padding (the constrained rows and columns are
zeroed and their diagonal set to one, so that the dimension remains a power of two; this is the
assembled comparator of the main paper, not the free-node matrix $K_0$). Let $D=M/2$,
$O=M-1$ and $N=2M^2$ (Q4) or $N=4M^3$ (H8). Then the number of Pauli strings satisfies
\begin{itemize}\itemsep1pt
\item Q4, penalty supports: $\T\ge(D+O)^2+1=\tfrac98N-O(\sqrt N)$;
\item Q4, identity-padded elimination: $\T\ge(M-2)(D+O)=\tfrac34N-O(\sqrt N)$;
\item H8, penalty supports: $\T\ge D^3+3DO^2+O^3+1=\tfrac{21}{32}N-O(N^{2/3})$;
\item H8, identity-padded elimination: $\T\ge(M-2)(M-1)(2M-1)=\tfrac12N-O(N^{2/3})$;
\end{itemize}
and $\T=O(N)$, so $\T=\Theta(N)$ in all four cases.
\end{proposition}
\begin{proof}
We count only the strings that are the identity on the component register. The coefficient of
$I\otimes R$ in $\sum_{c,c'}\ket c\!\bra{c'}\otimes K_{cc'}$ is the average over the components of the
coefficients of $R$ in the diagonal blocks $K_{cc}$; the off-diagonal blocks do not contribute.

\emph{Q4.} With exact quadrature, $K_{xx}=d_{11}L\otimes M^{(1)}+d_{33}M^{(1)}\otimes L$ and
$K_{yy}=d_{33}L\otimes M^{(1)}+d_{11}M^{(1)}\otimes L$, where $d_{11}=E/(1-\nu^2)$ and
$d_{33}=E/(2(1+\nu))$, so $K_{xx}+K_{yy}=(d_{11}+d_{33})(L\otimes M^{(1)}+M^{(1)}\otimes L)$. By
Lemma~\ref{lem:1dKM} the coefficient of $P\otimes Q$ in this sum is
$(d_{11}+d_{33})\,c_P(L)\,c_Q(L)\,(r_P+r_Q)$ for $P,Q\in\mathcal S$ and zero otherwise. Since
$r_P+r_Q\in\{\frac23,\frac16,-\frac13\}$ never vanishes, the free operator has exactly $(D+O)^2$
strings of the form $I\otimes P\otimes Q$.

\emph{Penalty supports} add $\beta\,I\otimes P_0^{\otimes b}\otimes I=\beta2^{-b}\sum_{\mathbf z}
I\otimes Z^{\mathbf z}\otimes I$. The $M/2$ strings with odd $|\mathbf z|$ are new; of those with even
$|\mathbf z|$, the identity cannot cancel (its coefficient is ${\rm tr}\,K/N>0$), so at most
$M/2-1$ strings are lost. Hence $\T\ge(D+O)^2+1$. (All $M/2-1$ coefficients are equal, so they cancel
together for one value of $\beta$ only; for every other $\beta$ the count of these strings is
$(D+O)^2+M/2$.)

\emph{Identity-padded elimination} changes $K$ only in rows or columns of the clamped nodes, and only in
entries that couple them to themselves or to their neighbours in direction 1, i.e.\ in entries whose
direction-1 node indices $(i,i')$ lie in $\{(0,0),(0,1),(1,0)\}$. The change therefore contains only
strings whose X-pattern on the direction-1 register is $0^b$ or $0^{b-1}1$, and the coefficients of
all strings $I\otimes P\otimes Q$ with $P\in\mathcal O$ of carry length $t\ge2$ are unchanged. There
are $\sum_{t=2}^b2^{t-1}=M-2$ such $P$, which gives $\T\ge(M-2)(D+O)$.

\emph{H8.} With $d_{11}=\lambda+2\mu$ and $d_{44}=\mu$, $K_{xx}+K_{yy}+K_{zz}=(d_{11}+2d_{44})(L\otimes
M^{(1)}\otimes M^{(1)}+M^{(1)}\otimes L\otimes M^{(1)}+M^{(1)}\otimes M^{(1)}\otimes L)$, and the
padding block adds only a multiple of the identity. The coefficient of $P\otimes Q\otimes R$ is
$(d_{11}+2d_{44})\,c_P(L)c_Q(L)c_R(L)\,(r_Qr_R+r_Pr_R+r_Pr_Q)$. The bracket is $\frac13$, $0$,
$-\frac1{12}$ and $\frac1{12}$ when three, two, one or none of $P,Q,R$ lie in $\mathcal D$, so
exactly the triples with two strings in $\mathcal D$ vanish, which leaves $D^3+3DO^2+O^3$ strings. The
penalty argument is the same as for Q4 (the strings $Z^{\mathbf z}\otimes I\otimes I$ have three
factors in $\mathcal D$). For identity-padded elimination take $P\in\mathcal O$ with $t\ge2$; then the bracket is
nonzero unless $Q,R\in\mathcal D$, which leaves $(D+O)^2-D^2=(M-1)(2M-1)$ pairs $(Q,R)$ for each of the
$M-2$ choices of $P$.

\emph{Upper bound.} Each block $K_{cc'}$ is a sum of at most $d$ tensor products of the 1D
matrices $L$, $M^{(1)}$, $G^{(1)}$ and $G^{(1)\mathsf T}$, each with at most $\frac32M$ strings; elimination modifies
each factor in $O(1)$ entries next to the clamped node, which adds $O(M)$ strings per factor; and the
penalty adds $M$ strings. The number of component blocks is fixed, so $\T=O(M^d)=O(N)$.
\end{proof}

The lower bounds count only strings with the identity on the component register; the actual counts
are larger. Table~\ref{tab:assembled-lb} compares them with full numerical Pauli expansions
(threshold $10^{-10}$) of the operators of Table~\ref{M-tab:counts} of the main paper and
Table~\ref{tab:costraw}: for generic $\beta$ the identity-sector count equals $(D+O)^2+M/2$ (Q4)
and $D^3+3DO^2+O^3+M/2$ (H8) for every tested $M$, and the same code reproduces the assembled totals of those tables.

\begin{table}[ht]
\centering\small
\caption{Assembled Pauli counts: total $\T$, number of strings with the identity on the component
register ($\T_I$), and the lower bound of Proposition~\ref{prop:assembled-lb}. Penalty supports with
$\beta=11.1$, $\nu=0.3$.}
\label{tab:assembled-lb}
\begin{tabular}{@{}cr rrr rrr@{}}
\toprule
 & & \multicolumn{3}{c}{penalty supports} & \multicolumn{3}{c}{identity-padded elimination}\\
\cmidrule(lr){3-5}\cmidrule(l){6-8}
 & $M$ & $\T$ & $\T_I$ & bound & $\T$ & $\T_I$ & bound\\
\midrule
Q4 & 4  & 64    & 27    & 26    & 100    & 40     & 10\\
   & 8  & 302   & 125   & 122   & 492    & 198    & 66\\
   & 16 & 1{,}306 & 537 & 530   & 2{,}164 & 874   & 322\\
   & 32 & 5{,}426 & 2{,}225 & 2{,}210 & 9{,}060 & 3{,}666 & 1{,}410\\
   & 64 & --    & --    & 9{,}026 & 37{,}060 & 15{,}010 & 5{,}890\\
H8 & 4  & 1{,}222 & 91  & 90    & 1{,}936 & 136   & 42\\
   & 8  & 12{,}990 & 999 & 996  & 21{,}172 & 1{,}570 & 630\\
\bottomrule
\end{tabular}
\end{table}

\begin{remark}[Rectangular elements]
For Q4 elements with aspect ratio $\rho=h_2/h_1\ne1$ the coefficient of $P\otimes Q$ becomes
proportional to $\rho r_Q+r_P/\rho$. The classes with $P,Q$ both in $\mathcal D$ or both in
$\mathcal O$ never vanish, and the two mixed classes vanish only for $\rho^2=2$ and $\rho^2=\frac12$,
respectively, so at most one of them is lost. Hence $\T\ge D^2+O^2+DO+1$ for every aspect ratio
(checked for $\rho=\sqrt2,1/\sqrt2,3$), and the assembled count remains $\Theta(N)$. The term $+1$
uses the identity string of the penalty-support case; for exact elimination with identity padding
the corresponding lower bound requires its own argument, which we do not give.
\end{remark}

The proposition concerns the standard Pauli basis with binary node indices on uniform grids. It
does not bound the number of terms of other decompositions of the assembled operator (e.g.\ the
shift-based linear combinations of Table~\ref{M-tab:shift} of the main paper, or the factorized encodings
cited in the main paper), nor the Pauli counts on curved or masked meshes, which are computed.

\paragraph{Contrast.}
In the assembly-free operator the element part is $I\otimes K_e$ and the coupling part
is a sum of one-register operators, so counts add over directions. In the assembled
operator the directions are combined through tensor products, so counts multiply. In
1D there is only one direction and the two coincide.

\section{Penalty error, conditioning and 1-norm}\label{sec:penalty}

\paragraph{One dimension.}
From~\eqref{eq:ugen}, the relative tip error is
$\frac k\alpha\frac{n-1}{n}+\frac{k}{n\beta}$: of order $1/\gamma$, uniformly in $n$, when
$\alpha=\beta=\gamma k$; otherwise the support term $k/(n\beta)$ must be retained separately.

\paragraph{Two and three dimensions.}
For Q4 plane stress the element stiffness scale is $k_0=Et$ (Eq.~\eqref{M-eq:gamma} of the main
paper), independent of $h$, so a
fixed $\gamma=\alpha/(Et)$ keeps the interface springs in proportion to the elements under
refinement. The resulting penalty errors for a point load are given in Table~\ref{tab:penalty}
below.

\paragraph{Conditioning.}
The largest eigenvalue of the assembly-free operator is dominated by the penalty
springs and scales with $\alpha$, while the smallest is governed by the elastic energy
of the softest mode, which is close to that of the assembled problem. This heuristic argument
suggests $\kappa_{\rm b}\approx c\,\gamma\,\kappa_{\rm a}$ for a modest constant $c$; it is a
numerical observation, not a proved bound. With Dirichlet DOFs eliminated in the assembled
problem, the observed $c$ ranged from $2.4$ to $3.4$ in 2D for $m=2$ to $16$, increasing slowly.

\paragraph{1-norm.}
$\lambda(A)=\lambda(K_e)+\lambda(A_{\rm c})+\lambda(A_{\rm bc})$ minus cancellations. The
dominant term for large $\gamma$ is $\lambda(A_{\rm c})=2\alpha(\log_2m+2-2/m)$ in 2D,
logarithmic in $m$ and linear in $\gamma$.

\paragraph{Point load: penalty error and condition numbers.}
Table~\ref{tab:penalty} gives the relative error of the loaded-corner displacement of the
cantilever (unit downward load at the far corner, face $x_1=0$ clamped) against the conforming
solution on the same mesh, in two and three dimensions, together with the condition numbers of the
unpreconditioned operators. The error is proportional to $1/\gamma$ to within one percent and grows
slowly with $m$: over the tested refinements the 2D coefficient $\gamma e$ ranges from about
$0.187$ ($m=2$) to $0.396$ ($m=16$); no limiting value is claimed. (For H8 the stiffness scale is
$k_0=Eh$.) Part of this growth arises
because the displacement under a point force is not bounded under refinement, so the table tests
the $1/\gamma$ behaviour rather than mesh independence. With the support penalty kept fixed at
$\beta=10^8$, these errors are those of the interface springs.

\begin{table}[ht]
\centering
\caption{Cantilever: relative error of the corner displacement caused by the penalty
($\alpha=\gamma$, $\beta=10^8$, nondimensional, i.e.\ $\alpha=\gamma k_0$ with $k_0=Et$ for Q4 and
$k_0=Eh$ for H8), and condition numbers. $\kappa_{\rm b}$: assembly-free,
$\alpha=\beta=10$; $\kappa_{\rm a}$: assembled with eliminated supports.}
\label{tab:penalty}
\footnotesize
\begin{tabular}{@{}cr|ccc|cc@{}}
\toprule
& & \multicolumn{3}{c|}{Relative error} & & \\
$d$ & $m$ & $\gamma=10$ & $\gamma=10^2$ & $\gamma=10^3$ & $\kappa_{\rm b}$ & $\kappa_{\rm a}$\\
\midrule
\multirow{4}{*}{2 (Q4)} & 2 & $1.87\times10^{-2}$ & $1.87\times10^{-3}$ & $1.87\times10^{-4}$ & $1.58\times10^{3}$ & $6.48\times10^{1}$\\
 & 4 & $2.99\times10^{-2}$ & $3.00\times10^{-3}$ & $3.00\times10^{-4}$ & $6.28\times10^{3}$ & $2.43\times10^{2}$\\
 & 8 & $3.62\times10^{-2}$ & $3.62\times10^{-3}$ & $3.62\times10^{-4}$ & $2.53\times10^{4}$ & $8.41\times10^{2}$\\
 & 16 & $3.95\times10^{-2}$ & $3.96\times10^{-3}$ & $3.96\times10^{-4}$ & $1.02\times10^{5}$ & $3.01\times10^{3}$\\
\midrule
\multirow{2}{*}{3 (H8)} & 2 & $7.92\times10^{-3}$ & $7.94\times10^{-4}$ & $7.95\times10^{-5}$ & $4.39\times10^{3}$ & $9.35\times10^{1}$\\
 & 4 & $1.06\times10^{-2}$ & $1.06\times10^{-3}$ & $1.07\times10^{-4}$ & $1.75\times10^{4}$ & $3.34\times10^{2}$\\
\bottomrule
\end{tabular}
\end{table}

\section{Invariance of term counts under qubit permutations}\label{sec:perm}
Let $\sigma$ permute the $q$ qubits and $U_\sigma\ket{b_1\cdots b_q}=\ket{b_{\sigma^{-1}(1)}\cdots b_{\sigma^{-1}(q)}}$.
For any string $P=P_1\otimes\cdots\otimes P_q$, $U_\sigma PU_\sigma^\dagger$ is the string with
permuted factors. Hence $c_P(A)=c_{U_\sigma PU_\sigma^\dagger}(U_\sigma AU_\sigma^\dagger)$, and $\T$,
$\lambda$ and $\mathcal X$ are unchanged. On a $2^{a'}\times2^{a'}$ node grid with index
$(i,j)$, lexicographic ordering uses the bit string $i_{a'-1}\cdots i_0\,j_{a'-1}\cdots j_0$
and Morton ordering $i_{a'-1}j_{a'-1}\cdots i_0j_0$; these differ by a qubit permutation.
Computed counts for both orderings were identical: $64,302,1306,5426$ (square) and
$100,516,2320,9816$ (L-shape) for $m'=3,7,15,31$. The argument covers permutations of the qubit
registers only; an arbitrary permutation of the node or degree-of-freedom indices is not a qubit
permutation and can change the Pauli sparsity.

\section{Masked domains: the L-shaped bracket}\label{sec:lshape}
\paragraph{Construction.}
On an $m\times m$ grid ($m=2^a$) the elements with $e_1\ge m/2$ and $e_2\ge m/2$ are void.
These are exactly the elements whose most significant bits $e_1^{\rm top}$, $e_2^{\rm top}$
are both one, so the void indicator is
$M=P_1\otimes P_1=\frac14(\Pa{II}-\Pa{IZ}-\Pa{ZI}+\Pa{ZZ})$ on these two qubits. Void
elements carry $c\,I_8$ and are not bonded. Then
\[
A_{\rm el}=(I-M)\otimes K_e+c\,M\otimes I_8 .
\]
$I-M$ has the same four strings as $M$, so $(I-M)\otimes K_e$ has $4\times10=40$ strings;
$c\,M\otimes I_8$ contributes strings of the form $(\cdot)\otimes\Pa{III}$, which already occur
because $K_e$ contains $\Pa{III}$. Hence $\T(A_{\rm el})=40$ for every $m$ (generic $\nu$, for which $K_e$ has ten strings, and
generic void regularization $c$).

\paragraph{Counts.}
Table~\ref{tab:lshape} lists the numerically evaluated Pauli term counts. Over the tested
refinements $m=4,\dots,64$, the compatibility and total counts are described by $17m-18$ and
$(35m+36)/2$, respectively; these expressions are empirically observed and have not been
established analytically. Removing the bonds inside and around
the void quadrant splits each direction's increment operator into pieces acting on
dyadic sub-blocks, controlled by the top bit of the other register; this is the origin
of the larger constant compared with $6m-5$, but a closed-form derivation along the lines of
Section~\ref{sec:2d} has not been carried out.

\begin{table}[ht]
\centering\small
\caption{L-shaped bracket: numerically evaluated term counts ($\alpha=7.3$, $\beta=11.1$, $c=1$).}
\label{tab:lshape}
\begin{tabular}{@{}rr|rrrr|rr@{}}
\toprule
$m$ & $N$ & $\T(A_{\rm el})$ & $\T(A_{\rm c})$ & $\T(A_{\rm bc})$ & $\T(A)$ & $m'$ (assembled) & $\T$\\
\midrule
4 & 128 & 40 & 50 & 8 & 88 & 3 & 100\\
8 & 512 & 40 & 118 & 16 & 158 & 7 & 516\\
16 & 2048 & 40 & 254 & 32 & 298 & 15 & 2320\\
32 & 8192 & 40 & 526 & 64 & 578 & 31 & 9816\\
64 & 32768 & 40 & 1070 & 128 & 1138 & -- & --\\
\bottomrule
\end{tabular}
\end{table}

\section{Graded meshes: numerical test}\label{sec:gradedsupp}

This section reports the test of a uniform penalty on a graded mesh that is summarized in
Section~\ref{M-sec:graded} of the main paper.
To test a uniform penalty on unequal elements, a cantilever of length 2 and height 1, clamped at
$x_1=0$ and loaded by a unit downward force at the free top corner, was meshed with $16\times16$ Q4
elements graded geometrically towards the clamped corner, with ratio $\rho$ between neighbouring
element widths in both directions. The element aspect ratio then varies over the mesh, up to 16
for $\rho=1.15$ and 102 for $\rho=1.3$. Table~\ref{tab:graded} gives the error of the tip
displacement against the conforming solution on the same graded mesh, with $\alpha=\beta=\gamma$.
It remains proportional to $1/\gamma$ and increases moderately with the grading, by factors of
about 1.7 and 2.8 for largest-to-smallest element ratios of 8 and 51, which reflects the large
aspect ratios. Because the Q4 stiffness depends on the aspect ratio but not on the size of an
element (Eq.~\eqref{M-eq:graded-q4}), the springs need not be rescaled with the element size in 2D
at a fixed aspect ratio; this statement concerns the element size only, since the stiffness does
depend on the aspect ratio, which varies over this mesh; in 3D the element stiffness scales with $h$, and the springs should be scaled with the local
element size.

\begin{table}[!htbp]
\centering
\caption{Graded Q4 cantilever ($2\times1$, $16\times16$ elements, uniform penalty
$\alpha=\beta=\gamma$): relative error of the tip displacement against the conforming solution on
the same mesh.}
\label{tab:graded}
\footnotesize
\begin{tabular}{@{}cccccc@{}}
\toprule
$\rho$ & $h_{\max}/h_{\min}$ & max.\ $h_1/h_2$ & $\gamma=10$ & $\gamma=10^2$ & $\gamma=10^3$\\
\midrule
1.00 & 1.0  & 2   & $3.52\times10^{-2}$ & $3.53\times10^{-3}$ & $3.53\times10^{-4}$\\
1.15 & 8.1  & 16  & $5.81\times10^{-2}$ & $5.85\times10^{-3}$ & $5.86\times10^{-4}$\\
1.30 & 51.2 & 102 & $9.63\times10^{-2}$ & $9.86\times10^{-3}$ & $9.89\times10^{-4}$\\
\bottomrule
\end{tabular}
\end{table}

\section{Curved meshes and Walsh compression}\label{sec:curved}
\paragraph{Geometries.}
Nodes are $\mathbf x(i/m,j/m)$ for a map $\mathbf x(\xi,\eta)$ of the unit square:
the curved bracket uses $\mathbf x=(1+\xi)(\cos\frac{\pi\eta}{2},\sin\frac{\pi\eta}{2})$;
the distorted plate uses
$x=\xi+0.12\sin(\pi\eta)(1+\xi)+0.05\eta$, $y=\eta(1+0.4\xi)+0.08\sin(2\pi\xi)\eta$.
The perturbed mesh moves each interior node of the distorted plate in the parameter
plane by independent uniform offsets of up to $\pm25\%$ of the spacing (seed 0). Element
matrices are isoparametric Q4 with $2\times2$ Gauss quadrature; the clamped edge is
$\xi=0$ and the point load acts in the $-y$ direction at $\mathbf x(1,1)$.

\paragraph{Why uncompressed counts grow approximately linearly over the tested refinements.}
With $A_{\rm el}=\sum_e\ket e\!\bra e\otimes K_e=\sum_{P}\big(\sum_e c_P(e)\ket e\!\bra e\big)\otimes P$,
each local string $P$ carries a diagonal field $c_P(e)$ whose Pauli strings are the
nonzero Walsh coefficients of $c_P$. For a generic field all $N_{\rm el}$ coefficients
are nonzero, so $\T(A_{\rm el})$ approaches $N_{\rm el}$ times the number of local strings.
On the uniform grid every $c_P$ is constant and only the zero coefficient survives.

\paragraph{Factorization.}
Stacking $\mathrm{vec}(K_e)\in\mathbb R^{64}$ as rows of $M\in\mathbb R^{N_{\rm el}\times64}$
and computing $M=U\Sigma V^{\mathsf T}$ gives $K_e=\sum_{k\le r}g_k(e)B_k$ with
$g_k=\Sigma_{kk}U_{:,k}$ and $B_k$ the $k$th row of $V^{\mathsf T}$, reshaped. The decomposition is exact only when all nonzero singular values are retained (in exact
arithmetic); the numerical rank was $r=12$ (curved bracket; the elements in a ring are rotated
copies) and $r=19$ (distorted plate). The compression below truncates the Walsh expansions of the
geometry factors $g_k$; this is a different coefficient representation from a direct truncation of
the assembled Pauli expansion, with which it is compared. Since every $B_k$ is a linear combination of element matrices, all of
which annihilate rigid translations, so does every $\tilde K_e$. Rigid rotations are
element-dependent and are preserved only approximately.

\paragraph{Algorithm and translation invariance.}
Algorithm~\ref{alg:compress} summarizes the compression of the element part used in the main
paper.

\begin{algorithm}[ht]
\caption{Geometry factorization and Walsh compression of the element part}
\label{alg:compress}
\begin{algorithmic}[1]
\Require element matrices $\{K_e\}_{e=1}^{N_{\rm el}}$; tolerance $\tau_{\rm c}$
\Ensure compressed element matrices $\{\tilde K_e\}$
\State stack $\mathrm{vec}(K_e)$ as rows of $M$ and compute $M=U\Sigma V^{\mathsf T}$; keep rank $r$
\State $g_k\gets\Sigma_{kk}U_{:,k}$, \quad $B_k\gets$ $k$th row of $V^{\mathsf T}$, reshaped \Comment{$K_e=\sum_kg_k(e)B_k$, exact if all nonzero singular values are kept (exact arithmetic)}
\For{$k=1,\dots,r$}
  \State $\hat g_k\gets\mathrm{FWHT}(g_k)/N_{\rm el}$ \Comment{transform over the element index}
\EndFor
\State set $\hat g_k(\mathbf s)\gets0$ wherever $|\hat g_k(\mathbf s)|<\tau_{\rm c}\max_{k,\mathbf s}|\hat g_k|$
\State $\tilde g_k\gets\mathrm{FWHT}(\hat g_k)$, \quad $\tilde K_e\gets\sum_k\tilde g_k(e)B_k$
\State \textbf{verify:} compute $\lambda_{\min}$ of the complete compressed operator (springs
included); if it is not positive, decrease $\tau_{\rm c}$ and repeat \Comment{positivity safeguard}
\State \Return $\{\tilde K_e\}$ \Comment{cost $O(rN_{\rm el}\log N_{\rm el})$ per pass, plus $O(N_{\rm el})$ for the SVD}
\end{algorithmic}
\end{algorithm}

The stated cost of Algorithm~\ref{alg:compress} excludes the computation of the smallest eigenvalue
of the complete compressed operator in the positivity safeguard, which is a global eigenvalue
problem, and any repetitions of the loop with a smaller tolerance.

\begin{proposition}[Rigid translations are preserved]\label{prop:transl}
If every $K_e$ annihilates a vector $\mathbf t$, then so does every $\tilde K_e$.
\end{proposition}
\begin{proof}
Each row of $V^{\mathsf T}=\Sigma^{-1}U^{\mathsf T}M$ is a linear combination of the rows of
$M$, so each $B_k$ is a linear combination of element matrices and $B_k\mathbf t=\mathbf 0$.
\end{proof}
Proposition~\ref{prop:transl} concerns rigid translations only. Positive semidefiniteness of the
individual $\tilde K_e$ is not preserved, and positivity of the complete operator was checked only
for the reported meshes and tolerances; it is not guaranteed for an arbitrary geometry or
threshold.

\paragraph{Results.}
For the curved bracket and the distorted plate, the uncompressed assembly-free counts grow linearly
but remain smaller by factors of 3.3 and 1.9, respectively, than the assembled counts at comparable resolution
(Table~\ref{tab:examples}). Algorithm~\ref{alg:compress} with $\gamma=100$ needed ranks
$r=12$ and $r=19$. At $\tau_{\rm c}=10^{-4}$ the compressed counts of the distorted plate grew by
factors of $1.8$ and $1.3$ over two refinements that each quadrupled $N$, with relative
displacement errors near $0.1\%$; for the curved bracket the factors were $1.6$ and $1.35$
at errors between $0.3\%$ and $0.6\%$.

\begin{figure}[ht]
\centering
\includegraphics{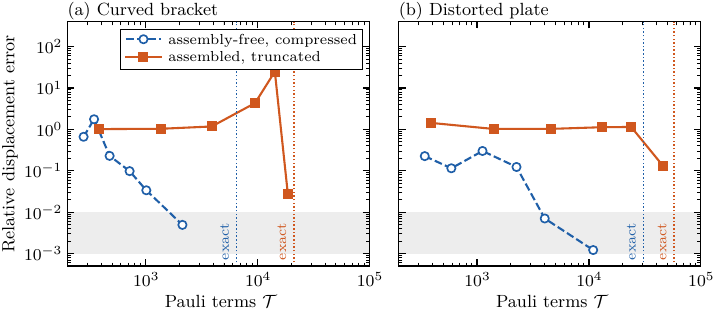}
\caption{Pauli term count versus relative displacement error (2-norm of the displacement vector)
under compression, at comparable resolution ($m=32$ assembly-free, $m'=31$ assembled). Assembly-free
(blue): Algorithm~\ref{alg:compress} with tolerances $10^{-1}$ to $10^{-4}$; assembled (orange):
truncation of the stiffness expansion with tolerances $10^{-2}$ to $10^{-5}$. Dotted vertical lines
mark the uncompressed counts; the shaded band is $10^{-3}$--$10^{-2}$.}
\label{fig:compression}
\end{figure}

Fig.~\ref{fig:compression} sweeps the tolerance at comparable resolution. The compressed
assembly-free operator reaches errors below $1\%$ with about $2{,}100$ strings (curved
bracket) and $4{,}000$ strings (distorted plate). Truncating the assembled expansion by the
same relative rule, applied to its stiffness part only, produces errors of order $100\%$
until almost all strings are kept: for the distorted plate, keeping $46{,}040$ of $57{,}824$
strings still leaves a $13\%$ error. On the randomly perturbed mesh, compression is
ineffective (at tolerance $10^{-4}$ the compressed operator has $32{,}892$ strings, more than the
$30{,}846$ of the uncompressed one, Table~\ref{tab:compr}), because the tested randomly perturbed geometry
does not exhibit sufficiently rapid Walsh-coefficient decay for effective compression. This case
marks a genuine limit of the approach.

\paragraph{Truncation and error.}
Walsh coefficients $\hat g_k=H g_k/N_{\rm el}$ below $\tau_{\rm c}\max|\hat g|$ are set to zero. The
error is the relative Euclidean norm of the difference between the displacements
computed with $\tilde A$ and with $A$ (same formulation, $\gamma=100$, $\beta=10^6$).
Table~\ref{tab:compr} lists all data. For the assembled operator, the same global
relative threshold was applied to the Pauli coefficients of the stiffness part, with the
Dirichlet penalty kept exact.

\begin{table}[ht]
\centering\small
\caption{Compressed term counts and relative displacement errors (assembly-free).}
\label{tab:compr}
\begin{tabular}{@{}llrrrr@{}}
\toprule
Problem & $m$ & $N$ & uncompressed $\T$ & $\tau_{\rm c}=10^{-3}$ & $\tau_{\rm c}=10^{-4}$\\
\midrule
Curved bracket & 16 & 2048 & 1{,}741 & 730 (0.99\%) & 1{,}344 (0.31\%)\\
 & 32 & 8192 & 6{,}485 & 1{,}011 (3.4\%) & 2{,}126 (0.49\%)\\
 & 64 & 32768 & 20{,}217 & 1{,}326 (3.9\%) & 2{,}863 (0.57\%)\\
Distorted plate & 16 & 2048 & 7{,}742 & 3{,}041 (0.51\%) & 6{,}076 (0.08\%)\\
 & 32 & 8192 & 30{,}668 & 4{,}026 (0.70\%) & 10{,}913 (0.12\%)\\
 & 64 & 32768 & 110{,}152 & 4{,}652 (0.87\%) & 14{,}574 (0.14\%)\\
Perturbed mesh & 32 & 8192 & 30{,}846 & 32{,}892 (1.5\%) & 32{,}892 (0.24\%)\\
\bottomrule
\end{tabular}
\end{table}

For the perturbed mesh the compressed operator is no smaller than the uncompressed one: with
rough geometry the truncation removes few coefficients but redistributes them among
the local strings. Uncompressed assembled counts at $m'=15,31$ were $4{,}080$ and $21{,}168$
(curved bracket) and $10{,}224$ and $57{,}824$ (distorted and perturbed). The full sweeps
are shown in Fig.~\ref{fig:compression}; at moderate thresholds the assembled truncation
errors were between $100\%$ and $2500\%$.

\paragraph{Term counts of the benchmark problems.}
Table~\ref{tab:examples} lists the uncompressed and compressed Pauli term counts quoted in
Section~\ref{M-sec:curved} of the main paper.

\begin{table}[ht]
\centering
\caption{Pauli term counts for the benchmark problems. Assembly-free: $m=32$ ($N=8{,}192$) unless
stated; assembled: $m'=31$ ($N=2{,}048$, comparable resolution). Compressed counts at tolerance
$10^{-4}$, with the relative displacement error in parentheses.}
\label{tab:examples}
\footnotesize\setlength{\tabcolsep}{4pt}
\begin{tabular}{@{}llrrr@{}}
\toprule
Problem & Mesh & Assembly-free & Assembly-free & Assembled\\
 & & uncompressed & compressed & uncompressed\\
\midrule
(a) Cantilever & uniform & 228 & -- & 5{,}426\\
(b) L-shaped bracket & masked & 578 & -- & 9{,}816\\
(b) L-shaped bracket, $m=64$ & masked & 1{,}138 & -- & --\\
(c) Curved bracket & polar & 6{,}485 & 2{,}126 (0.49\%) & 21{,}168\\
(c) Curved bracket, $m=64$ & polar & 20{,}217 & 2{,}863 (0.57\%) & --\\
(d) Distorted plate & mapped & 30{,}668 & 10{,}913 (0.12\%) & 57{,}824\\
(d) Distorted plate, $m=64$ & mapped & 110{,}152 & 14{,}574 (0.14\%) & --\\
(e) Perturbed mesh & random & 30{,}846 & ineffective & 57{,}824\\
\bottomrule
\end{tabular}
\end{table}

\section{Additional results for the operator study}\label{sec:opextra}
This section collects material that supports Section~\ref{M-sec:numerics} of the main paper.

\paragraph{Closed-form Pauli list.}
Proposition~\ref{thm:main} is constructive. Algorithm~\ref{alg:closed} lists the expansion of
the assembly-free operator directly, without forming any matrix.

\begin{algorithm}[ht]
\caption{Closed-form Pauli list of the assembly-free operator on a uniform grid}
\label{alg:closed}
\begin{algorithmic}[1]
\Require dimension $d$; $m=2^a$; element matrix $K_e$; penalties $\alpha,\beta$; projector $\Pi_{\mathsf D}$
\Ensure list $\mathcal L=\{(c_P,P)\}$ with $A=\sum c_PP$
\State expand $\Pi_{\mathsf D}=\sum_{k=1}^{s_\Pi}\pi_kQ_k$ in Pauli strings ($\Pi_{\mathsf D}=I$ for Q4 and for H8 with bonded padding); every ``add $c\,P\otimes\Pi_{\mathsf D}$'' below adds the $s_\Pi$ strings $c\,\pi_k\,P\otimes Q_k$
\State $\mathcal L\gets\{(c_\Lambda,\ I_E\otimes\Lambda)\}$ for the strings $\Lambda$ of $K_e$ \Comment{element part, Eq.~\eqref{eq:Ael}}
\For{$i=1,\dots,d$} \Comment{coupling part, Eq.~\eqref{eq:Ac}}
  \State add $\alpha(1-2^{-a})\,I\otimes\Pi_{\mathsf D}$
  \For{$\mathbf z\in\{0,1\}^a\setminus\{\mathbf 0\}$}
     \State add $-\alpha2^{-a}\,[Z^{\mathbf z}]_{E_i}\otimes[\Pa Z^{|\mathbf z|\bmod2}]_{\ell_i}\otimes\Pi_{\mathsf D}$ \Comment{Lemma~\ref{lem:proj}}
  \EndFor
  \For{$t=1,\dots,a$}
     \For{each even-$\Pa Y$ word $W\in\{\Pa X,\Pa Y\}^{t+1}$ with sign $\sigma_W$}
        \State add $-\alpha\,\sigma_W2^{-t}\,[\Pa I^{\otimes(a-t)}\otimes W]_{(E_i,\ell_i)}\otimes\Pi_{\mathsf D}$ \Comment{Lemmas~\ref{lem:inc}, \ref{lem:herm}}
     \EndFor
  \EndFor
\EndFor
\For{$\mathbf z\in\{0,1\}^a$, $b\in\{0,1\}$} \Comment{boundary part, Eq.~\eqref{eq:Abc}}
  \State add $\beta2^{-a-1}\,[Z^{\mathbf z}]_{E_1}\otimes[\Pa Z^b]_{\ell_1}\otimes\Pi_{\mathsf D}$
\EndFor
\State merge equal strings and discard zero coefficients
\State \Return $\mathcal L$ \Comment{$O(\T q)$ operations}
\end{algorithmic}
\end{algorithm}

\begin{figure}[ht]
\centering
\includegraphics{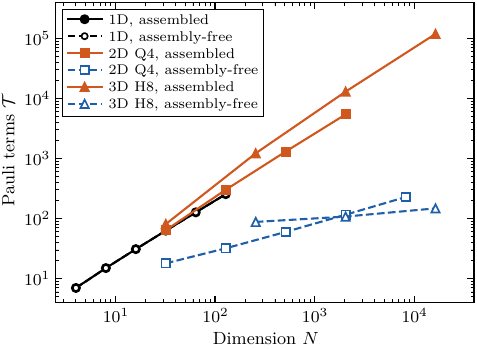}
\caption{Numerically evaluated Pauli term counts on uniform grids against the dimension $N$ ($\alpha=7.3$,
$\beta=11.1$; H8 with bonded padding). Assembled operators grow linearly; assembly-free operators
grow as $N^{1/2}$ in 2D and $N^{1/3}$ in 3D; in 1D both formulations give the same counts.}
\label{fig:scaling}
\end{figure}

\begin{figure}[ht]
\centering
\includegraphics{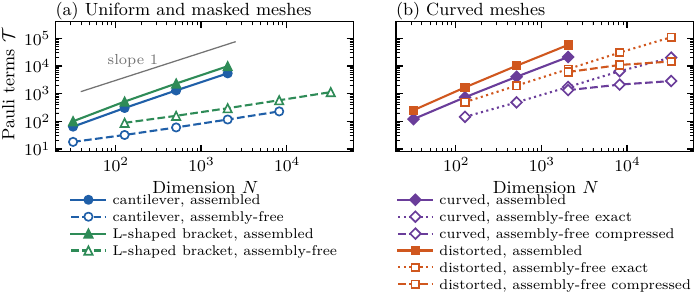}
\caption{Pauli term counts of the benchmark problems against the dimension $N$ ($\alpha=7.3$,
$\beta=11.1$). (a) Cantilever and L-shaped bracket; (b) curved bracket and distorted plate. Solid lines,
filled symbols: assembled operator; dashed lines, open symbols: assembly-free operator (compressed at
$\tau_{\rm c}=10^{-4}$ in (b)); dotted lines in (b): assembly-free operator without compression. The
thin line in (a) has slope one.}
\label{fig:scalingex}
\end{figure}

\paragraph{Term counts versus problem size.}
Figure~\ref{fig:scaling} plots the counts of Table~\ref{M-tab:counts} of the main paper, and
Fig.~\ref{fig:scalingex} those of the benchmark problems.

\begin{figure}[ht]
\centering
\includegraphics{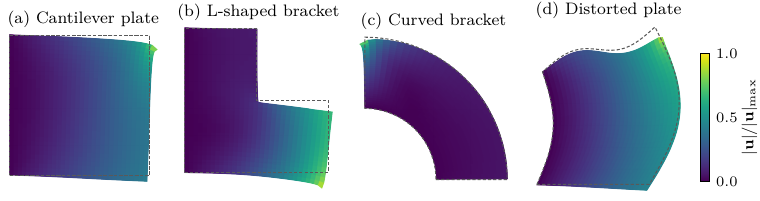}
\caption{Deformed configurations (magnified) of the conforming solutions on $32\times32$ meshes under
the point loads of Fig.~\ref{M-fig:problems} of the main paper, coloured by displacement magnitude. Dashed
lines: undeformed boundary.}
\label{fig:deformed}
\end{figure}

\paragraph{Deformed configurations.}
Figure~\ref{fig:deformed} shows the deformed configurations of the four smooth benchmark meshes.

\paragraph{Single-penalty cost proxy.}
Table~\ref{tab:proxy} weighs the reduced term count against the larger 1-norm and
conditioning using the proxy $G=\T\kappa_{\rm eff}$ at comparable resolution. On coarse grids
the assembly-free operator is more expensive, by up to a factor of about $12$. The ratio
falls by roughly a factor of two per refinement in 2D and four to eight in 3D, because the
ratio of term counts grows as $m^{d-1}$ while that of $\kappa_{\rm eff}$ stays nearly constant.
For $\gamma=10$ the crossover lies near $m=32$ in 2D and between $m=4$ and $m=8$ in 3D, as the
computed ratios $0.75$ and $0.25$ show. For $\gamma=100$ it moves one level finer in 2D and
stays between $m=4$ and $m=8$ in 3D. These ratios use a single penalty and an assembled
operator with a Dirichlet penalty; Section~\ref{M-sec:num-cost} gives a penalty-error-controlled comparison
with the exactly constrained assembled operator.

\begin{table}[ht]
\centering
\caption{Gate-cost proxy ratio $G_{\rm b}/G_{\rm a}$ at comparable resolution
(assembly-free $m$ versus assembled $m-1$, both with $\beta=\gamma k_0$). Values below one
favour the assembly-free operator.}
\label{tab:proxy}
\small
\begin{tabular}{@{}cr|cc@{}}
\toprule
$d$ & $m$ & $\gamma=10$ & $\gamma=100$\\
\midrule
\multirow{4}{*}{2} & 4 & 8.2 & 11.7\\
 & 8 & 3.6 & 6.4\\
 & 16 & 1.63 & 3.5\\
 & 32 & 0.75 & 1.88\\
\midrule
\multirow{3}{*}{3} & 2 & 10.3 & 10.5\\
 & 4 & 1.90 & 3.02\\
 & 8 & 0.25 & 0.59\\
\bottomrule
\end{tabular}
\end{table}

\section{Exchange form: details and verification}\label{sec:shiftsupp}

\paragraph{The exchange for $m=4$.}
In direction $i$ the eight copies $j=2e_i+\delta_i$, $j=0,\dots,7$, form a chain.
Element springs join $(0,1),(2,3),(4,5),(6,7)$, while compatibility bonds join
$(1,2),(3,4),(5,6)$.  Define the permutation $V_i$ by
\[
0\mapsto0,\quad 1\leftrightarrow2,\quad3\leftrightarrow4,\quad
5\leftrightarrow6,\quad7\mapsto7.
\]
Thus $V_i$ is an involution and is Hermitian.  On every compatibility pair its restriction is
$\bigl[\begin{smallmatrix}0&1\\1&0\end{smallmatrix}\bigr]$, whereas it is the identity on the two
exterior singletons.  Consequently
\[
 I-V_i=
 \sum_{e_i=0}^{m-2}
 (\ket{e_i,1}-\ket{e_i+1,0})
 (\bra{e_i,1}-\bra{e_i+1,0}),
\]
which is exactly the one-direction compatibility matrix.  This gives the identity
$A_c^{(i)}=\alpha(I-V_i)\otimes\Pi_{\mathsf D}$, i.e.\ Eq.~\eqref{M-eq:shift} of the main paper
when $\Pi_{\mathsf D}=I$ (Q4 and bonded H8), directly, without introducing a cyclic wrap pair
and a subsequent boundary correction.

\paragraph{Implementation as signed permutations.}
Every unitary used here is stored as a signed permutation
$U\mathbf e_j=\sigma_j\mathbf e_{\pi(j)}$.  The implementation stores $(\pi,\sigma)$ as integer arrays over
the full index space, composes them in $O(N)$ for verification, merges identical unitaries by
hashing the arrays, and converts the resulting LCU to a sparse matrix for an entry-by-entry
comparison with the operator built directly from the element, spring and support energies
(Eq.~\eqref{M-eq:split} of the main paper).  This verification route is
classical and intentionally independent of the circuit interpretation of $V_i$.  Pauli strings of
the fixed element matrix enter as $X^{\mathbf x}Z^{\mathbf z}$ on the local register. A single
$\Pa Y$ factor is complex, $\Pa Y=\mathrm i\,\Pa X\Pa Z$, but for the real symmetric operators
considered here every nonzero Pauli term contains an even number of $\Pa Y$ factors, so a string
equals $\pm X^{\mathbf x}Z^{\mathbf z}$; this sign is absorbed into the coefficient, and every
unitary of the decomposition is a real signed permutation.

\paragraph{Raw counts.}
For a uniform mesh the identity contribution of every $\alpha(I-V_i)$ merges with the identity
already present in the local element expansion.  A face support is
$\beta(I-Z_D)/2$ and therefore adds one non-identity reflection.  Hence
\[
 \U=\T(K_e)+d+1,
 \qquad
 \lambda=\lambda(K_e)+2d\alpha+\beta.
\]
Table~\ref{tab:shiftraw} lists the computed decompositions and direct matrix checks.  The count is the same
for every $m\ge2$.

\begin{table}[ht]
\centering\small
\caption{Exchange-form decompositions: distinct unitaries $\U$, coefficient
1-norm $\lambda$, and largest entry deviation from the sparse operator
($\alpha=7.3$, $\beta=11.1$).}
\label{tab:shiftraw}
\begin{tabular}{@{}llrrrr@{}}
\toprule
Operator & $m$ & $N$ & $\U$ & $\lambda$ & deviation\\
\midrule
Q4 cantilever, assembly-free & 2--128 & 32--131{,}072 & 13 & 42.03 & $1.1\times10^{-16}$\\
H8 cantilever, assembly-free & 2--16 & 256--131{,}072 & 81 & 57.44 & $7.1\times10^{-15}$\\
L-bracket, assembly-free & 4--64 & 128--32{,}768 & 23 & 42.54 & $1.8\times10^{-15}$\\
L-bracket with opening & 8--64 & 512--32{,}768 & 23 & 42.54 & $1.8\times10^{-15}$\\
Q4, assembled, penalty $\beta$ & $m'=3$--$127$ & 32--32{,}768 & 109 & 16.94 & $1.8\times10^{-15}$\\
H8, assembled, penalty $\beta$ & $m'=3$--$15$ & 256--16{,}384 & 2{,}776 & 28.19 & $4.4\times10^{-15}$\\
Q4, assembled, exact & $m'=3$--$63$ & 32--8{,}192 & 196 & 6.59 & $6.7\times10^{-16}$\\
H8, assembled, exact & $m'=3,7$ & 256--2{,}048 & 10{,}072 & 13.61 & $2.0\times10^{-15}$\\
\bottomrule
\end{tabular}
\end{table}

\paragraph{Masks.}
Let $\chi(e)\in\{0,1\}$ indicate whether element $e$ is active.  Define $V_i^\chi$ to exchange
the two copies across an interface if and only if both incident elements are active, and to fix
them otherwise.  It is again an involution, and the same two-by-two argument gives the complete
active-interface penalty as $\alpha(I-V_i^\chi)$.  For the element part, the active-element
indicator is encoded by the reflection $Z_\chi=I-2P_{\rm void}$ (Section~\ref{app:proofs}).  Thus a
generic Q4 mask requires at most
\[
 2\T(K_e)+d+1=23
\]
generalized unitaries: two branches for each local element string, one masked exchange per
direction, and one support reflection, with identity terms merged.  Both the L-shaped void and
the additional dyadic opening attain 23 terms in the implementation.  The complexity of the
predicate implementing $Z_\chi$ or $V_i^\chi$ is a separate circuit-cost question and is not
identified with this branch count.

\paragraph{Assembled operator in shift form.}
On a node grid with $n=2^{a'}$ nodes per direction and elements with origins $0,\dots,n-2$,
\[
K=\sum_{s,t}\sum_{\mathbf e}\ket{\mathbf e+\boldsymbol\delta_s}\!\bra{\mathbf e+\boldsymbol\delta_t}\otimes K_e[s,t]
=\sum_{s,t}C^{\boldsymbol\delta_s}\,\Pi_{\rm o}\,C^{-\boldsymbol\delta_t}\otimes K_e[s,t],\qquad
\Pi_{\rm o}=\prod_{i=1}^d\tfrac12(I+R_{{\rm top},i}),
\]
where $s,t$ run over the local nodes, $C^{\boldsymbol\delta}$ are cyclic shifts of the node
registers, and the projector $\Pi_{\rm o}$, built from the reflections $R_{{\rm top},i}$ that mark
the last node in direction $i$, excludes the last element origin in each direction.  The Dirichlet
penalty adds $\beta(I-Z_D)/2$, with $Z_D$ the reflection of the clamped nodes.  Exact elimination of
the Dirichlet degrees of freedom with identity padding is represented by
\[
K_{\rm pad}=P_FKP_F+(I-P_F),\qquad P_F=\tfrac12(I+Z_D),
\]
where $P_F$ projects onto the free degrees of freedom: projecting each stiffness branch on both sides
turns it into four reflected branches, and the identity on the constrained subspace adds the
branches of $(I-Z_D)/2$. Without the term $I-P_F$ the projected matrix would be singular.  In 3D the padding
identity adds the four strings of $\ket{11}\!\bra{11}$ on the slot register.

\section{Penalty expansion, extrapolation and controlled cost comparison}\label{sec:penextra}
This section collects the penalty analysis summarized in Section~\ref{M-sec:extrap} of the main
paper.  Let $\varepsilon=1/\gamma$ and $\alpha=\beta=\gamma k_0$.  Then
$A(\varepsilon)=A_{\rm el}+\varepsilon^{-1}B$ with $B=\gamma^{-1}(A_{\rm c}+A_{\rm bc})$
independent of $\varepsilon$ and positive semidefinite. We assume, as in
Proposition~\ref{M-prop:penalty-bound} of the main paper, that the copies of every node are connected by springs, so
that the kernel of $B$ on the active physical sector (excluding the decoupled, unloaded void slots
of masked domains and the H8 padding) consists exactly of the conforming displacements satisfying
the essential conditions, and that the conforming stiffness $K_0$ is positive definite; both hold on the complete
Cartesian copy graphs and on the masked domains used here, but not, e.g., where two elements share
only a corner.

\begin{proposition}[Penalty expansion]\label{prop:extrap}
The solution $\mathbf u(\varepsilon)=A(\varepsilon)^{-1}\mathbf f$ is analytic at $\varepsilon=0$:
$\mathbf u(\varepsilon)=\mathbf u_0+\varepsilon\mathbf u_1+\varepsilon^2\mathbf u_2+\cdots$ for
$|\varepsilon|<\varepsilon^\ast$ (the physically relevant values are $0<\varepsilon<\varepsilon^\ast$), where $\mathbf u_0$ is the assembled solution with exact essential
conditions, written in assembly-free form (every copy carries its nodal value).
\end{proposition}

\paragraph{Proof of Proposition~\ref{prop:extrap}}
Let $P$ be the orthogonal projector onto $\ker B$ and $Q=I-P$, and write $A_{\rm el}$ and
$\mathbf f$ in blocks. $B_{QQ}$ is positive definite, and $A_{PP}$ is positive definite because
it is the assembled stiffness with essential conditions written in assembly-free form. Eliminating
$\mathbf u_Q=(A_{QQ}+\varepsilon^{-1}B_{QQ})^{-1}(\mathbf f_Q-A_{QP}\mathbf u_P)$, where
$(A_{QQ}+\varepsilon^{-1}B_{QQ})^{-1}=\varepsilon B_{QQ}^{-1}(I+\varepsilon A_{QQ}B_{QQ}^{-1})^{-1}$ is
analytic in $\varepsilon$ for $\varepsilon\|A_{QQ}B_{QQ}^{-1}\|<1$ and vanishes at
$\varepsilon=0$, gives a Schur complement $A_{PP}+O(\varepsilon)$ and a right-hand side
$\mathbf f_P+O(\varepsilon)$, both analytic. Hence $\mathbf u_P$ and $\mathbf u_Q$ are analytic,
with $\mathbf u_Q(0)=\mathbf 0$ and $A_{PP}\mathbf u_P(0)=\mathbf f_P$, which is the assembled
problem. The combination~\eqref{eq:extrap} cancels the term linear in $\varepsilon$. \qed

Richardson extrapolation \cite{Richardson1927} therefore
applies. For two penalties $\gamma_1<\gamma_2$,
\begin{equation}\label{eq:extrap}
\tilde{\mathbf u}=\frac{\gamma_2\,\mathbf u(\gamma_2)-\gamma_1\,\mathbf u(\gamma_1)}{\gamma_2-\gamma_1}
=\mathbf u_0+O\big((\gamma_1\gamma_2)^{-1}\big),
\end{equation}
and in one dimension, where the error is exactly affine in $1/\gamma$ (Sections~\ref{sec:4el} and~\ref{sec:penalty}),
$\tilde{\mathbf u}=\mathbf u_0$. On a quantum computer the combination is applied to estimated
observables, for instance a linear functional $\mathbf g^{\mathsf T}\mathbf u$, from two
independent solves. The weights $w=(-\gamma_1,\gamma_2)/(\gamma_2-\gamma_1)$ amplify
statistical errors by $\sum_k|w_k|=(\gamma_2+\gamma_1)/(\gamma_2-\gamma_1)$, which equals
$5/3$ for the pair $(\gamma_1,\gamma_2)=(2,8)$ used below. The purpose of extrapolation is to
permit small penalties, for which the conditioning is close to that of the assembled
operator, while cancelling the leading $O(1/\gamma)$ penalty error.

\paragraph{Extrapolation on the benchmarks.}
Table~\ref{tab:extrap} reports the relative error in the displacement of the loaded node
against the assembled solution with exact essential conditions on the same mesh, with
$\alpha=\beta=\gamma$ in nondimensional form, i.e.\ $\alpha=\beta=\gamma k_0$ in dimensional
variables ($k_0=Et$ for Q4, $k_0=Eh$ for H8). (With $\beta=10^8$, as in Table~\ref{tab:penalty}, the single-penalty
errors are slightly smaller.) A single penalty leaves errors proportional to $1/\gamma$. The
extrapolation~\eqref{eq:extrap} from $(\gamma_1,\gamma_2)=(2,8)$ reduces them to between
$1.2\times10^{-4}$ and $3.3\times10^{-3}$, below those of a single solve at $\gamma=100$. The H8
cantilever (unit cube, corner load, row (e)) behaves in the same
way, with extrapolated errors of $2.0\times10^{-4}$ and $1.6\times10^{-4}$ at $m=4$ and 8, well below
the two-dimensional worst case.
The small penalties substantially reduce the additional penalty-induced conditioning, although
the raw assembly-free operator is still appreciably worse conditioned than the assembled system. The
pair $(10,100)$ reaches $10^{-6}$--$10^{-4}$. The curved bracket is the most challenging benchmark; its
extrapolated error grows slowly with $m$, whereas the others decrease.

\begin{table}[ht]
\centering
\caption{Relative penalty error in the displacement of the loaded node, single penalties
and extrapolation~\eqref{eq:extrap}; $\alpha=\beta=\gamma$ (nondimensional; $\alpha=\beta=\gamma k_0$ in dimensional variables).}
\label{tab:extrap}
\footnotesize
\begin{tabular}{@{}lr|cc|cc@{}}
\toprule
Problem & $m$ & $\gamma=10$ & $\gamma=100$ & extrap.\ $(2,8)$ & extrap.\ $(10,100)$\\
\midrule
(a) Cantilever & 16 & $4.30\times10^{-2}$ & $4.30\times10^{-3}$ & $1.78\times10^{-4}$ & $3.45\times10^{-6}$\\
 & 64 & $4.33\times10^{-2}$ & $4.33\times10^{-3}$ & $1.23\times10^{-4}$ & $2.35\times10^{-6}$\\
(b) L-bracket & 16 & $4.80\times10^{-2}$ & $4.80\times10^{-3}$ & $2.98\times10^{-4}$ & $5.96\times10^{-6}$\\
 & 64 & $4.65\times10^{-2}$ & $4.65\times10^{-3}$ & $1.21\times10^{-4}$ & $2.38\times10^{-6}$\\
(c) Curved bracket & 16 & $8.88\times10^{-2}$ & $8.93\times10^{-3}$ & $2.48\times10^{-3}$ & $5.18\times10^{-5}$\\
 & 64 & $9.61\times10^{-2}$ & $9.67\times10^{-3}$ & $3.29\times10^{-3}$ & $6.95\times10^{-5}$\\
(d) Distorted plate & 16 & $3.83\times10^{-2}$ & $3.84\times10^{-3}$ & $4.35\times10^{-4}$ & $9.35\times10^{-6}$\\
 & 64 & $3.70\times10^{-2}$ & $3.71\times10^{-3}$ & $3.11\times10^{-4}$ & $6.46\times10^{-6}$\\
\midrule
(e) H8 cantilever & 4 & $1.32\times10^{-2}$ & $1.33\times10^{-3}$ & $2.02\times10^{-4}$ & $3.62\times10^{-6}$\\
 & 8 & $1.22\times10^{-2}$ & $1.22\times10^{-3}$ & $1.61\times10^{-4}$ & $2.82\times10^{-6}$\\
\bottomrule
\end{tabular}
\end{table}

\phantomsection\label{sec:extrapsupp}
\paragraph{One dimension.}
From the tip error $P[(n-1)/\alpha+1/\beta]$ of Section~\ref{sec:penalty}, with
$\alpha=\beta=\gamma k$, the tip displacement is $u(\gamma)=nP/k+nP/(\gamma k)$, which is affine in
$1/\gamma$. The two-point combination~\eqref{eq:extrap} therefore returns the exact
value $nP/k$ for any $\gamma_1\neq\gamma_2$. In several dimensions higher powers of $1/\gamma$
appear, and the combination removes only the leading one.

\paragraph{Weights and amplification.}
For penalties $\gamma_1<\dots<\gamma_s$, the extrapolated value is $\sum_kw_ku(\gamma_k)$, where
$w$ solves the Vandermonde system that reproduces polynomials of degree $s-1$ in $1/\gamma$ and
returns their value at $0$. For two points $w=(-\gamma_1,\gamma_2)/(\gamma_2-\gamma_1)$; for
$(2,4,8)$, $w=(\frac13,-2,\frac83)$ with $\sum|w_k|=5$. If each $o_k$ is estimated with error
$\epsilon_k$ at cost $G_k/\epsilon_k$, then minimizing $\sum_kG_k/\epsilon_k$ subject to
$\sum_k|w_k|\epsilon_k=\epsilon$ gives $\epsilon_k\propto\sqrt{G_k/|w_k|}$ and the total cost
$\big(\sum_k\sqrt{|w_k|G_k}\big)^2/\epsilon$ used for the cost ratios below.

\paragraph{Choice of scheme.}
Table~\ref{tab:schemes} compares schemes by their worst penalty error over the four
two-dimensional benchmark problems of the main paper at $m=8,16,32,64$, and by the cost ratio $R$ of
Section~\ref{M-sec:num-cost} of the main paper. The pair $(2,8)$ gives the lowest cost among schemes with worst error below
$10^{-2}$; $(4,16)$ is the least expensive with worst error below $10^{-3}$. Three-point schemes are more
accurate but their larger weights make them more expensive. For the H8 cantilever at $m=4$ and 8,
whose cost ratios are listed in the last two columns, the largest errors of each scheme (third
column) are all below the corresponding two-dimensional worst errors, so the stated worst errors
also bound the three-dimensional cases.

\begin{table}[ht]
\centering\small
\caption{Extrapolation schemes: worst relative penalty error (loaded-node displacement, four
two-dimensional benchmarks, $m=8$--$64$), worst error of the H8 cantilever ($m=4$, 8), amplification $\sum|w_k|$, and cost ratios $R$ (Pauli\,/\,exchange form)
against the exactly constrained assembled operator.}
\label{tab:schemes}
\setlength{\tabcolsep}{3.5pt}
\begin{tabular}{@{}lcccccccc@{}}
\toprule
Scheme & worst error & H8, $m\le8$ & $\sum|w_k|$ & 2D, $m=16$ & 2D, $m=32$ & 2D, $m=64$ & 3D, $m=4$ & 3D, $m=8$\\
\midrule
$(2,4)$ & $6.1\times10^{-3}$ & $3.9\times10^{-4}$ & 3.00 & 2.55\,/\,4.50 & 1.05\,/\,4.58 & 0.46\,/\,4.62 & 4.87\,/\,0.47 & 0.37\,/\,0.59\\
$(2,8)$ & $3.3\times10^{-3}$ & $2.0\times10^{-4}$ & 1.67 & 2.19\,/\,3.77 & 0.91\,/\,3.84 & 0.40\,/\,3.88 & 4.21\,/\,0.40 & 0.32\,/\,0.50\\
$(4,16)$ & $9.7\times10^{-4}$ & $5.4\times10^{-5}$ & 1.67 & 4.12\,/\,6.96 & 1.71\,/\,7.11 & 0.75\,/\,7.18 & 7.98\,/\,0.75 & 0.62\,/\,0.94\\
$(8,32)$ & $2.6\times10^{-4}$ & $1.4\times10^{-5}$ & 1.67 & 7.99\,/\,13.37 & 3.33\,/\,13.66 & 1.46\,/\,13.81 & 15.5\,/\,1.46 & 1.22\,/\,1.83\\
$(2,4,8)$ & $4.3\times10^{-4}$ & $1.1\times10^{-5}$ & 5.00 & 8.99\,/\,15.49 & 3.72\,/\,15.79 & 1.63\,/\,15.94 & 17.3\,/\,1.65 & 1.33\,/\,2.06\\
$(4,8,16)$ & $7.0\times10^{-5}$ & $1.5\times10^{-6}$ & 5.00 & 16.9\,/\,28.57 & 7.03\,/\,29.18 & 3.09\,/\,29.49 & 32.8\,/\,3.09 & 2.55\,/\,3.87\\
$(10,100)$ & $6.9\times10^{-5}$ & $3.6\times10^{-6}$ & 1.22 & -- & -- & -- & -- & --\\
\bottomrule
\end{tabular}
\end{table}

\paragraph{Penalty-error-controlled cost proxy.}
We now compare with the assembled operator with exactly eliminated Dirichlet degrees of
freedom, which has no penalty error.  We use the LCU-size-weighted proxy
$G=\T\kappa_{\rm eff}$ in Pauli form and $G=\U\kappa_{\rm eff}$ in exchange form, where $\U$ is
the number of unitaries and $\kappa_{\rm eff}=\lambda\|A^{-1}\|_2$.  This is not
asserted to be a universal gate complexity: structured SELECT/PREPARE circuits can have costs
that differ substantially from the number of terms. The comparison controls the penalty error but
not the discretization, inversion and readout errors, and it is not an end-to-end runtime
prediction; ``matched'' below means a penalty error bounded by the stated worst case, not equal
total errors. For the extrapolated
estimate $\sum_kw_ko_k$, allocating the precision optimally among the solves gives the cost
$\big(\sum_k\sqrt{|w_k|G_k}\big)^2/\epsilon$. We report the ratio $R$ of this cost to that of the
exactly constrained assembled operator at comparable resolution ($m'=m-1$), for the pair
$(2,8)$, whose worst penalty error over all benchmarks is $3.3\times10^{-3}$
(Table~\ref{tab:costmatched}).

In Pauli form the assembly-free operator has a smaller value of the cost proxy from $m=32$ in 2D ($R=0.91$, then $0.40$
at $m=64$) and from $m=8$ in 3D ($R=0.32$), and $R$ falls by factors of about $2.3$ (2D) and
$13$ (3D) per refinement.  In exchange form both formulations have mesh-independent numbers
of unitaries, so the raw proxy ratio is nearly constant: $R=3.6$--$3.9$ in 2D, where the exactly
constrained assembled operator retains the smaller proxy value because its smaller 1-norm and
larger smallest eigenvalue outweigh its $15$ times larger number of unitaries, and $R=0.40$--$0.50$ in
3D, where the assembly-free operator has the smaller proxy value.  These raw comparisons deliberately precede the BPX and sum/difference transformation; after robust
preconditioning the relevant solver parameter is instead that of the factor $\mathcal H$ in
Eq.~\eqref{M-eq:factor-qsvt}. With the
tighter pair $(4,16)$, whose worst error is $9.7\times10^{-4}$, the Pauli ratios at the finest
grids rise to $0.75$ (2D, $m=64$) and $0.62$ (3D, $m=8$) but remain below one, and the 3D
exchange-form ratios rise to $0.75$--$0.94$.

\begin{table}[ht]
\centering
\caption{Penalty-error-controlled cost proxy: ratio $R$ of the extrapolated assembly-free cost
(penalties $2$ and $8$) to that of the exactly constrained assembled operator ($m'=m-1$).
$\T$, $\U$: Pauli strings and unitaries, assembly-free\,/\,assembled. $R<1$ favours the
assembly-free operator.}
\label{tab:costmatched}
\footnotesize
\begin{tabular}{@{}cr|rr|rr@{}}
\toprule
& & \multicolumn{2}{c|}{Pauli form} & \multicolumn{2}{c}{Exchange form}\\
$d$ & $m$ & $\T$ & $R$ & $\U$ & $R$\\
\midrule
\multirow{4}{*}{2} & 8 & $60\,/\,492$ & $5.96$ & $13\,/\,196$ & $3.62$\\
 & 16 & $116\,/\,2{,}164$ & $2.19$ & $13\,/\,196$ & $3.77$\\
 & 32 & $228\,/\,9{,}060$ & $0.91$ & $13\,/\,196$ & $3.84$\\
 & 64 & $452\,/\,37{,}060$ & $0.40$ & $13\,/\,196$ & $3.88$\\
\midrule
\multirow{2}{*}{3} & 4 & $108\,/\,1{,}936$ & $4.21$ & $81\,/\,10{,}072$ & $0.40$\\
 & 8 & $148\,/\,21{,}172$ & $0.32$ & $81\,/\,10{,}072$ & $0.50$\\
\bottomrule
\end{tabular}
\end{table}

\paragraph{L-shaped bracket with an opening: mesh convergence.}
Fig.~\ref{M-fig:lhole-conv}(a) and Table~\ref{tab:lhole} give the operator sizes. At $m=8$ the
opening is a single element and the assembly-free Pauli count ($1{,}003$) exceeds the assembled
one ($458$). From $m=16$ the assembly-free count is smaller, by a factor of $10.3$ at $m=64$
($5{,}861$ against $60{,}658$), and it grows by about $1.9$ per refinement against $4.3$ for the
assembled operator. The exchange form requires $23$ unitaries at every resolution.

Fig.~\ref{M-fig:lhole-conv}(b) shows why extrapolation matters on fine meshes. The
discretization error falls from $12\%$ at $m=8$ to $0.13\%$ at $m=256$, at a rate limited by
the re-entrant corner. A single penalty $\gamma=10$ leaves a penalty error near $5\%$ that does
not decrease: it is comparable to the discretization error at $m=16$ and exceeds it from
$m=32$ on, so that a mesh-convergence study would stall. With the extrapolation from $(2,8)$ the penalty error falls from
$7.2\times10^{-4}$ to $5.7\times10^{-5}$ and stays $24$--$170$ times below the discretization
error at every resolution; with $(10,100)$ it lies between $1.2\times10^{-6}$ and $1.4\times10^{-5}$. The assembly-free solution thus
reproduces the convergence of the assembled one.

\begin{table}[ht]
\centering
\caption{L-shaped bracket with an opening. Pauli counts (assembly-free $m$, assembled
$m'=m-1$) and relative errors in the compliance: discretization error of the assembled
solution against $m=512$, and penalty errors against the assembled solution on the same
mesh.}
\label{tab:lhole}
\footnotesize
\setlength{\tabcolsep}{4pt}
\begin{tabular}{@{}r|rr|cccc@{}}
\toprule
$m$ & $\T$ free & $\T$ assembled & discretization & $\gamma=10$ & extrap.\ $(2,8)$ & extrap.\ $(10,100)$\\
\midrule
8 & $1{,}003$ & $458$ & $1.23\times10^{-1}$ & $5.65\times10^{-2}$ & $7.17\times10^{-4}$ & $1.43\times10^{-5}$\\
16 & $1{,}697$ & $3{,}018$ & $5.83\times10^{-2}$ & $5.31\times10^{-2}$ & $4.02\times10^{-4}$ & $8.14\times10^{-6}$\\
32 & $3{,}085$ & $14{,}118$ & $2.56\times10^{-2}$ & $5.11\times10^{-2}$ & $2.24\times10^{-4}$ & $4.58\times10^{-6}$\\
64 & $5{,}861$ & $60{,}658$ & $1.08\times10^{-2}$ & $5.00\times10^{-2}$ & $1.30\times10^{-4}$ & $2.66\times10^{-6}$\\
128 & -- & -- & $4.28\times10^{-3}$ & $4.94\times10^{-2}$ & $8.09\times10^{-5}$ & $1.65\times10^{-6}$\\
256 & -- & -- & $1.34\times10^{-3}$ & $4.91\times10^{-2}$ & $5.69\times10^{-5}$ & $1.16\times10^{-6}$\\
\bottomrule
\end{tabular}
\end{table}

\section{Cost model: raw data}\label{sec:costsupp}

Table~\ref{tab:costraw} lists the quantities behind Table~\ref{tab:costmatched}: term counts,
1-norms and smallest eigenvalues of the exactly constrained assembled operators and of the
assembly-free operators at the two penalties of the scheme $(2,8)$, with $\alpha=\beta=\gamma$
(nondimensional; $\alpha=\beta=\gamma k_0$ in dimensional variables).
Smallest eigenvalues were computed by shift-invert Lanczos iterations. For comparison only, single solves at $\gamma=2$ without extrapolation give cost ratios of $0.06$
(2D, $m=64$, Pauli) and $0.04$--$0.11$ (3D, $m=8$); the penalty errors of these individual
low-penalty solves are between $18\%$ and $47\%$, whereas the extrapolated errors are those of
Table~\ref{tab:extrap}.

\begin{table}[ht]
\centering\small
\caption{Raw data of the penalty-error-controlled cost comparison. $\lambda_{\rm P}$, $\lambda_{\rm S}$:
1-norms of the Pauli and exchange forms; $\mu$: smallest eigenvalue.}
\label{tab:costraw}
\setlength{\tabcolsep}{3.5pt}
\begin{tabular}{@{}cr|rrrrr|rrrr|rr@{}}
\toprule
& & \multicolumn{5}{c|}{Assembled, exact ($m'=m-1$)} & \multicolumn{4}{c|}{Assembly-free, $\gamma=2$} & \multicolumn{2}{c}{$\gamma=8$}\\
$d$ & $m$ & $\T$ & $\lambda_{\rm P}$ & $\U$ & $\lambda_{\rm S}$ & $\mu$ & $\T$ & $\lambda_{\rm P}$ & $\lambda_{\rm S}$ & $\mu$ & $\lambda_{\rm P}$ & $\mu$\\
\midrule
2 & 8 & 492 & 13.73 & 196 & 6.59 & $6.24\times10^{-3}$ & 60 & 20.98 & 11.73 & $1.35\times10^{-3}$ & 78.73 & $1.60\times10^{-3}$\\
2 & 16 & 2{,}164 & 20.85 & 196 & 6.59 & $1.62\times10^{-3}$ & 116 & 25.36 & 11.73 & $3.41\times10^{-4}$ & 96.23 & $4.00\times10^{-4}$\\
2 & 32 & 9{,}060 & 28.23 & 196 & 6.59 & $4.14\times10^{-4}$ & 228 & 29.54 & 11.73 & $8.56\times10^{-5}$ & 112.98 & $9.99\times10^{-5}$\\
2 & 64 & 37{,}060 & 36.05 & 196 & 6.59 & $1.05\times10^{-4}$ & 452 & 33.64 & 11.73 & $2.15\times10^{-5}$ & 129.36 & $2.50\times10^{-5}$\\
3 & 4 & 1{,}936 & 13.52 & 10{,}072 & 13.61 & $1.84\times10^{-2}$ & 108 & 24.04 & 16.54 & $3.11\times10^{-3}$ & 88.54 & $3.46\times10^{-3}$\\
3 & 8 & 21{,}172 & 36.18 & 10{,}072 & 13.61 & $5.67\times10^{-3}$ & 148 & 31.29 & 16.54 & $7.75\times10^{-4}$ & 117.54 & $8.54\times10^{-4}$\\
\bottomrule
\end{tabular}
\end{table}

The exchange-form 1-norm at $\gamma=8$ is $41.73$ (2D) and $58.54$ (3D) for every $m$, and the number of
generalized unitaries is $13$ (2D) and $81$ (3D).

\section{L-shaped bracket with an opening}\label{sec:lholesupp}

\paragraph{Set-up.}
The background grid has $m\times m$ square elements of size $h=1/m$, $m=2^a\ge8$. An element
$(e_1,e_2)$ is void if $e_1,e_2\ge m/2$ (upper-right quadrant) or if
$5m/8\le e_1<6m/8$ and $m/8\le e_2<2m/8$ (opening). Copies of the nodes on $x_1=0$ are clamped. A
uniform traction $q=-2$ in $x_2$, of total magnitude one, acts on the edge $x_1=1$,
$0\le x_2\le\frac12$; each element on that edge receives the consistent nodal loads $qh/2$ at its
two nodes with $\delta_1=1$. In the assembly-free formulation these element loads are kept on
the element's own copies, so no assembly of the load vector is needed either. For the Pauli
counts of the assembled operator the mask is evaluated on the background grid of $m=m'+1$ cells per direction,
while the assembled mesh has $m'=m-1$ elements per direction, as for the L-shaped bracket of Section~\ref{sec:lshape}; nodes that touch no
active element receive a unit diagonal.

\paragraph{Raw data.}
The X-pattern counts are $14,16,18,20$ (assembly-free, $m=8,\dots,64$) and $32,50,72,98$
(assembled). The assembled compliances on the same mesh are $17.9281$, $19.2455$, $19.9139$,
$20.2159$, $20.3500$ and $20.4102$ for $m=8,\dots,256$, and $20.4375$ for the reference $m=512$.
The successive reductions of the discretization error, $2.1$, $2.3$, $2.4$ and $2.5$ per halving
of $h$ up to $m=128$, reflect the re-entrant corner. The penalty errors of
Table~\ref{tab:lhole} are measured against the assembled solution on the same mesh, so they
isolate the effect of the penalty. Deformed shapes were obtained from the extrapolated assembly-free
solution by keeping each element's own copies.

\section{Preconditioning raw data in two and three dimensions}\label{sec:bpxsupp}
Table~\ref{tab:bpxraw} lists the extreme eigenvalues behind
Table~\ref{M-tab:bpx-q4h8} of the main paper ($E=1$, $\nu=0.3$, $\alpha=\beta=10$).  All operators are in the nondimensional form of
Section~\ref{M-sec:uniform-preconditioning} of the main paper: stiffnesses are divided by the stiffness
scale $k_0=Eh^{d-2}$ of the finest elements and lengths are measured in units of the finest element
size. The H8 data therefore use bonded padding and unit-size finest elements; since the H8 element
matrix scales with $Eh=k_0$, this is the same as refining a fixed domain, not enlarging it. In these
units the level-$\ell$ BPX frame weight is $2^{(\ell-L)/2}$, which is the weight
$(h_\ell/h)^{(2-d)/2}$ of the main paper, and for H8 $\bar k=\|K_e^{\rm phys}\|_2=1.2500$ and
$\eta=0.1250$. The Q4 conforming and assembly-free entries for $m\ge128$, including the results
for $m=512$, were computed with an independent Lanczos implementation 
that reproduces all Q4 entries for $m\le64$ to the printed digits.

\begin{table}[ht]
\centering\small
\caption{Smallest positive and largest eigenvalues of the BPX-preconditioned scalar,
conforming elasticity and complete assembly-free operators.}
\label{tab:bpxraw}
\begin{tabular}{@{}lr|rr|rr|rr@{}}
\toprule
& & \multicolumn{2}{c|}{scalar Q1} & \multicolumn{2}{c|}{conforming} & \multicolumn{2}{c}{assembly-free}\\
element & $m$ & $\lambda_{\min}^+$ & $\lambda_{\max}$ & $\lambda_{\min}^+$ & $\lambda_{\max}$ & $\lambda_{\min}^+$ & $\lambda_{\max}$\\
\midrule
Q4 & 2 & 0.6937 & 2.908 & 0.1090 & 2.654 & 0.1020 & 2.669\\
Q4 & 4 & 0.7747 & 4.832 & 0.1115 & 4.578 & 0.1048 & 4.608\\
Q4 & 8 & 0.8036 & 6.261 & 0.1128 & 6.272 & 0.1065 & 6.310\\
Q4 & 16 & 0.8128 & 7.371 & 0.1135 & 7.589 & 0.1074 & 7.632\\
Q4 & 32 & 0.8155 & 8.267 & 0.1138 & 8.669 & 0.1079 & 8.716\\
Q4 & 64 & 0.8163 & 9.005 & 0.1139 & 9.557 & 0.1081 & 9.607\\
Q4 & 128 & 0.8165 & 9.622 & 0.1140 & 10.30 & 0.1082 & 10.35\\
Q4 & 256 & 0.8166 & 10.14 & 0.1140 & 10.92 & 0.1082 & 10.97\\
Q4 & 512 & -- & -- & 0.1140 & 11.44 & 0.1082 & 11.49\\
H8 & 2 & 0.3216 & 2.774 & 0.06214 & 2.774 & 0.06117 & 2.797\\
H8 & 4 & 0.3783 & 4.845 & 0.0695 & 5.072 & 0.06727 & 5.096\\
H8 & 8 & 0.3843 & 6.279 & 0.07188 & 7.234 & 0.06974 & 7.263\\
H8 & 16 & 0.3847 & 7.367 & 0.07294 & 8.892 & 0.07089 & 8.926\\
H8 & 32 & 0.3848 & 8.241 & 0.07333 & 10.26 & 0.07134 & 10.30\\
\bottomrule
\end{tabular}
\end{table}

\paragraph{Normalization diagnostics of the energy factor.}
The Gram identity~\eqref{M-eq:Hhat-gram} was verified numerically with explicitly formed matrices;
for all such cases ($m\le8$) it is reproduced to below $4\times10^{-14}$. This verifies the
algebraic factorization, not an encoded circuit. Two normalization diagnostics were
also evaluated
(Table~\ref{tab:direct-q4}). The first, $\Gamma_{\rm ideal}$, uses the numerically evaluated spectral norm of each
level energy block; it is the ideal norm-based normalization for the chosen block construction, not
necessarily the minimum over all block encodings of $\mathcal H$. The second, $\Gamma_{\rm explicit}$, uses the Cartesian bound~\eqref{eq:level-gamma-bound}.
For $E=1$, $\nu=0.3$ and $\alpha=\beta=10$, $\eta=1/[10(1-\nu)]\simeq0.143$.

\begin{table}[ht]
\centering
\caption{Normalization diagnostics of the Q4 preconditioned energy factor ($E=1$, $\nu=0.3$,
$\alpha=\beta=10$). $\Gamma=\omega(\mathcal H)^2$ from the numerically evaluated level-block norms (ideal) and from
the explicit bound~\eqref{eq:level-gamma-bound}. Last column: explicit normalization
$\sqrt{\Gamma_{\rm explicit}}$ divided by the computed $\|\mathcal H\|_2$.}
\label{tab:direct-q4}
\small
\begin{tabular}{@{}rrrrrr@{}}
\toprule
$m$ & $L$ & $\kappa$ & $\Gamma_{\rm ideal}$ &
$\Gamma_{\rm explicit}$ & $\sqrt{\Gamma_{\rm explicit}}/\|\mathcal H\|_2$\\
\midrule
2  &1&26.2&4.65 &24.0&3.00\\
4  &2&44.0&8.37 &35.4&2.77\\
8  &3&59.3&12.55&46.9&2.73\\
16 &4&71.0&16.88&58.3&2.76\\
32 &5&80.8&21.26&69.7&2.83\\
64 &6&88.9&25.66&81.1&2.91\\
\bottomrule
\end{tabular}
\end{table}

Here $\|\mathcal H\|_2=\sqrt{\lambda_{\max}(F^{\mathsf T}AF)}$, so the last column compares the
proved normalization with the actual norm of the factor; it is an empirical diagnostic, not a
circuit bound. It stays between 2.7 and 3.0: the growth of the explicit normalization with the
number of levels follows the growth of $\|\mathcal H\|_2$ itself. The corresponding ratio for
$\Gamma_{\rm ideal}$ lies between $1.32$ and $1.63$.

These finite-level values validate the implemented frame; the refinement-independent spectral
bound comes from Theorem~\ref{M-thm:full-pre-final}, not from the table. Together with the Gram
identity~\eqref{M-eq:Hhat-gram}, Theorem~\ref{M-thm:direct-precond-block} shows that the
preconditioner enters the quantum solver without the extra condition-number factor of a separately
encoded preconditioner.

\paragraph{Penalty and Poisson's-ratio sweeps.}
Table~\ref{tab:sweeps} gives the data behind the two sweeps of Section~\ref{M-sec:numerics} of the
main paper (Q4 cantilever, $E=1$, $\alpha=\beta=\gamma$, combined sum/difference scaling and BPX
transformation). The condition number decreases monotonically as the springs are made stiffer,
with both extreme eigenvalues converging, consistent with the decreasing theoretical upper bound of
Theorem~\ref{M-thm:full-pre-final}; it
increases mildly with $\nu$ in plane stress, through the largest eigenvalue.

\begin{table}[ht]
\centering\small
\caption{Condition numbers and extreme positive eigenvalues of the preconditioned Q4 operator.
Left: penalty sweep at $m=32$, $\nu=0.3$, with $\eta=\bar k/p_{\min}=1.4286/\gamma$. Right:
Poisson's-ratio sweep at $m=16$, $\gamma=10$ (plane stress).}
\label{tab:sweeps}
\begin{tabular}{@{}rrrrr@{\qquad}rrrr@{}}
\toprule
\multicolumn{5}{c}{penalty sweep ($m=32$, $\nu=0.3$)} & \multicolumn{4}{c}{Poisson sweep ($m=16$, $\gamma=10$)}\\
$\gamma$ & $\eta$ & $\kappa$ & $\lambda_{\min}^+$ & $\lambda_{\max}$ & $\nu$ & $\kappa$ & $\lambda_{\min}^+$ & $\lambda_{\max}$\\
\midrule
2    & 0.714   & 99.30 & 0.0897 & 8.906 & 0     & 61.59 & 0.1134 & 6.983\\
5    & 0.286   & 85.40 & 0.1026 & 8.764 & 0.1   & 63.31 & 0.1112 & 7.039\\
10   & 0.143   & 80.79 & 0.1079 & 8.716 & 0.2   & 66.35 & 0.1092 & 7.246\\
20   & 0.0714  & 78.48 & 0.1108 & 8.693 & 0.3   & 71.04 & 0.1074 & 7.632\\
50   & 0.0286  & 77.09 & 0.1126 & 8.679 & 0.4   & 78.02 & 0.1059 & 8.259\\
100  & 0.0143  & 76.63 & 0.1132 & 8.674 & 0.45  & 82.70 & 0.1051 & 8.695\\
1000 & 0.00143 & 76.21 & 0.1138 & 8.670 & 0.49  & 87.22 & 0.1046 & 9.122\\
     &         &       &        &       & 0.499 & 88.35 & 0.1045 & 9.229\\
\bottomrule
\end{tabular}
\end{table}

\section{Block encoding of the preconditioned factor and transformed loads}\label{sec:circuitsupp}
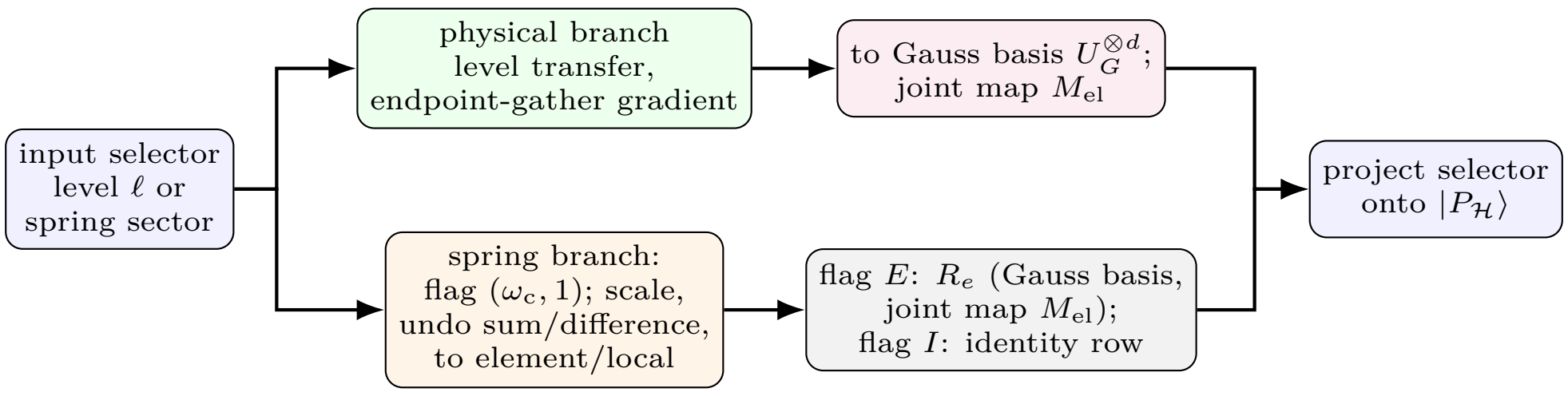
\begin{figure}[ht]
\centering
\begin{tikzpicture}[>=Latex,
  box/.style={draw,rounded corners,minimum height=8mm,align=center,font=\scriptsize,inner sep=3pt},
  line/.style={-Latex,thick}]
\node[box,fill=blue!6] (prep) at (0,0) {input selector\\level $\ell$ or\\spring sector};
\node[box,fill=green!7] (bpx) at (3.6,1.0) {physical branch\\level transfer,\\endpoint-gather gradient};
\node[box,fill=orange!9] (walsh) at (3.6,-1.0) {spring branch:\\flag $(\omega_{\rm c},1)$; scale,\\undo sum/difference,\\to element/local};
\node[box,fill=purple!7] (mat) at (7.3,1.0) {to Gauss basis $U_G^{\otimes d}$;\\joint map $M_{\rm el}$};
\node[box,fill=gray!10] (id) at (7.3,-1.0) {flag $E$: $R_e$ (Gauss basis,\\joint map $M_{\rm el}$);\\flag $I$: identity row};
\node[box,fill=blue!6] (unprep) at (10.9,0) {project selector\\onto $\ket{P_{\mathcal H}}$};
\draw[line] (prep.east) -- ++(0.35,0) |- (bpx.west);
\draw[line] (prep.east) -- ++(0.35,0) |- (walsh.west);
\draw[line] (bpx) -- (mat);
\draw[line] (walsh) -- (id);
\draw[line] (mat.east) -| ($(unprep.west)+(-0.45,0)$) -- (unprep.west);
\draw[line] (id.east) -| ($(unprep.west)+(-0.45,0)$) -- (unprep.west);
\end{tikzpicture}
\caption{Logical structure of one block-encoding call of the rectangular preconditioned factor
$\mathcal H$ (Section~\ref{app:normalization}). The selector is part of the input; the spring branch
prepares an output flag with amplitudes proportional to $\omega_{\rm c}$ and 1; projecting the
selector onto $\ket{P_{\mathcal H}}$ of Eq.~\eqref{eq:factor-prep-state} yields $\mathcal
H/\sqrt\Gamma$. The physical branch uses the multilevel transfer and gradient circuits; the
assembly-free overhead is confined to the node-local spring-scaling branch. Conceptual structure,
not a compiled circuit.}
\label{fig:H-circuit}
\end{figure}

This section gives the derivations behind the block encoding summarized in
Section~\ref{M-sec:circuit-resources} of the main paper: the selection circuit of the preconditioned
factor (Fig.~\ref{fig:H-circuit}), its two fixed sub-circuits, the register sizes and the
validation of the transformed point load.

\paragraph{Selection and output projection.}
Write the rectangular factor of Section~\ref{M-sec:direct-block-precond} as
\begin{equation}\label{eq:H-expanded-circuit}
 \mathcal H=
 \begin{bmatrix}
 H_0&H_1&\cdots&H_L&H_{\rm c}\\
 0&0&\cdots&0&I_{\rm c}
 \end{bmatrix},
 \qquad
 H_\ell=RF_\ell,\quad H_{\rm c}=RF_{\rm c}.
\end{equation}
For plane-stress Q4 the level maps may all be normalized by the same constant
\begin{equation}
 \omega_{\rm p}=2\sqrt{\frac{d}{1-\nu}},\qquad d=2,
\end{equation}
and the spring energy map by $\omega_{\rm c}\le\sqrt{\eta}$, with
$\eta=\|K_e\|_2/p_{\min}$ and $p_{\min}=\min\{\beta,2\alpha\}$. Following
Section~\ref{app:normalization}, the input register carries the selector $(\mathrm p,\ell)$ or
$\mathrm c$ together with the coefficient vector of that block, and no preparation is applied on
the input side. After the selection stages below, the selector is projected onto
\begin{equation}\label{eq:factor-prep-state}
 \ket{P_{\mathcal H}}=
 \frac{1}{\sqrt{\Gamma}}
 \left[
 \omega_{\rm p}\sum_{\ell=0}^{L}\ket{\mathrm p,\ell}
 +\Omega_{\rm c}\ket{\mathrm c}
 \right],
 \qquad
 \Omega_{\rm c}^2=\omega_{\rm c}^2+1,\qquad
 \Gamma=(L+1)\omega_{\rm p}^2+\omega_{\rm c}^2+1,
\end{equation}
by the inverse of a unitary that prepares this state, and the result is $\mathcal H/\sqrt\Gamma$.
With $r=\lceil\log_2(L+1)\rceil$, a contiguous uniform superposition over the valid level codes
requires at most $r$ nontrivial one-qubit rotation angles and one further angle separates the
physical and spring sectors. The selection is factored into the following stages.
\begin{enumerate}
\item \textbf{Physical BPX branch.}  Conditional on $(\mathrm p,\ell)$, apply the level
transfer $T_{\ell,L}$ and the quadrature-weighted Q1 gradient $C_\ell$, using the overlapping
level-transfer/gradient network of Deiml--Peterseim \cite{Deiml2026}; the displacement
component is an additional spectator/control qubit. The orthonormal discontinuous-Q1 gradient
coefficients are converted to quadrature-weighted Gauss-point values by the fixed orthogonal map
$U_G^{\otimes d}$ of Eq.~\eqref{eq:gauss-basis}, so that both branches use the common output basis
of Section~\ref{app:normalization}, and the joint strain/material map $M_{\rm el}$ then converts
the four displacement-gradient channels into the three weighted plane-stress strains.
\item \textbf{Spring branch.}  Prepare an output flag in $(\omega_{\rm c}\ket{E}+\ket{I})/\Omega_{\rm
c}$. If the flag is $E$, compute the copy-mode penalty eigenvalue,
block-encode $D_c$, i.e.\ the factor $\lambda(n,z)^{-1/2}$ of Eq.~\eqref{eq:lambda-circuit}, by a
single multiplexed rotation, perform the inverse local sum/difference transformation and map
physical-node/copy coordinates to element/local coordinates (with the boundary copy bits and
validity projectors described below). The result is a vector of element-local nodal
displacements, not of gradients, so the branch then applies the complete element energy factor
$R_e$ of Section~\ref{app:normalization}, $R_e^{\mathsf T}R_e=K_e$, whose rows are
$\sqrt{w^g|J^g|}\,(D/E)^{1/2}B^g$ for the quadrature points $g$: the quadrature-weighted
displacement gradient followed by the same joint strain/material map, in the same Gauss-point
output basis as the physical branch. If the flag is $I$, route the spring-mode coefficient
unchanged to the lower identity row of~\eqref{eq:H-expanded-circuit}. The flag is part of the
output and labels the two row blocks; in the physical branch it remains $\ket E$.
\end{enumerate}
This is a factor-selection circuit, not the
13-term LCU of the unpreconditioned stiffness: the input selector has $L+2$ branches ($L+1$ BPX
levels and the spring sector), and the spring sector has two output row blocks labelled by the
flag, while the expensive mesh dependence remains confined to reversible coordinate
arithmetic and the BPX transfer network.

\paragraph{Structure of one block-encoding call.}
Figure~\ref{fig:H-circuit} summarizes the stages; it is the diagram referred to in
Section~\ref{M-sec:direct-block-precond} of the main paper.

\paragraph{Exact sum/difference scaling circuit.}
For Q4 encode a spring-mode coefficient by a physical node
$(n_x,n_y)$, displacement component $s$, and copy-mode bits $(z_x,z_y)$.  Let
\begin{equation}
 a_i=\mathbf 1_{\{0<n_i<m\}},\qquad
 b=\mathbf 1_{\{n_x=0\}},\qquad
 w=a_xz_x+a_yz_y.
\end{equation}
The penalty eigenvalue is exactly
\begin{equation}\label{eq:lambda-circuit}
 \lambda(n,z)=\beta b+2\alpha w.
\end{equation}
Codes with $\lambda=0$ are the free compatible modes and belong to the BPX branch; the
spring branch contains precisely the codes with $\lambda>0$.  With
$p_{\min}=\min\{\beta,2\alpha\}$, one ancilla rotation
\begin{equation}
 \ket{0}\mapsto
 \sqrt{\frac{p_{\min}}{\lambda(n,z)}}\ket{0}
 +\sqrt{1-\frac{p_{\min}}{\lambda(n,z)}}\ket{1}
\end{equation}
block-encodes $D_c$ with normalization $p_{\min}^{-1/2}$.  With one clamped face, Q4 has at most
four nonzero penalty classes,
\begin{equation}
 \beta,\quad 2\alpha,\quad 4\alpha,\quad \beta+2\alpha,
\end{equation}
so at most three classes require a nontrivial synthesized angle because one class equals
$p_{\min}$.  The mode bits are then converted to copy bits $c_i$ by a Hadamard on $z_i$ iff
$a_i=1$, i.e.\ only in interior directions. On a boundary direction ($a_i=0$) the node has a single
copy, the mode bit is unused and must be $z_i=0$, and the copy bit is fixed by the position of the
node:
\begin{center}
\begin{tabular}{@{}lccc@{}}
\toprule
node coordinate & $n_i=0$ & $0<n_i<m$ & $n_i=m$\\
\midrule
valid copy bit $c_i$ & 0 & 0 or 1 & 1\\
\bottomrule
\end{tabular}
\end{center}
It is set by $c_i=z_i\oplus\mathbf 1_{\{n_i=m\}}$ when $a_i=0$. Codes with $z_i=1$ on a boundary
direction, or with $\lambda(n,z)=0$ in the spring branch, are invalid and are removed by validity
projectors (flag qubits computed from the endpoint tests and projected onto their valid value).
Finally
\begin{equation}
 e_i=n_i-c_i,\qquad \delta_i=c_i,
\end{equation}
converts node/copy coordinates back to the element-major basis; with the boundary copy bits above,
$0\le e_i\le m-1$ for every valid code (without them, $n_i=m$, $c_i=0$ would give $e_i=m$, outside
the mesh). Together with the validity projectors this makes the map from valid spring-mode codes
to element/local coordinates an isometry at the level of the circuit model.  Thus all penalty-dependent
arithmetic is independent of the number of elements except for $O(L)$-bit endpoint tests and
controlled increments/decrements.

\paragraph{Fixed plane-stress strain/material map.}
The physical BPX gradient circuit produces four Q1-gradient channels,
\begin{equation}
 (\partial_xu_x,\;\partial_yu_x,\;\partial_xu_y,\;\partial_yu_y),
\end{equation}
and a constant-size strain mixer maps them to
\begin{equation}
 \varepsilon=(\partial_xu_x,\ \partial_yu_y,\ \partial_yu_x+\partial_xu_y)^{\mathsf T}.
\end{equation}
For plane stress, in the nondimensional form (divided by $E$),
\begin{equation}
 D/E=\frac{1}{1-\nu^2}
 \begin{bmatrix}1&\nu&0\\\nu&1&0\\0&0&(1-\nu)/2\end{bmatrix}.
\end{equation}
In the normal-strain $+$/$-$ basis its eigenvalues are
$1/(1-\nu)$ and $1/(1+\nu)$, while the shear eigenvalue is
$1/[2(1+\nu)]$.  After normalization by the largest singular value
$\sqrt{1/(1-\nu)}$ of its square root, the material square root therefore has only two nontrivial amplitudes,
\begin{equation}
 \rho_1=\sqrt{\frac{1-\nu}{1+\nu}},\qquad
 \rho_2=\sqrt{\frac{1-\nu}{2(1+\nu)}}.
\end{equation}
A Hadamard changes between $(\varepsilon_{xx},\varepsilon_{yy})$ and the $+$/$-$ basis; two
controlled rotations apply $\rho_1$ and $\rho_2$. These rotations illustrate the constitutive square
root; the normalization argument uses the joint map below. The strain mixer and the constitutive square root
are not encoded as two separate block encodings, whose normalizations would multiply
($\|S\|_2=\sqrt2$); they are encoded jointly as the constant-size map
$M_{\rm el}=(D/E)^{1/2}S$, by one unitary dilation normalized by
$\|M_{\rm el}\|_2=(1-\nu)^{-1/2}$, consistent with Theorem~\ref{M-thm:direct-precond-block}.
Consequently the elasticity extension of the scalar BPX gradient adds only a fixed number of gates
and a fixed number of material-dependent rotation angles, independent of $m$.

\paragraph{Register sizes.}
The Q4 assembly-free space has $8m^2=2^{2L+3}$ rows. The complete BPX-plus-spring frame has
\begin{equation}
 N_F=\frac{26m^2+6m-8}{3}
\end{equation}
columns, while the rectangular factor has $14m^2-2m$ output rows when $R$ is taken as the
square element-local root. Both fit in $2L+4$ data qubits. The strain/quadrature realization
of Fig.~\ref{fig:H-circuit} has $12m^2<2^{2L+4}$ physical output rows (three strains at four
Gauss points per element); the identity rows are distinguished by a spring-row flag qubit. These
are data-register sizes only, not a complete qubit count: the selector register
($\lceil\log_2(L+2)\rceil$ qubits), the validity tests, the output flags and the dilation ancillas
of the rotations and of the transfer network require further qubits. Gate counts, rotation
synthesis and the costs of a complete solver depend on the quantum algorithm and its compilation
and are not analysed here.

\paragraph{Transformed point load.}
For the downward point load at the far corner, the physical node has one element copy and the
spring-mode component is exactly zero.  The BPX-transformed vector has one unit
basis entry per level, so its norm is $\sqrt{L+1}$ and its normalized state is a uniform level
superposition.  The following data are generated directly from the assembled frame matrices.

\begin{table}[ht]
\centering\small
\caption{Validation of the transformed far-corner Q4 point load.  The minimum and maximum
level norms are taken over all BPX levels.}
\label{tab:loadraw}
\begin{tabular}{@{}rrrrr@{}}
\toprule
$m$ & $L$ & $\|F_c^Tf\|_2$ & min. level norm & max. level norm\\
\midrule
2&1&0&1&1\\
4&2&0&1&1\\
8&3&0&1&1\\
16&4&0&1&1\\
32&5&0&1&1\\
\bottomrule
\end{tabular}
\end{table}

For $m=2$ and $4$, direct dense pseudoinverse evaluation of
$\tilde f^T(F^TAF)^+\tilde f$ agrees with $f^TA^{-1}f$ to relative errors
$4.4\times10^{-15}$ and $2.6\times10^{-14}$, respectively.  This validates both the transformed
load and the compliance identity used in the main paper.

\section{Traction benchmark: discretization and penalty accuracy}\label{sec:tractionsupp}
This section gives the discretization and penalty data of the cantilever of
Section~\ref{M-sec:num-penalty} and Table~\ref{M-tab:traction} of the main paper. The benchmark is a unit square cantilever with the
left edge clamped and a uniform downward traction of total magnitude one on the right edge.

\paragraph{Why the point load is not used for continuum accuracy.}
The far-corner point force of Section~\ref{sec:circuitsupp} is useful for algebraic validation because its transformed
load state is exceptionally simple.  In two-dimensional elasticity, however, a point load
introduces a singular field and its discrete compliance keeps changing under mesh refinement.
It is therefore unsuitable for statements such as ``one-percent continuum compliance'' unless a
load regularization is specified.  For the accuracy study we replace it by a uniform
downward traction on the right edge, normalized to total force one.  This does not change the
operator or preconditioner.

The traction load is still quantum-friendly.  On a mesh of width $h=1/m$, the physical nodal
weights on the loaded edge are $h/2$ at the two endpoints and $h$ at the interior nodes.  Equal
sharing among element copies eliminates every spring-mode component exactly.  On a
BPX level with $m_\ell=2^\ell$ elements, restriction of the finest-level consistent traction is
again the consistent traction on that level.  Thus each level state is a one-dimensional edge
state with only an endpoint correction, and the outer level amplitudes are known analytically;
the whole transformed load is preparable with $O(L)$ controlled rotations and $O(L)$-bit
arithmetic, with no $O(N)$ data loading.

\paragraph{Discretization.}
Exactly constrained Q4 computations for $m=4,\ldots,256$ fit
\begin{equation}\label{eq:traction-fit}
 C_h=C_\infty-a m^{-p},
 \qquad
 C_\infty=7.03864,\quad p=1.672,
\end{equation}
when the two coarsest meshes are omitted from the fit. The fitted exponent is an effective,
pre-asymptotic rate. The clamped--free corners of the cantilever have the leading Williams
exponent $\lambda_1=0.758$ (plane stress, $\nu=0.3$), so the asymptotic compliance rate is
$2\lambda_1\approx1.52$; consistently, the local rates of successive differences on
$m=4$--$256$ peak at $1.68$ and decrease to $1.65$, and the rate that includes the $m=512$ solve is
$1.63$. The fitted law and a two-term model with the corner exponent both indicate that $m=32$, 128 and 512
are sufficient for relative discretization errors of the compliance below $2.5\times10^{-3}$,
$2.5\times10^{-4}$ and $2.5\times10^{-5}$, respectively; these are model estimates, not certified
continuum-error bounds;
a direct solve at $m=512$, reported below, is consistent with the last value.

\paragraph{Discretization convergence.}
Table~\ref{tab:e2e-compliance} lists the compliances of the exactly constrained assembled model.

\begin{table}[ht]
\centering\small
\caption{Compliance of the exactly constrained Q4 cantilever under the edge traction (direct
solves), and the directly computed relative difference from the fitted limit
$C_\infty=7.03864429$.}
\label{tab:e2e-compliance}
\begin{tabular}{@{}rrr@{}}
\toprule
$m$ & compliance & direct difference from fitted limit\\
\midrule
4   & 6.61222838 & $6.06\times10^{-2}$\\
8   & 6.90079701 & $1.96\times10^{-2}$\\
16  & 6.99536475 & $6.15\times10^{-3}$\\
32  & 7.02507131 & $1.93\times10^{-3}$\\
64  & 7.03436669 & $6.08\times10^{-4}$\\
128 & 7.03730022 & $1.91\times10^{-4}$\\
256 & 7.03823740 & $5.78\times10^{-5}$\\
512 & 7.03854067 & $1.47\times10^{-5}$\\
\bottomrule
\end{tabular}
\end{table}

A nonlinear fit on $m\ge16$ to $C_h=C_\infty-a m^{-p}$ gives
$C_\infty=7.03864429$ and $p=1.6718848$.  The local rates
$\log_2[(C_{2m}-C_m)/(C_{4m}-C_{2m})]$ for $m=4,8,16,32,64$ (meshes up to $m=256$) are $1.610$,
$1.671$, $1.676$, $1.664$ and $1.646$, and $1.628$ for $m=128$ with the $m=512$ value of
Table~\ref{tab:e2e-compliance}, so the fitted exponent is pre-asymptotic.  The leading Williams exponent of the clamped--free right-angle
corner (plane stress, $\nu=0.3$; Kolosov--Muskhelishvili potentials $\Phi=Az^\lambda$,
$\Psi=Bz^\lambda$, traction-free on $\theta=0$, clamped on $\theta=\pi/2$) is
$\lambda_1=0.7583$; the numerical root-finding procedure also reproduces $0.5445$ and $0.9085$ for the free--free
$270^\circ$ corner. For a wedge of opening angle $\theta_0$ with one face clamped and the other
traction-free, the exponents of the displacement field $u\propto r^\lambda$ are the roots of
\citep{Williams1952}
\begin{equation}\label{eq:williams}
\kappa^2+1+2\kappa\cos(2\lambda\theta_0)-4\lambda^2\sin^2\theta_0=0,\qquad
\kappa=\frac{3-\nu}{1+\nu}\ \ \text{(plane stress)},
\end{equation}
and $\lambda_1$ is the root with the smallest positive real part. For $\theta_0=\pi/2$ and
$\nu=0.3$ it is real, $\lambda_1=0.7583$; the next roots are complex,
$1.719\pm0.448\,\mathrm i$. For $\theta_0=\pi$ Eq.~\eqref{eq:williams} gives the known
$\lambda=\frac12\pm\mathrm i\ln\kappa/(2\pi)$. The corner geometry is that of the cantilever, i.e.\
the two ends of the clamped edge, where a clamped and a traction-free face meet at a right angle;
the stresses there behave as $r^{\lambda_1-1}$.  The asymptotic compliance rate is therefore $2\lambda_1=1.517$.  The
two-term model $C_\infty-a m^{-2\lambda_1}-b m^{-2}$ fitted to $m=64,128,256$ gives
$C_\infty=7.03870$. Table~\ref{tab:e2e-models} compares the two error models.

\begin{table}[ht]
\centering\small
\caption{Relative discretization error of the compliance predicted by the convergence models:
pre-asymptotic fitted law and two-term model with the corner exponent $2\lambda_1=1.517$. These are
model predictions, not direct differences (cf.\ Table~\ref{tab:e2e-compliance}; at $m=512$ the direct
difference is $1.47\times10^{-5}$ against the fitted-law prediction $1.87\times10^{-5}$).}
\label{tab:e2e-models}
\begin{tabular}{@{}rcc@{}}
\toprule
$m$ & fitted law & two-term (corner exponent)\\
\midrule
32 & $1.93\times10^{-3}$ & $1.95\times10^{-3}$\\
64 & $6.06\times10^{-4}$ & $6.15\times10^{-4}$\\
128 & $1.90\times10^{-4}$ & $1.98\times10^{-4}$\\
256 & $5.97\times10^{-5}$ & $6.51\times10^{-5}$\\
512 & $1.87\times10^{-5}$ & $2.17\times10^{-5}$\\
1024 & $5.88\times10^{-6}$ & $7.33\times10^{-6}$\\
\bottomrule
\end{tabular}
\end{table}

The conforming model was also solved directly at $m=512$ with an independent implementation
: the compliance is $7.03854067$, and the relative
discretization error estimated with the corner exponent is $2.1$--$2.3\times10^{-5}$, in agreement
with the corner-exponent column of Table~\ref{tab:e2e-models}.

\paragraph{Penalty error and the compliance bound.}
Let $C(\alpha,\beta)$ be the assembly-free compliance, $C_\gamma=C(\gamma,\gamma)$, and $C_0^{(h)}$ the
conforming compliance on the same mesh, and $e_\gamma=(C_\gamma-C_0^{(h)})/C_0^{(h)}$. The product
$\gamma e_\gamma$ is almost constant: at $m=32$ the pairs $(\gamma,\gamma e_\gamma)$ are
\[
(10,0.47485),\ (20,0.47496),\ (50,0.47503),\ (100,0.47506),
\]
\[
(200,0.47507),\ (500,0.47508),\ (1000,0.47508),
\]
and across $m=16,32,64$ the largest observed coefficient is below $0.489$.

The first-order coefficient follows from the conforming solution alone. With $\varepsilon=1/\gamma$,
the Schur complement of the spring modes (Section~\ref{sec:penextra}) gives
$C_\gamma=C_0^{(h)}+\varepsilon\,\mathbf w^{\mathsf T}\hat A_{\rm s}^+\mathbf w+O(\varepsilon^2)$, where
$\mathbf w=A_{\rm el}\mathbf u_0$ is the element force of the conforming solution in copy form and
$\hat A_{\rm s}=A_{\rm s}/\gamma$; $\hat A_{\rm s}^+$ is applied exactly by the node-local
sum/difference transformation. For the consistently distributed copy load of this traction
benchmark, $\mathbf f\in\ker A_{\rm s}=\operatorname{range}J_0$ (equal sharing among the copies, no
load on clamped copies), and consequently $A_{\rm s}^+\mathbf f=\mathbf 0$. Hence
$\mathbf w^{\mathsf T}\hat A_{\rm s}^+\mathbf w=\gamma\,\mathbf r^{\mathsf T}A_{\rm s}^+\mathbf r$
with the residual $\mathbf r=A_{\rm el}\mathbf u_0-\mathbf f$ of
Proposition~\ref{M-prop:penalty-bound}, and the first-order term equals the bound of the main
paper. This is a property of the benchmark load, not a restriction of the formulation: for an
element-local load with a nonzero spring-mode component, $\mathbf w$ is replaced by $\mathbf r$,
and the first-order term is still $\varepsilon\,\mathbf r^{\mathsf T}\hat A_{\rm s}^+\mathbf r$. The coefficient $c_1=\mathbf w^{\mathsf T}\hat A_{\rm s}^+\mathbf w/C_0^{(h)}$ reproduces
the coefficients of Table~\ref{M-tab:traction} of the main paper ($0.488$, $0.475$, $0.468$,
$0.465$ for $m=16$--$128$) and gives
\[
c_1=0.47508\ (m=32),\quad 0.46473\ (m=128),\quad 0.46297\ (m=256),\quad 0.46208\ (m=512).
\]
Direct solves of the assembly-free system agree with $c_1$: $\gamma e_\gamma=0.47507$, $0.47508$,
$0.47508$ at $m=32$ for $\gamma=196$, 1{,}960 and 19{,}600, and $0.46471$, $0.46451$ at $m=128$ for
$\gamma=1{,}960$ and 19{,}600 (the last value is limited by the conditioning of the unpreconditioned
direct solve). For this traction benchmark the second-order term is therefore negligible for
$\gamma\ge10$.

Table~\ref{tab:penbound} tests the two-sided bound
$0\le C(\alpha,\beta)-C_0^{(h)}\le\mathbf r^{\mathsf T}A_{\rm s}^+\mathbf r$ of
Proposition~\ref{M-prop:penalty-bound} on the same cantilever, including soft and unequal springs.
The bound holds in every case and is sharp: the computed error is $99.1$--$99.95\%$ of it, and the
ratio tends to one as both penalties are increased proportionally (for a nonzero leading error
coefficient), as the first-order expansion predicts.

\begin{table}[ht]
\centering\small
\caption{Relative penalty error $e=(C(\alpha,\beta)-C_0^{(h)})/C_0^{(h)}$ of the Q4 cantilever under the
edge traction, the bound $b=\mathbf r^{\mathsf T}A_{\rm s}^+\mathbf r/C_0^{(h)}$ of
Proposition~\ref{M-prop:penalty-bound}, and their ratio.}
\label{tab:penbound}
\begin{tabular}{@{}rrrrrr@{}}
\toprule
$m$ & $\alpha$ & $\beta$ & $e$ & $b$ & $e/b$\\
\midrule
8  & 0.5 & 0.5  & $1.0136$ & $1.0231$ & 0.9907\\
8  & 2   & 2    & $2.548\times10^{-1}$ & $2.558\times10^{-1}$ & 0.9963\\
8  & 10  & 10   & $5.111\times10^{-2}$ & $5.116\times10^{-2}$ & 0.9991\\
8  & 7.3 & 11.1 & $6.433\times10^{-2}$ & $6.439\times10^{-2}$ & 0.9991\\
8  & 100 & 3    & $4.377\times10^{-2}$ & $4.404\times10^{-2}$ & 0.9940\\
16 & 0.5 & 0.5  & $9.699\times10^{-1}$ & $9.762\times10^{-1}$ & 0.9936\\
16 & 2   & 2    & $2.434\times10^{-1}$ & $2.441\times10^{-1}$ & 0.9974\\
16 & 10  & 10   & $4.878\times10^{-2}$ & $4.881\times10^{-2}$ & 0.9994\\
16 & 7.3 & 11.1 & $6.390\times10^{-2}$ & $6.394\times10^{-2}$ & 0.9994\\
16 & 100 & 3    & $2.469\times10^{-2}$ & $2.483\times10^{-2}$ & 0.9944\\
32 & 0.5 & 0.5  & $9.453\times10^{-1}$ & $9.502\times10^{-1}$ & 0.9949\\
32 & 2   & 2    & $2.371\times10^{-1}$ & $2.375\times10^{-1}$ & 0.9980\\
32 & 10  & 10   & $4.749\times10^{-2}$ & $4.751\times10^{-2}$ & 0.9995\\
32 & 7.3 & 11.1 & $6.356\times10^{-2}$ & $6.360\times10^{-2}$ & 0.9995\\
32 & 100 & 3    & $1.478\times10^{-2}$ & $1.484\times10^{-2}$ & 0.9958\\
\bottomrule
\end{tabular}
\end{table}

\paragraph{Transformed traction load.}
The consistent edge traction is copy-compatible, so its spring-mode component
vanishes.  The norms of its BPX level blocks obey
\begin{equation}
 \|f_\ell\|_2^2=\frac{2^{\ell+1}-1}{2^{2\ell+1}},
 \qquad \ell=0,\ldots,L,
\end{equation}
which is reproduced by the explicit frame matrices; hence $\|F^{\mathsf T}f\|_2^2\to\tfrac43$.
Table~\ref{tab:e2e-load} lists the norms of the transformed load together with the smallest
positive eigenvalue of the preconditioned operator at $\gamma=10$; both are inputs to the
analysis of a particular quantum solver.

\begin{table}[ht]
\centering\small
\caption{Norm of the transformed edge-traction load and smallest positive eigenvalue of the
preconditioned operator ($\gamma=10$).}
\label{tab:e2e-load}
\begin{tabular}{@{}rrr@{}}
\toprule
$m$ & $\|F^{\mathsf T}f\|_2$ & $\lambda_{\min}^+$\\
\midrule
2  & 0.935414 & 0.1020\\
4  & 1.045825 & 0.1048\\
8  & 1.100426 & 0.1065\\
16 & 1.127601 & 0.1074\\
32 & 1.141160 & 0.1079\\
64 & 1.147932 & 0.1081\\
\bottomrule
\end{tabular}
\end{table}

\bibliographystyle{unsrtnat}
\bibliography{refs}